%% file: clt.tex
\documentclass{article}

\usepackage{amsfonts}
\usepackage{amsbsy,amssymb,amsmath}
\usepackage{url}
\usepackage{color}
\usepackage{mathpartir}
\usepackage{isabelle,isabellesym}

\usepackage{amsthm}
\newtheorem*{theorem}{Theorem}
\newtheorem*{definition}{Definition}

\usepackage[draft]{fixme}
\usepackage[usenames]{xcolor}
\usepackage{comment}

\newcommand{\NN}{\mathbb{N}}
\newcommand{\ZZ}{\mathbb{Z}}

\newcommand{\RR}{\mathbb{R}}
\newcommand{\CC}{\mathbb{C}}
\newcommand{\ennRR}{\overline{\mathbb{R}}_{\ge 0}}
\newcommand{\eRR}{\overline{\mathbb{R}}}

\newcommand{\fn}[1]{\mathrm{#1}} 
\newcommand{\mdl}[1]{{\mathcal #1}} 
\newcommand{\ph}{\varphi}

\newcommand{\sinc}{\mathop{\fn{sinc}}\nolimits}

\fxsetup{targetlayout=colorcb}
\fxsetface{target}{\itshape}
\fxsetface{margin}{\tiny \color{red}}
\definecolor{fxnote}{rgb}{1,0,0}

\newcommand{\Snippet}[1]{\csname snippet--#1\endcsname}

\newcommand{\DefineSnippet}[2]{%
  \expandafter\newcommand\csname snippet--#1\endcsname{%
    \begin{quote}
    \begin{isabelle}
    #2
    \end{isabelle}
    \end{quote}}}

\def\isadelimproof{}
\def\endisadelimproof{}
\def\isatagproof{}
\def\endisatagproof{}

\input{snippets-generated}

\begin{document}

\title{A formally verified proof of the \\ Central Limit Theorem}


\author{Jeremy Avigad, Johannes H\"olzl, and Luke Serafin}


\maketitle

\begin{abstract}
We describe a proof of the Central Limit Theorem that has been formally verified in the Isabelle proof assistant. Our formalization builds upon and extends Isabelle's libraries for analysis and measure-theoretic probability. The proof of the theorem uses \emph{characteristic functions}, which are a kind of Fourier transform, to demonstrate that, under suitable hypotheses, sums of random variables converge weakly to the standard normal distribution. We also discuss the libraries and infrastructure that supported the formalization, and reflect on some of the lessons we have learned from the effort.
\end{abstract}

\section{Introduction}
\label{section:introduction}

If you roll a fair die many times and compute the average number of spots showing, the result is likely to be close to 3.5, and the odds that the average is far from the expected value decreases roughly as the area under the familiar bell-shaped curve. Something similar happens if the measurement is continuous rather than discrete, such as when you repeatedly toss a needle on the ground and measure the angle it makes with respect to a fixed reference line. Even if the die is not a fair die or the geometry of the needle and the ground makes some angles more likely than others, the distribution of the average still approaches the area under a bell-shaped curve centered on the expected value. The width of the bell depends on both the variance of the random measurement and the number of times it is performed. Made precise, this amounts to a statement of the Central Limit Theorem.

The Central Limit Theorem lies at the heart of modern probability. Many generalizations and variations have been studied, some of which either relax the requirement that the repeated measurements are independent of one another and identically distributed (cf.~in particular, the results of Lyapunov and Lindberg \cite{billingsley:95}), while others provide additional information on the rate of convergence.

Here we report on a formalization of the Central Limit Theorem that was carried out in the Isabelle proof assistant. This result is noteworthy for a number of reasons. Not only is the CLT fundamental to probability theory and the study of stochastic processes, but so is the machinery developed to prove it, ranging from ordinary calculus to the properties of real distributions and characteristic functions. There is a pragmatic need to subject statistical claims made in engineering, risk analysis, and financial computation to formal verification, and our formalization along with the surrounding infrastructure can support such practical efforts.

The formalization is also a good test for Isabelle's libraries, proof language, and automated reasoning tools. As we will make clear, the proof draws on a very broad base of facts from analysis, topology, measure theory, and probability theory, providing a useful evaluation of the robustness and completeness of the supporting libraries. Moreover, the concepts build on one another. For example, a measure is a function from a class of sets to the reals, and reasoning about convergence of measures involves reasoning about sequences of such functions. The operation of forming the characteristic function is a functional taking a measure to a function from the reals to the complex numbers, and the convergence of such functionals is used to deduce convergence of measures. The conceptual underpinnings are thus as deep as they are broad, and working with them tests Isabelle's mechanisms for handling abstract mathematical notions.

In Section~\ref{section:overview} we provide an overview of the Central Limit Theorem and the proof that we formalized, following the textbook presentation of Billingsley~\cite{billingsley:95}. In Section~\ref{section:isabelle} we describe the Isabelle proof assistant, and the parts of the library that supported our formalization. In Section~\ref{section:formal} we describe the formal proof itself, and in Section~\ref{section:reflections} we reflect on what we have learned from the effort.

Our formalization is currently part of the Isabelle library, which can be found online at \url{https://isabelle.in.tum.de/}.\footnote{The probability library in particular can be found at\\ \url{https://isabelle.in.tum.de/dist/library/HOL/HOL-Probability/index.html}.} A preliminary, unpublished report on the formalization can be found on arXiv \cite{avigad:hoelzl:serafin:14}. Our presentation also draws heavily on Serafin's Carnegie Mellon MS thesis \cite{serafin:15}, which provides additional information.

\emph{Acknowledgments.} We are grateful to Tobias Nipkow, Lawrence Paulson, Makarius Wenzel, and the entire Isabelle team for the ongoing development of Isabelle. We are especially grateful to Tobias for steadfast encouragement and support. We thank our two anonymous referees for a very careful reading and helpful comments. Avigad and Serafin's work has been partially supported by NSF grant DMS-1068829, and Avigad's work has been partially supported by AFOSR grants FA9550-12-1-0370 and FA9550-15-1-0053. H\"olzl's work has been partially supported by DFG projects Ni 491/15-1 and Ni 491/16-1.

\section{Overview of the Central Limit Theorem}
\label{section:overview}

For our formalization we followed Billingsley's textbook, \emph{Probability and Measure} \cite{billingsley:95}, which provides an excellent introduction to these topics. Here we provide some historical background, briefly review the key concepts, give a precise statement of the Central Limit Theorem, and present an outline of the proof.

\subsection{Historical background}
\label{subsection:historical}

In 1733, De Moivre privately circulated a proof that, as $n$ approaches infinity, the distribution of $n$ flips of a fair coin converges to a normal distribution. This material was later published in the 1738 second edition of his book {\em The Doctrine of Chances,} the first edition of which was published in 1712. That book is widely regarded as the first textbook on probability theory. De Moivre also considered the case of what we would call a biased coin, that is, an event which has value one with probability $p$ and zero with probability $1-p$ for some $p \in (0,1)$. He showed that his convergence theorem continues to hold in that case.

De Moivre's result was generalized by Laplace in the period between about 1776 and 1812 to sums of random variables with various other distributions, such as the uniform distribution on an interval. Over the next three decades Laplace developed conceptual and analytical tools to extend this convergence theorem to sums of independent and identically distributed random variables with ever more general distributions, and this work culminated in his treatise {\em Th\'eorie analytique des probabilit\'es}. This included the development of the method of characteristic functions to study the convergence of sums of random variables, a move which firmly established the usefulness of analytic methods in probability theory.

Laplace's theorem later became known as the Central Limit Theorem, a designation due to P\'olya, stemming from its importance both in the theory and applications of probability. In modern terms, the theorem states that the normalized sum of a sequence of independent and identically distributed random variables with finite, nonzero variance converges to a normal distribution. All of the main ingredients of the proof of the CLT are present in the work of Laplace, though of course the theorem was refined and extended as probability underwent the radical changes necessitated by its move to measure-theoretic foundations in the first half of the twentieth century.

Gauss was one of the first to recognize the importance of the normal distribution to the estimation of measurement errors. The usefulness of the normal distribution in this context is largely a consequence of the Central Limit Theorem, since errors occurring in practice are frequently the result of many independent factors which sum to an overall error in a way which can be regarded as approximated by a sum of independent and identically distributed random variables. The normal distribution also arose with surprising frequency in a wide variety of empirical contexts, from the heights of men and women to the velocities of molecules in a gas. This gave the CLT the character of a natural law, as seen in the following poetic quote from Sir Francis Galton in 1889 \cite{galton:89}:
\begin{quote}
 I know of scarcely anything so apt to impress the imagination as the wonderful form of cosmic order expressed by the ``Law of Frequency of Error.'' The law would have been personified by the Greeks and deified, if they had known of it. It reigns with serenity and in complete self-effacement, amidst the wildest confusion. The huger the mob, and the greater the apparent anarchy, the more perfect is its sway. It is the supreme law of Unreason. Whenever a large sample of chaotic elements are taken in hand and marshaled in the order of their magnitude, an unsuspected and most beautiful form of regularity proves to have been latent all along.
\end{quote}
More details on the history of the Central Limit Theorem and its proof can be found in \cite{fischer:11}.

\subsection{Background from measure theory}
\label{subsection:background}

A \emph{measure space} $(\Omega, \mdl F)$ consists of a set $\Omega$ and a \emph{$\sigma$-algebra $\mdl F$ of subsets of $\Omega$}, that is, a collection of subsets of $\Omega$ containing the empty set and closed under complements and countable unions. Think of $\Omega$ as the set of possible states of affairs, or possible outcomes of an action or experiment, and each element $E$ of $\mdl F$ as representing the set of states or outcomes in which some \emph{event} occurs --- for example, that a card drawn is a face card, or that Spain wins the World Cup. A \emph{probability measure} $\mu$ on this space is a function that assigns a value $\mu(E)$ in $[0, 1]$ to each event $E$, subject to the following conditions:
\begin{enumerate}
 \item $\mu(\emptyset) = 0$,
 \item $\mu(\Omega) = 1$, and
 \item $\mu$ is countably additive: if $(E_i)$ is any sequence of disjoint events in $\mdl F$, then $\mu(\bigcup_i E_i) = \sum_i \mu(E_i)$.
\end{enumerate}
Intuitively, $\mu(E)$ is the ``probability'' that $E$ occurs.

The collection $\mdl B$ of \emph{Borel subsets} of the real numbers is the smallest $\sigma$-algebra containing all intervals $(a, b)$. A \emph{random variable} $X$ on the measure space $(\Omega, \mdl F)$ is a measurable function from $(\Omega, \mdl F)$ to $(\RR, \mdl B)$. Saying $X$ is measurable means that for every Borel subset $B$ of the real numbers, the set $\{ \omega \in \Omega \; | \; X(\omega) \in B \}$ is in $\mdl F$. Think of $X$ as some real-valued measurement that one can perform on the outcome of the experiment, in which case, the measurability of $X$ means that if we are given any probability measure $\mu$ on $(\Omega, \mdl F)$, then for any Borel set $B$ it makes sense to talk about ``the probability that $X$ is in $B$.'' In fact, if $X$ is a random variable, then any measure $\mu$ on $(\Omega, \mdl F)$ gives rise to a measure $\nu$ on $(\RR, \mdl B)$, defined by $\nu(B) = \mu ( \{ \omega \in \Omega \; | \; X (\omega) \in B \})$. A probability measure on $(\RR, \mdl B)$ is called a \emph{real distribution}, or, more simply, a \emph{distribution}, and the measure $\nu$ just described is called \emph{the distribution of $X$}.

If $X$ is a random variable, the \emph{mean} or \emph{expected value} of $X$ with respect to a probability measure $\mu$ is $\int X d\mu$, the integral of $X$ with respect to $\mu$. If $c$ is the mean, the \emph{variance} of $X$ is $\int (X - c)^2 d\mu$, a measure of how far, on average, we should expect $X$ to be from its mean.

Note that in passing from $X$ to its distribution $\nu$ (with respect to $\mu$), instead of worrying about the probability that some abstract event occurs, we focus more concretely on the probability that some measurement on the outcome lands in some set of real numbers. In fact, many theorems of probability theory do not really depend on the abstract space $(\Omega, \mdl F)$ on which $X$ is defined, but rather the associated distribution on the real numbers. Nonetheless, it is often more intuitive and convenient to think of the real distribution as being the distribution of a random variable (and, indeed, any real distribution can be represented that way).

One way to define a real distribution is in terms of a \emph{density}. For example, in the case where $\Omega = \{1, 2, 3, 4, 5, 6\}$, we can specify a probability on all the subsets of $\Omega$ by specifying the probability of each of the events $\{1\}, \{2\}, \ldots, \{6\}$. More generally, we can specify a distribution $\mu$ on $\RR$ by specifying a function $f$ such that for every interval $(a, b)$, $\mu((a, b)) = \int_a^b f(x) \; \mathit{dx}$. The measure $\mu$ is then said to be the real distribution with density $f$. In particular, the \emph{normal distribution} with mean $c$ and variance $\sigma^2$ is defined to be the real distribution with density function
\[
f(x) = \frac{1}{\sigma \sqrt{2 \pi}} e^\frac{-(x - c)^2}{2 \sigma^2}.
\]
The graph of $f$ is the bell-shaped curve centered at $c$. When $c = 0$ and $\sigma = 1$, the associate real distribution is called the \emph{standard normal distribution}.

Let $X_0, X_1, X_2, \ldots$ be any sequence of independent random variables, each with the same distribution $\mu$, mean $c$, and variance $\sigma^2$. Here ``independent'' means that the random variables $X_0, X_1, \ldots$ are all defined on the same measure space $(\Omega, \mdl F)$, but they represent independent measurements, in the sense that for any finite sequence of events $B_1, B_2, \ldots, B_k$ and any sequence of distinct indices $i_1, i_2, \ldots, i_k$, the probability that $X_{i_j}$ is in $B_j$ for each $j$ is just the product of the individual probabilities that $X_{i_j}$ is in $B_j$. For each $n$, let $S_n = \sum_{i < n} X_i$ be the sum of the first $n$ random variables in the sequence. Notice that each $S_n$ is itself a measurable function on $(\Omega, \mdl F)$ (which is to say it is a random variable), and so it is natural to ask how its values are distributed. We can shift the expected value of $S_n$ to $0$ by subtracting $n c$, and scale the variance to $1$ by dividing by $\sqrt{ n \sigma^2}$. The Central Limit Theorem says that the corresponding quantity,
\[
 \frac{S_n - nc}{\sqrt{n \sigma^2}},
\]
approaches the standard normal distribution as $n$ approaches infinity.

All that remains to do is to make sense of the assertion that a sequence of distributions $\mu_0, \mu_1, \mu_2, \ldots$ ``approaches'' a distribution, $\mu$. For distributions that are defined in terms of densities, the intuition is that over time the graph of the density should look more and more like the graph of the density of the limit. For example, if you flip a coin a number of times and graph all the possible values of the average number of ones, the discrete points plotted over the possibilities $0, 1/n, 2/n, 3/n, \ldots, 1$ start to look like a bell-shaped curve centered on $1 / 2$. The notion of \emph{weak convergence} makes the notion of ``starts to look like'' precise.

If $\mu$ is any real distribution, then the function $F_\mu(x) = \mu((-\infty, x])$ is called the \emph{cumulative distribution function} of $\mu$. In words, for every $x$, $F_\mu(x)$ returns the likelihood that a real number chosen randomly according to the distribution is at most $x$. Clearly $F_\mu(x)$ is nondecreasing, and it is not hard to show that $F_\mu$ is right continuous, approaches $0$ as $x$ approaches $-\infty$, and approaches $1$ as $x$ approaches $\infty$. Conversely, one can show that any such function is the cumulative distribution function of a unique measure. Thus there is a one-to-one correspondence between functions $F$ satisfying the properties above and real distributions.

The notion of weak convergence can be defined in terms of the cumulative distribution function:
\begin{definition}
 Let $(\mu_n)$ be a sequence of real distributions, and let $\mu$ be a real distribution. Then \emph{$\mu_n$ converges weakly to $\mu$}, written $\mu_n \Rightarrow \mu$, if $F_{\mu_n}(x)$ approaches $F_\mu(x)$ at each point $x$ where $F_\mu$ is continuous.
\end{definition}

To understand why we need to exclude the points of discontinuity of $F_\mu$, consider for each $n$ the probability measure $\mu_n$ that puts all its ``weight'' on $1 / n$, which is to say, for any Borel set $B$, $\mu(B) = 1$ if and only if $B$ contains $1 / n$. Then $F_{\mu_n}$ is the function that jumps from $0$ to $1$ at $1 / n$. Intuitively, it makes sense to say that $\mu_n$ approaches the real distribution $\mu$ that puts all its weight at $0$. But for every $n$, $F_{\mu_n}(0) = 0$, while $F_\mu(0) = 1$, which explains why we want to exclude the point $0$ from consideration. Notice that since $F_\mu$ is a monotone function, it can have at most countably many points of discontinuity, so we are excluding only countably many points.

The fact that weak convergence is a robust notion is evidenced by the fact that it has a number of equivalent characterizations, as discussed in Section~\ref{subsection:weak:convergence} below.

With this background in place, we can now state the Central Limit Theorem precisely, as follows:
\begin{theorem}
\label{theorem:clt}
Let $X_0, X_1, X_2, \ldots$ be a sequence of independent random variables with mean $c$, strictly positive variance $\sigma^2$, and common distribution $\mu$. Let $S_n = X_0 + X_1 + \ldots + X_{n-1}$. Then the distribution of $(S_n - n c) / \sqrt{n \sigma^2}$ converges weakly to the standard normal distribution.
\end{theorem}
This is Theorem 27.1 in Billingsley's book \cite{billingsley:95}. Our formulation
in Isabelle is as follows:
\Snippet{clt}
Here, \isa{M} denotes the underlying probability space, that is, a triple $(\Omega,\mathcal F,\mu)$ with the requisite properties. We present a formal proof of the mean zero case in the appendix to this paper, and then derive the version above as a corollary (applying the mean zero case to the shifted random variables $X_i - c$).

\subsection{An overview of the proof}
\label{subsection:overview}

Contemporary proofs of the Central Limit Theorem rely on the use of \emph{characteristic functions}, a powerful method that dates back to Laplace. If $\mu$ is a real-valued distribution, its characteristic function $\ph(t)$ is defined by
\[
\ph(t) = \int_{-\infty}^{\infty} e^{itx} \mu(dx).
\]
In words, $\ph(t)$ is the integral of the function $f(x) = e^{itx}$ over the whole real line, with respect to the measure $\mu$. Notice that for each $t \neq 0$, the function $e^{itx}$ is periodic with period $2 \pi / t$. It might be helpful to think of $e^{itx}$ as like a sine or cosine; indeed, $e^{itx}= \cos (t x) + i \sin (t x)$. Notice that $\ph(0)$ is equal to $1$, the measure of the entire real line. The characteristic function of a real distribution $\mu$ is a Fourier transform of the measure $\mu$, and when $t \neq 0$, $\ph(t)$ ``detects'' periodicity in the way that the real distribution $\mu$ distributes its ``weight'' over different parts of the real line.

A key property of characteristic functions is the fact that if $X_1$ and $X_2$ are independent random variables, then the characteristic function of $X_1 + X_2$ is the product of the characteristic function of $X_1$ and the characteristic function of $X_2$. Of course, this extends to sums with any finite number of terms, and the resulting products are often convenient to work with.

The \emph{L\'evy Uniqueness Theorem} asserts that if $\mu_1$ and $\mu_2$ have the same characteristic function, then $\mu_1 = \mu_2$. In other words, a measure $\mu$ can be ``reconstructed'' from its characteristic function, and the characteristic function of a measure determines the measure uniquely.  Let $(\mu_n)$ be a sequence of distributions, where each $\mu_n$ has characteristic function $\ph_n$, and let $\mu$ be a distribution with characteristic function $\ph$. The \emph{L\'evy Continuity Theorem} states that $\mu_n$ converges to $\mu$ weakly if and only if $\ph_n(t)$ converges to $\ph(t)$ for every $t$.

Remember that the CLT asserts that if $(X_n)$ is a sequence of random variable satisfying certain hypotheses, and $\mu_n$ is for each $n$ a certain distribution defined in terms of $X_1, \ldots, X_n$, then $\mu_n$ converges weakly to the standard normal distribution. The L\'evy Continuity Theorem provides a straightforward strategy to prove the theorem: if we let $\ph_n$ denote the characteristic function of $\mu_n$ for each $n$, we need only show that $\ph_n$ approaches the characteristic function of the standard normal distribution pointwise.

Implementing this strategy requires two key ingredients. First, one needs to know that the characteristic function of the standard normal distribution is $\ph(t) = e^{-t^2/2}$. Second, one needs to compute the characteristic functions of the distributions $\mu_n$, which are defined in terms of finite sums of the independent random variables $X_0, X_1, \ldots$, and show that they have the desired behavior. This is where the key property of characteristic functions of sums of independent random variables comes into play.

Once all these components were in place, putting the pieces together was not hard. Given the continuity theorem, the characteristic function of the standard normal distribution, the result on the characteristic functions of sums of random variables, and suitable approximations to the complex exponential function, the proof of the Central Limit Theorem is quite short. In our formalization, it is only about 120 lines long, and is presented in full in the appendix.

\section{Isabelle and its libraries}
\label{section:isabelle}

When we began our project, a good deal of infrastructure was already available in the Isabelle libraries, but we had to add to it substantially. The formalization thus provided a stress test, allowing us to fill in gaps in the library and ensure its practical efficacy. In this section, we will describe those features of Isabelle and its libraries that were most relevant to the formalization, and indicate some of our contributions to the latter.

\subsection{The Isabelle proof assistant}

The Isabelle proof assistant \cite{nipkow:et:al:02} is based on classical simple type theory \cite{church:40}, with variables ranging polymorphically over types, and a Hilbert choice operator (\isa{SOME}) which returns an indeterminate element satisfying a given predicate, if there is one. Given a type $\alpha$ and a predicate $P$ on $\alpha$, one can introduce a new abstract type representing the elements of $\alpha$ satisfying $P$, using a \isa{typedef} command. In addition to the references given in this section, one should consult the documentation available on the Isabelle web site\footnote{\texttt{https://isabelle.in.tum.de/}} for the most up-to-date information.

Isabelle is an LCF-style theorem prover. This means that stating a theorem amounts to introducing a proof goal, and one can then construct proofs by applying \emph{tactics} that reduce that goal to other goals. Layered on top of that, the Isabelle system includes the \emph{Isar} proof language \cite{wenzel:02}, which provides a natural, declarative way of writing structured proofs. Although in some places our proofs resort to sequences of tactic applications, for the most part we relied on Isar to make our proofs more robust, readable, and maintainable. 

One attractive feature of Isabelle is the strength of its automation. We relied extensively on built-in procedures such as its term rewriter (\isa{simp}), its generic theorem provers (such as \isa{auto}), and its procedure for linear arithmetic (\isa{arith}). Occasionally we relied on Isabelle's \emph{sledgehammer} command \cite{paulson:10}, which invokes external theorem provers and then reconstructs the results in Isabelle.

Isabelle has two mechanisms for reasoning algebraically and generically. The first, axiomatic type classes \cite{wenzel:97}, constitutes a conservative extension of the axiomatic framework. Type variables are allowed to range over types with associated functions and relations, satisfying specified axioms. Theorems can be proved generically within a type class, and then instantiated to concrete structures that have been shown to satisfy the given axioms. This is used, for example, to develop facts about arithmetic, sums, products, and orderings that are shared among various number classes, including the natural numbers, integers, rationals, reals, and complex numbers. In the Isabelle library, they are also used to associate a topological structure to a type. Thus, our reasoning about topological aspects of the reals, described in Section~\ref{subsection:topology} below, made use of the associated type classes. There are also classes for various types of normed spaces, of which the reals, finite powers of the reals, and the complex numbers are instances. We describe our use of this structure on the real and complex numbers in Section~\ref{subsection:real:analysis}.

Type classes are limited by the fact that they can only be parameterized by a single type parameter. An even more serious limitation is that, in simple type theory, types cannot depend on elements of other types. For example, there is no way of using type classes to reason about $\ZZ_n$, the integers modulo a parameter $n$, as an instance of a ring. For that purpose, Isabelle has a more flexible mechanism, \emph{locales} \cite{ballarin:06}, which, however, cannot take advantage of all of the benefits of the ambient type theory. Locales do \emph{not} constitute an axiomatic extension; in terms of the underlying logic, a locale is nothing more than a predicate on some data. But Isabelle provides mechanisms for reasoning ``in'' such a locale, that is, fixing some data and the locale assumptions, and reasoning on that basis. Locales can also introduce notation that implicitly depends on the locale parameters. Isabelle provides mechanisms for instantiating locales, either with fixed types, or on the fly in a proof: one shows (typically with the help of automation) that some data satisfy the locale axioms, at which point all the definitions and theorems of the locale are made available, pre-instantiated with the relevant data and facts. For example, once we show that a relation satisfies the axioms of the \isa{partial\_order} locale, we can use notation and facts about partial orders freely for that relation, without having to repeatedly cite the fact that the relation is a partial order.

In the measure theory library, as described in Sections~\ref{subsection:measure:theory} and \ref{subsection:probability}, locales are used to reason about algebras and $\sigma$-algebras, for example. Moreover, once the type of measurable spaces has been introduced, locales are used to introduce extra hypotheses, for example, the hypothesis that a measure space is finite, or is a probability space.

\subsection{Topology and Limits}
\label{subsection:topology}

Isabelle's extensive library for topological spaces includes properties of open and closed sets, limits, compactness, continuity, and so on. The library is described in detail by H\"olzl, Immler, and Huffman in \cite{hoelzl:et:al:13}. Topological notions interact with measure-theoretic notions in various ways. For example, a real distribution is a measure on the real numbers that measures the \emph{Borel sets}, the smallest $\sigma$-algebra containing the open sets. Continuous functions are therefore measurable. Topological notions come into the statements of many measure-theoretic theorems described below, notions including the points of continuity of a function or the boundary of a set. Proving Skorohod's theorem required showing that the set of points of continuity of an arbitrary function from reals to reals is Borel; this is done in a four-line footnote in Billingsley (\cite[page 334]{billingsley:95}), and requires characterizing the set of discontinuities as a union of an intersection of open sets.

Conventional reasoning about limits was ubiquitous in our formalization. Everyday mathematics requires one to deal with expressions such as the following:
\begin{itemize}
 \item[$\bullet$] $\lim_{x \to a} f(x) = b$
 \item[$\bullet$] $\lim_{n \to \infty} a_n = a$
 \item[$\bullet$] $\lim_{x \to \infty} f(x) = b$
 \item[$\bullet$] $\lim_{x \to a^-} f(x) = b$
 \item[$\bullet$] $\lim_{x \to a} f(x) = \infty$
\end{itemize}
Here, the source and target spaces can be any topological space, including metric spaces or the natural numbers with the order topology. One can consider limits as $x$ approaches a value $a$, or $\infty$, or $-\infty$. One can also restrict the allowed values for $x$ and consider the limit as $x$ approaches $a$ within a set $s$; saying $x$ approaches $a$ from the left (where $x$ and $a$ are real-valued, for example) is equivalent to saying that $x$ approaches $a$ within the interval $(-\infty, a)$. There is a similar range of variations on the output: $f(x)$ can approach a value, $b$, or $\infty$, or $-\infty$; and it can approach the value from the left, or from the right, or within any subset of the range of $f$. Not only does this threaten a combinatorial explosion of definitions, but also redundancy. For example, assuming $f(x)$ and $g(x)$ converge as $x$ approaches $a$, we have the identity $\lim_{x \to a} (f(x) + g(x)) = \lim_{x \to a} f(x) + \lim_{x \to a} g(x)$, but this also holds under all the variations of convergence in the source.

To handle the many instances of convergence that arose in the formalization, we used Isabelle's elegant library for dealing with limits via filters \cite{hoelzl:et:al:13}. The idea is that when dealing with any notion of limit, the relevant notions of convergence in the source and the target can be represented by \emph{filters}. A {\em filter} over $X$ is a nonempty set $\mathcal F \subseteq \mathcal P(X)$ such that if $A \subseteq B$ and $A \in \mathcal F$, then $B \in \mathcal F$, and if $A, B \in \mathcal F$, then $A \cap B \in \mathcal F$. The general notion of limit in Isabelle, {\tt filterlim f F1 F2}, says, roughly, that the function $f$ converges in the sense of $F_2$ as the input converges in the sense of $F_1$. By specializing $F_1$ and $F_2$ appropriately, we obtain all the variations described in the last paragraph, and more. In addition, theorems can be proved at the appropriate level of generality. For example, we have:
\Snippet{tendstoadd}
Here \isamath{(f \xrightarrow{\hphantom{AAA}} x)~F} is an abbreviation for {\tt filterlim f F (nhds x)}, where {\tt nhds x} is the filter of topological neighborhoods of $x$. This avoids the need to formalize endless variations of the same theorem; we only need to instantiate the general version to the relevant filters. Details can be found in \cite{hoelzl:et:al:13}.

In the Isabelle library, topological facts are found in the standard \texttt{HOL} library, in files such as \texttt{Filter}, \texttt{Topological\_Spaces}, and \texttt{Limits}. Topological notions for the reals, and vector spaces over the reals, can be found in \texttt{Real\_Vector\_Spaces}.

\subsection{Measure theory and integration}
\label{subsection:measure:theory}

Our formalization required the fundamentals of measure theory and integration, as described in any introductory textbook on the subject (including Billingsley). The fundamental development of the subject is well-described in the paper ``Three chapters of measure theory'' \cite{hoelzl:heller:11}. Therefore we only summarize key features of this development here, and indicate some of the ways the library has changed as a result of our formalization.

Measure theory requires the use of the \emph{extended nonnegative reals} $\ennRR$, obtained by restricting the usual reals $\RR$ to their non-negative part and adding the value $\infty$. A \emph{$\sigma$-algebra} is represented in Isabelle as a record \isa{S} depending on a type $\alpha$, which specifies an underlying subset of $\alpha$, a set of elements \isa{space S}, and a collection of subsets \isa{sets S} of \isa{space S} that contains the empty set and is closed under countable unions and complements. These assumptions are specified as a locale.

A \emph{measure space} \isa{M} extends the notion of a $\sigma$-algebra with a function \isa{emeasure M} from subsets of $\alpha$ to the extended nonnegative reals, satisfying the usual axioms: the measure of the empty set is $0$, and the measure of a countable disjoint union of sets in the underlying $\sigma$-algebra is equal to sum of the measures of each set in the union (which might be $\infty$). The underlying $\sigma$-algebra corresponds to the usual notion of the collection of \emph{measurable subsets} corresponding to the measure. In Isabelle, for any type $\alpha$, the \isa{typedef} mechanism is used to specify a new type, $\isa{measure} \; \alpha$, consisting of measure spaces on some subset of $\alpha$.

If $\mdl M$ and $\mdl N$ are two measure spaces, a \emph{measurable function} $f$ from $\mdl M$ to $\mdl N$ (written $f \in \mdl{M} \rightarrow_M \mdl{N}$) is a function between the underlying sets that has the property that the inverse image of any measurable subset of the codomain is a measurable subset of the domain. Note that, in fact, the property of being measurable has nothing to do with the measure; it is really a property of the function with respect to the two associated $\sigma$-algebras. Given a measurable function $f : \mdl M \to \mdl N$, a measure $\mu$ on $\mdl M$ gives rise to a new measure $\nu$ on $\mdl N$, defined by $\nu(A) = \mu(f^{-1} A)$. This is sometimes called a \emph{pushforward measure}, but in Isabelle it is denoted \isa{distr M N f}, for reasons that are explained in Section~\ref{subsection:probability}. It is defined formally as follows:
\Snippet{distr}

Another way to define a measure $\nu$ in terms of a measure $\mu$ on $\mdl M$ is to take a measurable function $f$ from $\mdl M$ to $\ennRR$ and define, for every set $A$, $\nu(A) = \int^+ f \chi_A \; d\mu$. This is defined formally in the Isabelle library as follows:
\Snippet{density}
The function $\chi_A$ is the \emph{characteristic function} of $A$, also called the \emph{indicator function}. The integral $\int^+ f \; d\mu$ is the \emph{nonnegative Lebesgue integral}, defined for functions into $\ennRR$. For measurable functions it has the expected properties: it is closed under addition, constant multiplication, and monotone convergence. It is monotone even for non-measurable functions, which simplifies certain proofs, since measurability is not always easy to prove. Because the nonnegative Lebesgue integral takes values in $\ennRR$, it is well-defined for all measurable functions, even when the integral is infinite.

The library includes a construction of the Borel sets in any topology, and the Carath\'eodory extension theorem. In Isabelle, the Lebesgue measure on the reals was initially constructed from the gauge integral, which is discussed in Section~\ref{subsection:real:analysis}. After our formalization, however, the construction was replaced by the more common textbook definition as the extension via the Carath\'eodory theorem of the usual measure on finite intervals, as described in Section~\ref{subsection:distribution:functions} below. The measure space consisting of the Lebesgue measure on the Borel subsets of the reals is denoted \isa{lborel} in the Isabelle library.

The fundamentals of measure theory can be found in the \texttt{HOL-Probability} library, including the files \texttt{Sigma\_Algebra}, \texttt{Measure\_Space}, \texttt{Caratheodory}, and \texttt{Lebesgue\_Measure}.

\subsection{Bochner integration}
\label{subsection:bochner}

Our initial formalization of the Central Limit Theorem relied on the theory of Lebesgue integration, described in \cite{hoelzl:heller:11}. This provides a notion of integration for suitable functions $f : X \to \RR$, where $X$ is any space on which a measure is defined. After we completed the proof, however, the second author, H\"olzl, generalized the construction to the \emph{Bochner integral}. This provides a theory of integration for functions $f : X \to B$, where now $B$ is any second-countable Banach space. In particular, $B$ can be any of the spaces $\RR^n$, or the complex numbers, $\CC$. Our formalization made extensive use of integration of functions from $\RR$ to $\CC$, as discussed in Section~\ref{subsection:real:analysis}.

Similar to the Lebesgue integral, the Bochner integral approximates a function $f$ by a sequence of simple functions $s$. Each simple function has a finite range, and hence its integral can be expressed by finite summation. When a function can be approximated by simple functions, its  integral is the limit of the integrals of those simple functions. Whereas approximations for Lebesgue integration are taken with respect to the pointwise order on $\ennRR$, approximations for Bochner integration are taken with respect to the $L^1$-norm, defined using the Lebesgue integral by $\lVert f \rVert = \int^+ |f| d\mu$.

More formally, a function $s$ is \emph{simple Bochner-integrable} if $s$ is Borel-measurable on $\mdl M$, has a finite range $f[\mdl M]$, and a support $\{x \in \mdl M \mid f(x) \not= 0\}$ with finite measure. The integral of a simple Bochner function $s$ is a finite sum over the vectors of the range of $s$ times the measure of their support:
\[ \int s~d\mu = \sum_{ y \in f[\mdl M]} \mu~(f^{-1}[y]) \cdot y\]
A Borel-measurable function $f$ is Bochner-integrable if there is a sequence $(s_i)$ of simple Bochner-integrable functions such that:
\begin{enumerate}
 \item $f$ is the limit of $(s_i)$ in the $L^1$ norm, i.e.~$\lim_{i \rightarrow \infty} \lVert s_i - f \rVert = 0$; and
 \item the sequence of integrals of the functions $s_i$ converges, i.e.~$\lim_{i \rightarrow \infty} \int s_i ~d\mu$ exists.
\end{enumerate}
In that case, the Bochner integral, denoted \isa{LINT x|M. f x} in Isabelle, is defined by
\[ \int f~d\mu = \lim_{i \rightarrow \infty} \int s_i ~d\mu \]
The notation \isa{LINT} is a holdover from Lebesgue integration, but since Bochner integration functions in similar ways, the notation is still a useful mnemonic.

From the definition it follows that each Bochner-integrable function $f$ is Borel-measurable and has a finite $L^1$-norm. In the other direction we prove that each Borel-measurable function is approximated pointwise by a sequence of simple Bochner functions. Then it follows that a function $f$ is Bochner-integrable if and only if $f$ is measurable and the $L^1$-norm of $f$ is finite (which is equivalent to saying that $f$ is absolutely integrable).
\Snippet{integrablebounded}

As one would expect of an integral, the Bochner integral respects scalar multiplication and addition. As with the Lebesgue integral, we obtain a version of the dominated convergence theorem:
\Snippet{dominatedconvergence}
Here the quantifier \isa{AE x in M} expresses that the subsequent statement holds for almost every element $x$ of the measure space $M$, which is to say, the set of examples where it doesn't hold has measure zero. We have the monotone convergence theorem, which applies to sequences of functions taking values in the real numbers. We also obtain Fubini's theorem. These are all staples of the theory of integration, and were used throughout our formalization.

If $f : X \to B$ is any measurable function on a space $X$ with measure $\mu$, and $S$ is any measurable set, one can define the integral over the set $S$ by $\int_S f \, d\mu = \int f \chi_S \, d\mu$. Rather than introduce a new definition, we took notation for integration over sets to be an abbreviation for the definition in terms of indicators, with the notation \isa{LINT x:S|M. f x}. But because reasoning about integrals over sets is so fundamental, we found it helpful to develop a small library to support it. For example, the following is a consequence of the dominated convergence theorem:
\Snippet{lebesgueintegralcountableadd}

The theory of Bochner integration is included in the \texttt{HOL-Probability} library, in \texttt{Bochner\_Integration} and \texttt{Set\_Integral}.

\subsection{Probability}
\label{subsection:probability}

Modern probability is based on measure theory, although probabilists and statisticians tend to adopt their own distinct terminology. A \emph{probability space} is simply a measure space in which the measure of the entire space is equal to 1. You should think of the space as the space of possible outcomes of a random event. A \emph{random variable} on such a space is a measurable function from that space to the reals; think of it as a real number that depends on the outcome of the random event. The \emph{expectation} of a random variable is the integral of the function over the entire space. Thus talk of probability spaces, random variables, and expectations is really talk of measure spaces, measurable functions, and integrals in disguise.

The Isabelle library defines a locale for \emph{finite measures}, which are simply measures for which the measure of the entire space, \isa{emeasure M (space M)}, is not infinity. For such spaces, one can work more conveniently with the associated real-valued function, \isa{measure}, which casts the value of \isa{emeasure} to a real. There is also a locale for \emph{probability measures}, which are finite measures where the measure of the entire space is equal to $1$. When working with cumulative distribution functions, as described in Section~\ref{subsection:distribution:functions}, we found it convenient to define a locale for \emph{real distributions}; a real distribution is a probability space in which the space is the set of real numbers and the measurable sets consist of exactly the Borel subsets of the reals. To capture the language of informal probability theory, the library defines all of the following abbreviations:
\Snippet{probspace}

If $X$ is a random variable on a measure space $\mdl M$ with measure $\mu$, the distribution of $X$, as described in the previous section, has the following interpretation: it is the measure $\nu$ on the Borel sets of $\RR$ such that for every $A$, $\nu(A)$ is the probability that $X$ takes a value in $A$. Even though we think of $X$ as depending on some underlying source of randomness, represented by $\mdl M$, often we only care about the induced probability on the real numbers that is given by its distribution. Notice that the word ``distribution'' is used in probability theory in at least three distinct but related ways. In addition to the uses of the term described in this paragraph and the previous one, one also often speaks of the (cumulative) distribution function of a real distribution, as described in Section~\ref{subsection:distribution:functions}. Thus, if $X$ is a random variable, its distribution is a real distribution, which in turn has a distribution function.

In probability theory, real distributions are often specified as densities, as described in the previous section. Thus the normal distribution with mean $\mu$ and variance $\sigma$ is defined formally as follows:
\Snippet{normaldistr}

Various notions of independence are used in probability. Perhaps the most general is the following: suppose that for every $i$ in some index set $I$, $F_i$ is a collection of events (measurable sets) from some fixed measure space. Then the sequence $(F_i)_{i \in I}$ is said to be \emph{independent} if for every finite subset $J \subseteq I$ and every choice of a set $A_j \in F_j$ for each $j$, the probability of the intersection $\bigcap_{j \in J} A_j$, i.e.~the probability that all of the $A_j$'s occur, is the product of the individual probabilities.
\Snippet{indepsets}
If now $(A_i)_{i \in I}$ is a sequence of \emph{events} (rather than collections of events), saying that the sequence $(A_i)$ is independent amounts to saying that the sequence of singletons $(\{A_i\})_{i \in I}$ is an independent sequence of collections.
\Snippet{indepevents}
Finally, if $(X_i)_{i \in I}$ is a sequence of random variables with inputs in one measure space, $\mdl M$, and values in another space, $\mdl M'$ (typically, but not necessarily, the reals), saying that the sequence $X_i$ is independent amounts to saying that the sequence of collections of measurable sets
\[
(\{ X_i^{-1}(A) \; | \; \mbox{$A$ is a measurable subset of $\mdl M'$}\})_{i \in I}.
\]
is independent.
\Snippet{indepvars}
In probabilistic terms, this means that given any finite $J \subseteq I$ and any finite sequence $A_{j_1}, \ldots, A_{j_n}$ of events, the probability that each $X_{j_u}$ is in $A_{j_u}$ is just the product of the individual probabilities. Of course, we can say that any \emph{two} events, or random variables, or collections of events, are independent by taking $I$ to be any two-element type, such as the Booleans. Isabelle's library defines the binary notions as well, and develops basic properties of independent sets, events, and random variables.

In the Isabelle 2016 distribution, these definitions are in the \texttt{HOL-Probability} library, including \texttt{Probability\_Measure}, \texttt{Independent\_Family}, \texttt{Convolution}, and \texttt{Distributions}.

\subsection{Real analysis and complex-valued functions}
\label{subsection:real:analysis}

Isabelle has an extensive library for real multivariate analysis, which is again well-described in \cite{hoelzl:et:al:13}. In Isabelle, the reals are instantiated as a complete ordered field, and as a conditionally complete lattice, which means that nonempty bounded sets have sups and infs. The library also includes definitions of transcendental functions like the sine, cosine, and exponential functions. In fact, the exponential function is defined generically for any Banach space, including the complex numbers. Of course, we have the relation $e^{i x} = \cos x + i \sin x$ for real $x$.

Isabelle's general notion of the derivative is the \emph{Fr\'echet derivative}, which makes sense for functions $f$ between any two Banach spaces. As with limits, the notion of Fr\'echet derivative supports multiple modes of convergence; the expression \isa{(f has\_derivative D) F} means that the function $f$ has the bounded linear functional $D$ as derivative ``at'' the filter $F$. In practice, $F$ is usually the filter expressing that $D$ is the derivative at a point $x$, or that $D$ is the derivative at a point $x$ when we restrict attention to a subset $S$ of the source. The more familiar notion of the scalar derivative for functions from the reals to reals (or, more generally, from one normed field to another) is derived from the Fr\'echet derivative as a special case. So is the notion of a vector derivative for functions from $\RR$ to $\RR^n$.

The characteristic function of a measure is a function from the reals, $\RR$, to the complex numbers, $\CC$. The theory of such functions is much simpler than the theory of functions from $\CC$ to $\CC$, which is the subject of complex analysis. One can view a function $f : \RR \to \CC$ as essentially two functions from $\RR$ to $\RR$, $f^\mathit{re}$ and $f^\mathit{im}$, the first returning the real part and the second returning the imaginary part of the output. Integrals and derivatives of such functions can be understood in terms of the integrals and derivatives of these two parts.

In fact, for differentiation, we did not have to define a new notion of derivative: if we view the complex numbers as a two-dimensional real Banach space, the derivative we need is nothing more than the Fr\'echet derivative.

For integration, the story is more involved. Isabelle's library now has two forms of the integral. The multivariate analysis library generally relies on the gauge integral, which is defined for functions from $\RR^n$ to $\RR$. When we consider the reals, $\RR$, with the usual Lebesgue measure, the Bochner integral and the gauge integral agree on finite intervals, but otherwise the gauge integral is slightly more general: for a function $f : \RR \to \RR$ to be Bochner-integrable, both the positive and negative parts of $f$ have to have a finite Bochner integral, whereas the gauge integral can accommodate some functions whose positive and negative parts cancel each other out in a suitable fashion. Nonetheless, for the vast majority of applications, the Bochner integral is quite sufficient. Since our formalization required integration with general measures and spaces in addition to the usual integration over $\RR^n$, we used the Bochner integral throughout.

With the Bochner integral, as with the Fr\'echet derivative, integrating functions taking values in $\CC$ is no different from integrating functions taking values in $\RR$. Indeed, this was the primary motivation for generalizing from the Lebesgue integral to the Bochner integral.

\subsection{Calculus}
\label{subsection:calculus}

Our formalization required extensive use of calculus at an undergraduate level, including integration by parts, Taylor series approximations, changes of variable, and so on. For example, the calculation of moments of the normal distribution required the following estimate on the complex exponential:
\[
 \left| e^{ix} - \sum_{k=0}^n \frac{(ix)^k}{k!} \right| \le \min\left(\frac{|x|^{n+1}}{(n+1)!}, \frac{2 x^n}{n!}\right).
\]
We followed Billingsley~\cite[Section 26]{billingsley:95} in obtaining this using an inductive argument and integration by parts. Notice that this involves reasoning about functions from the real to complex numbers; as explained in the previous section, the relevant properties generally follow from the corresponding properties for real-valued functions, upon splitting functions to the real and imaginary parts. In addition, we have the general inequality $\| \int_A f \; d\mu \| \le \int_A \| f \| d\mu$, which allows us to bound the modulus of a complex integral by bounding the real-valued integral of the norm. This is an instance of a more general fact about the Bochner integral:
\Snippet{integralnormbound}

Textbook results from calculus involve integrals $\int_a^b f(x) \; dx$ over the interval $(a,b)$. These can be viewed as ordinary integrals over the set $(a,b)$, with the following two caveats:
\begin{itemize}
 \item Textbooks allow $a$ to be $-\infty$ and allow $b$ to be $\infty$, which is to say, $a$ and $b$ should be taken to be extended real numbers (i.e. the reals extended with $\pm\infty$).
 \item It is convenient to adopt the convention that if $b < a$, then \\ $\int_a^b f(x) \; dx = -\int_b^a f(x) dx$.
\end{itemize}
We thus defined a notion of ``interval integral'' along these lines, together with supporting the notation \isa{LBINT x=a..b. f x}. We could then state the first fundamental theorem of calculus in the following form, for finite intervals:
\Snippet{ftcfinite}
The following version, for arbitrary intervals, makes sense when the limits are infinite:
\Snippet{ftcintegrable}
Similarly, we could state the second fundamental theorem of calculus, where the variable bound to the integral can be before or after the fixed endpoint:
\Snippet{ftc2}
The use of such an integral was a mixed blessing. It simplified many of our theorems and proofs, but at the expense of introducing yet another notion of integral, which required another library of supporting facts, as well as, at times, translations to and from the other notions of integral.

Many textbook integration arguments require a change of variable, sometimes known as ``integration by substitution.'' It was not hard to prove that if a function $g$ from $\RR$ to $\RR$ has a continuous derivative (and hence is continuous itself) on a closed interval $[a,b]$, and $f$ is continuous on the image of $[a, b]$ under $g$, then $\int_a^b f(g(x)) g'(x) \; dx = \int_{g(a)}^{g(b)} f(x) \; dx$.
\Snippet{substfinite}
Manuel Eberl later generalized this to arbitrary Borel measurable functions $f$, but with the added hypothesis that $g'$ is nonnegative on $[a, b]$. However, we also needed a version of the theorem for intervals with potentially infinite endpoints. This requires using either the monotone convergence theorem or the dominated convergence theorem to pass from finite interval approximations to the full interval. In fact, we proved two versions. The following one requires showing independently that both $f(x)$ and $f(g(x)) g'(x)$ are integrable over the relevant intervals:
\Snippet{substintegrable}
Another version assumes instead that $f$ is nonnegative, and concludes that $f$ is therefore integrable.

As an example where various uses of these components came together, consider the \emph{sine integral function}. The function $\sin x / x$ is undefined at $0$, but it can be made continuous at $0$ by giving it the value $1$ there. The resulting function is called $\sinc$. The sine integral function is (confusingly) defined to be the indefinite integral of the $\sinc$ function, starting at $0$:
\[
\fn{Si}(t) = \int_0^t \sinc x \, dx.
\]
The proof of the L\'evy inversion formula uses the fact that
\[
\lim_{t \rightarrow \infty} \fn{Si}(t) = \frac{\pi}{2}.
\]
A textbook proof (sketched in \cite[Example 18.4]{billingsley:95}) runs as follows. By the fundamental theorem of calculus, we can verify that
\[
\int_0^t e^{-ux} \sin x \, dx = \frac{1}{1+u^2}[1 - e^{-ut}(u \sin t + \cos t)]
\]
by taking the derivative of both sides. Calculating, we can also show that
\[
\int_0^t \left( \int_0^\infty |e^{-ux} \sin x| \, du\right) \, dx = \int_0^t x^{-1} |\sin x| \, dx \le t.
\]
The fact that the double-integral on the left is finite means that Fubini's theorem may be used to change the order of integration of $e^{-ux} \sin x$ over $(0,t) \times (0, \infty)$. So we have
\begin{align*}
\int_0^t \frac{\sin x}{x} \, dx &= \int_0^t \sin x \left(\int_0^\infty e^{-ux} \, du\right) \, dx \\
                                &= \int_0^\infty \left(\int_0^t e^{-ux} \sin x \, dx\right) \, du \\
                                &= \int_0^\infty \frac{du}{1+u^2} - \int_0^\infty \frac{e^{-ut}}{1+u^2} (u \sin t + \cos t) \, du.
\end{align*}
Substituting $u = \tan x$ in the first term yields
\[
 \int_0^\infty \frac{du}{1+u^2} = \int_0^{\pi/2} \frac{1}{1 + \tan^2 x} (1 + \tan^2 x) \, dx = \pi/2,
\]
and the change of variable $v = ut$ can be used to show that the second integral converges to $0$ as $t \rightarrow \infty$. Hence
\[
\lim_{t \rightarrow \infty} \fn{Si}(t) = \lim_{t \rightarrow \infty} \int_0^t \frac{\sin x}{x} \, dx = \frac{\pi}{2},
\]
as required.

Proving this result required a tremendous amount of formal machinery: not only suitable forms of substitution, but also Fubini's theorem, the fundamental theorem of calculus, integration by parts, integral comparisons, properties of limits, and properties of the tangent function. It also required a lot of work, establishing that the relevant functions were continuous, integrable, and so on. It was somewhat demoralizing that a small calculus exercise required so much effort, but it is a good illustration of the infrastructure that is needed to carry out the kinds of calculus computations that come up routinely in engineering, modeling, and the sciences. We faced a similar calculus exercise in computing the moments of the normal distribution, as described in Section~\ref{subsection:characteristic}.

In the Isabelle 2016 distribution, the formulation of the fundamental theorem of calculus that we used and the substitution theorems described above are in \texttt{Interval\_Integral}. Eberl's generalization can be found in the file \texttt{Lebesgue\_Integral\_Substitution}. The calculation concerning $\fn{Si}$ is in the file \texttt{Sinc\_Integral}.

\subsection{Distribution functions and the Lebesgue-Stieltjes measure}
\label{subsection:distribution:functions}

Every measure on $\RR$ gives rise to the real-valued function which, at each input $x$, returns the amount of ``mass'' below that argument:

\begin{definition}
Let $\mu$ be a finite measure on $\RR$. The \emph{cumulative distribution function} $F_\mu$ is defined by $F_\mu(x) = \mu (-\infty, x]$.
\end{definition}
The cumulative distribution function (or \emph{cdf}) is sometimes also called, more simply, the \emph{distribution function} of the measure. In Isabelle, the definition is rendered as follows:

\Snippet{cdf}

It is not hard to see that the distribution function $F_\mu$ of a finite Borel measure~$\mu$ is nondecreasing and right-continuous, and satisfies $\lim_{x \rightarrow -\infty} F_\mu(x) = 0$.

\Snippet{cdfprop1}
\Snippet{cdfprop2}
\Snippet{cdfprop3}

Conversely, it turns out that any function with these properties is the distribution of a Borel measure on $\RR$. The requisite measure $\mu$ is constructed by defining $\mu (a,b] = F(b) - F(a)$ and extending this to the Borel $\sigma$-algebra using the Carath\'eodory extension theorem. To that end, we defined an operation, \isa{interval\_measure}, that generates a measure from a nondecreasing, right-continuous function. To use the Carath\'eodory extension theorem, the key property that needs to be verified is that if a half-open interval $(a, b]$ is written as a disjoint union of countably many intervals $(a_i, b_i]$, then $b - a = \sum_i (b_i - a_i)$. This is trickier than it sounds. For example, the interval $(0, 1]$ can be written as a countable union of intervals $(1/2^{i+1}, 1/2^i]$, and any one of \emph{those} intervals could similarly be replaced by a countable union. It is not hard to show that the infinite sum $\sum_i (b_i - a_i]$ is bounded by $b - a$. In the other direction, one picks a small $\varepsilon$, enlarges each interval $(a_i, b_i]$ to a slightly larger interval $(a_i - \varepsilon/2^i, b_i + \varepsilon / 2^i)$, argues that the union of the enlargements covers the closed interval $[a, b]$, and then appeals to the compactness of $[a,b]$. The measure associated to a right-continuous, nondecreasing function in this way is called the \emph{Lebesgue-Stieltjes measure}. When the function $F(x)$ is the identity function, we obtain the Lebesgue-Borel measure \isa{lborel}, and, in fact, this now serves as the definition of Lebesgue-Borel measure in the Isabelle library.

In the case of a probability measure, we have the additional property that $\lim_{x \rightarrow \infty} F_\mu(x) = 1$:
\Snippet{cdfprop4}
Conversely, any function $F$ satisfying all four properties is a probability measure:
\Snippet{intervalmeasure}
Recall that \isa{real\_distribution} is the name of the locale for probability measures on the Borel subsets of the reals. So, for any function \isa{F} satisfying the four properties above, \isa{interval\_measure F} is the measure whose cdf is exactly \isa{F}. The use of the word ``the'' is justified by the fact that the association is unique, in the sense that if two real distributions have the same cumulative distribution function, then they are equal:
\Snippet{cdfunique}
Thus one can pass freely between talk of measures on $\RR$ and of their distribution functions, a key fact in the proof of the CLT.

In the Isabelle 2016 distribution, the construction of the Lebesgue measure on the reals as a Lebesgue-Stieltjes measure is in the theory \texttt{Lebesgue\_Measure}, and the correspondence between measures and their distribution functions is developed in \texttt{Distributions}.

\subsection{Automation}
\label{subsection:automation}

To improve automation, Isabelle's multivariate analysis library provides a large set of introduction rules, to establish things like openness or closedness of sets or continuity of functions. Continuity is nicely reduced by compositionality; if we know that two functions are continuous, their composition is again continuous. Applying this as a rule requires matching terms of the form $f~(g~x)$ where both $f$ and $g$ are variables. But this is often not the right choice. The straightforward way to express that functions like multiplication and $\ln$ are continuous is to write $\isa{continuous\_on}~(\RR \times \RR)~(\lambda (x, y).~x * y)$ and $\isa{continuous\_on}~(0, \infty)~\ln$. The composition rule is then not sufficient to prove continuity of $\lambda x.~\ln (1 + x * x)$, because it does not accommodate binary operations like $+$ and $*$. In addition, the composition rule does not allow for the fact that the domain of $\ln$ has to be restricted to the positive reals.

A simple solution to these two problems is to state continuity rules \emph{precomposed} with arbitrary continuous functions. For example, we can state the following rules for arbitrary $f$ and $g$:
\begin{mathpar}
  \inferrule{\isa{continuous\_on}~A~f \\ \forall x \in A.~0 < f~x}%
  {\isa{continuous\_on}~A~(\lambda x.~\ln~(f~x))}
\\
  \inferrule{\isa{continuous\_on}~A~f \\ \isa{continuous\_on}~A~g}%
  {\isa{continuous\_on}~A~(\lambda x.~f~x + g~x))}
\end{mathpar}
Now, to prove $\lambda x.~\ln (1 + x * x)$, we just apply rules like these. We ultimately end up with the goal $\forall x.~0 < 1 + x * x$, which is proved by the simplifier. This idea goes back to a paper by Gottliebsen~\cite{gottliebsen:00}, which describes an implementation in PVS.

Rules for establishing openness and closedness of sets are not as important, but nonetheless helpful. Besides the usual rules for intersections and unions, we also have rules working on logical connectives and relations in set-comprehension. For example, the set $\{x \mid f~x < g~x \}$ is open whenever $f$ and $g$ are continuous functions into real numbers.

Isabelle's automation for measurability uses precomposed rules in a similar way. A difference is that measurability is also integrated as a special-purpose simplification procedure (in Isabelle terminology, a \emph{simproc}), called \isa{measurable}. To use the measurability prover, the user needs to annotate all the relevant measurability assumptions with the \isa{[measurable]} attribute. This measurability prover then tries to massage all added assumptions into the right form, and proves measurability statements by applying them as introduction rules. The massaging also includes destructions of certain compositions, e.g.~the assumption that $\lambda x.~(f~x, g~x)$ is $X \times Y$-measurable is replaced by the fact that $f$ is $X$-measurable and $g$ is $Y$-measurable. Such destructions are important for higher-order proof steps like induction. As a special case it also allows us to decompose subterms with a countable range, since the measurability of $f~(g~x)~x$ can be reduced to the measurability of $g$ and the measurability of $f~c~x$ for all $c$ in the range of $g$. It is also important to add measurability rules for logical connectives, including quantifiers over countable sets. As a result, predicates can also be proved measurable, and therefore expressions that depend on case distinctions.

An example of the power of such rule sets is given by the proof that the predicate ``$f$ is continuous at $x$'' is measurable in $x$ for a function $f$ on metric spaces. We can express the continuity of $f$ at $x$ in the following way:
\[ \forall i>0.~ \exists j>0.~ \forall y\,z.~ d(x, y) < \frac{1}{j} \land
  d(x, z) < \frac{1}{j} \implies d(f(y), f(z)) \le \frac{1}{i}~. \]
Then the proof that this is measurable is a straightforward application of rules, as follows: (1) the quantifiers over $i$ and $j$ are countable, hence measurable; (2) we get into a closed set by eliminating the quantifiers over $y$ and $z$; (3) for the implication, the right-hand side is constant, hence closed; and (4) the left-hand side is open, as it is a strict inequality between two continuous functions.

\section{The proof of the Central Limit Theorem}
\label{section:formal}

\subsection{Weak convergence}
\label{subsection:weak:convergence}

Recall from Section~\ref{subsection:background} that if $(\mu_n)$ is a sequence of real distributions and $\mu$ is a real distribution, then $(\mu_n)$ converges weakly to $\mu$, written $\mu_n \Rightarrow \mu$, if $F_{\mu_n}(x)$ approaches $F_\mu(x)$ at each point $x$ where $F_\mu$ is continuous. In Isabelle, this is expressed by the following two definitions:

\Snippet{weakconv}
In words, a sequence of functions $(F_n)_{n \in \NN}$ converges weakly to $F$ if $(F_n(x))_{n \in \NN}$ converges to $F(x)$ for each point $x$ where $F$ is continuous, and the sequence of measures $(\mu_n)$ converges weakly to $\mu$ if the corresponding cumulative distribution functions converge weakly.

That the notion of weak convergence is robust is supported by the fact that there are a number of equivalent characterizations. The following theorem is sometimes known as the \emph{Portmanteau Theorem}:
\begin{theorem}
The following are equivalent:
\begin{enumerate}
 \item $\mu_n \Rightarrow \mu$.
 \item $\int f \; d\mu_n$ approaches $\int f \; d\mu$ for every bounded function $f$ that is continuous almost everywhere.
 \item $\int f \; d\mu_n$ approaches $\int f \; d\mu$ for every bounded, continuous function $f$.
 \item If $A$ is any Borel set, $\partial A$ denotes the topological boundary of $A$, and $\mu(\partial A) = 0$, then $\mu_n(A)$ approaches $\mu(A)$.
\end{enumerate}
\end{theorem}
The theorem is interesting in that it combines measure-theoretic notions (measures and the integral) with topological notions (continuity and topological boundaries). The proof from 1 to 2 uses Skorohod's theorem. This states that if $(\mu_n)$ is a sequence of real distributions that converges to a real distribution $\mu$, there is a sequence $(Y_n)$ of random variables and another random variable $Y$, all defined on a common probability space, such that each $Y_n$ has distribution $\mu_n$, $Y$ has distribution $\mu$, and $Y_n$ converges to $Y$ pointwise. In other words, Skorohod's theorem tells us that $(\mu_n)$ and $\mu$ can be represented in a particularly nice way.
\Snippet{skorohod}
Proving Skorohod's theorem formally presented a number of technical challenges. One was that we needed to choose a continuity point of an arbitrary probability measure in an arbitrary open interval, that is, a real number $x$ in an open interval $I$ such that the measure of $\{x\}$ is zero. To that end, we showed that the number of atoms of a measure (that is, points $x$ such that $\{x\}$ has strictly positive measure) is countable:
\Snippet{countableatoms}
The result then follows from the fact that any open interval in the reals is uncountable.

Returning to the proof of the portmanteau theorem, the implication from 2 to 3 is immediate. Notice that 1 is equivalent to saying that for every point $x$ of continuity of the measure $\mu$, $\int \chi_{(-\infty,x]} \; d\mu_n$ approaches $\int \chi_{(-\infty,x]} \; d\mu$, where $\chi_{(-\infty,x]}$ is the characteristic function of the interval $(-\infty,x]$. The implication from 3 to 1 is obtained by approximating this characteristic function by continuous step functions whenever $x$ is a point of continuity. The implication from 4 to 1 is easy, noticing that $(-\infty,x]$ is a set of the specified type, whenever $x$ is a point of continuity of $\mu$. To complete the proof of the theorem, it is enough to prove that 2 implies 4. This implication is also not hard, once we show that the characteristic function $\chi_A$ is bounded and continuous at any point not on the boundary.

The results discussed in this section are found in the theory \texttt{Weak\_Convergence}.

\subsection{Characteristic functions}
\label{subsection:characteristic}

Recall that the \emph{characteristic function} $\ph$ of a probability measure $\mu$ on the real line is defined by
\[
\ph(t) = \int_{-\infty}^{\infty} e^{itx} \mu(dx).
\]
If $X$ is a random variable, the characteristic function of $X$ is defined to be the characteristic function of its distribution.
In our formalization, the characteristic function of a measure is defined as follows:
\Snippet{char}
The characteristic function of a random variable \isa{X} defined on a measure space \isa{M} is then written \isa{char (distr M borel X)}, since \isa{distr M borel X} denotes the distribution of \isa{X} with respect to the usual Borel measure on the real numbers.

The characteristic function $\ph$ of a measure is continuous, and satisfies $\ph(0) = 1$ and $|\ph(t)| \leq 1$ for every $t$:
\Snippet{charprop1}
\Snippet{charprop2}
\Snippet{charprop3}
As noted above, a key property of characteristic functions is this: if $X_1$ and $X_2$ are independent random variables, the characteristic function of $X_1 + X_2$ is the product of the individual characteristic functions. Because we used the Bochner integral, which allows us to integrate complex-valued functions directly, our final proof of this fact is even simpler than the one in Billingsley \cite{billingsley:95}. The calculation runs as follows:
\begin{align*}
\ph_{X_1 + X_2}(t) &= \int e^{i t (X_1 + X_2)} \, dM  \\
                   &= \int e^{i t X_1} e^{i t X_2} \, dM \\
                   &= \left(\int e^{i t X_1} \, dM\right) \left(\int e^{i t X_2} \, dM\right) \\
                   &= \ph_{X_1}(t) \ph_{X_2}(t).
\end{align*}
Here, $X_1$ and $X_2$ are really functions over the underlying probability space $M$, and the third equation follows from the independence of $X_1$ and $X_2$. We reproduce our formal proof in full:
\Snippet{chardistrsum}
By induction, we have that for any finite set $A$ and any sequence $(X_i)_{i \in A}$ of mutually independent random variables,
\[
\ph_{\sum_{i \in A} X_i}(t) = \prod_{i \in A} \ph_{X_i}(t).
\]
The formal proof is as follows:
\Snippet{chardistrsetsum}
(We do not require finiteness of \isa{A}: by definition, if \isa{A} is infinite, the sum over \isa{A} is $0$ and the product over \isa{A} is $1$, and the equation still holds.)

We also needed explicit approximations to the characteristic functions of a random variable, obtained using the calculation described at the beginning of Section~\ref{subsection:calculus}. One of the results we used is as follows:
\Snippet{charapprox}

Finally, we needed to compute the characteristic function $\ph$ of the standard normal distribution, which means showing $\ph(t) = e^{-t^2/2}$. Establishing this fact took more work than we thought it would. Many textbook proofs of this invoke facts from complex analysis that were unavailable to us. Billingsley \cite[page 344]{billingsley:95} sketches an elementary proof, which required calculating the moments and absolute moments of the standard normal distribution. This is where the calculations of $\int_{-\infty}^\infty x^k e^{-x^2 / t} \; dx$, mentioned in Section~\ref{subsection:calculus}, were needed. Specifically, we have for even $k$,
\Snippet{momenteven}
and for odd $k$,
\Snippet{momentodd}
A prior calculation by Sudeep Kanav covered the cases $k = 0, 1$, which provide the base cases for an inductive proof. Filling in the details involved carrying out careful computations with integrals and power series approximations to $e^x$.

In the Isabelle 2016 distribution, characteristic functions are defined in the theory \texttt{Characteristic\_Functions}, and the properties cited above are proved there. The calculation of the moments of the normal distribution is found in the theory \texttt{Distributions}.

\subsection{L\'evy Inversion and Uniqueness}

In Fourier analysis, an ``inversion theorem'' says that a function can be recovered from its Fourier transform, under suitable hypotheses and in a suitable sense. Along those lines, the L\'evy Inversion and Uniqueness Theorems say that a measure can be recovered from its characteristic function.

More precisely, the L\'evy Inversion Theorem states the following:
\begin{theorem}
Let $\mu$ be a probability measure, and $\ph$ be the characteristic function of $\mu$. If $a$ and $b$ are continuity points of $\mu$ and $a < b$, then
\[
\mu (a,b] = \lim_{T \rightarrow \infty} \frac{1}{2\pi} \int_{-T}^T \frac{e^{-ita} - e^{-itb}}{it} \ph(t) \, dt.
\]
\end{theorem}
By definition, saying that a point $p$ is a continuity point of a measure $\mu$ means that $\mu(\{p\}) = 0$.

The proof is a long and subtle calculation. Let $I(T)$ denote the expression after the limit. Expanding the definition of $\ph(t)$ and appealing to Fubini's theorem to switch the order of the two integrals, we obtain
\[
I(T) = \frac{1}{2\pi} \int_{-\infty}^\infty \int_{-T}^T \frac{e^{it(x-a)} - e^{it(x-b)}}{it} \, dt \, \mu(dx).
\]
The idea is that as $T$ approaches $\infty$, the inner integral approaches a step function which jumps from $0$ to $1$ at $a$ and then back down to $0$ at $b$. This is shown by expanding the complex exponential in terms of $\sin$ and $\cos$, using properties of the sine integral, and manipulating integrals and limits.

It is not hard to show that fixing the values of a measure on intervals $(a, b]$ as above is enough to determine the measure on all Borel sets. Thus the Inversion Theorem has the following result, known as the Uniqueness Theorem, as an important corollary:
\begin{theorem}
If $\mu_1$ and $\mu_2$ are probability measures and $\ph_{\mu_1} = \ph_{\mu_2}$, then $\mu_1 = \mu_2$.
\end{theorem}
In our formalization, this is expressed simply as follows:
\Snippet{levyunique}

\subsection{The L\'evy Continuity Theorem}

Let $(\mu_n)$ be a sequence of distributions, where each $\mu_n$ has characteristic function $\ph_n$, and let $\mu$ be a distribution with characteristic function $\ph$. The \emph{L\'evy Continuity Theorem} states that $\mu_n$ converges to $\mu$ weakly if and only if $\ph_n(t)$ converges to $\ph(t)$ for every $t$. In our formalization, it is expressed as follows:
\Snippet{levycont}

Proving the ``only if'' direction is easy, using the Portmanteau Theorem of Section~\ref{subsection:weak:convergence}, since $e^{itx}$ is bounded and continuous. In fact, in our formalization, it has a one-line proof:
\Snippet{levycont1}
The other direction is a lot harder. Here is an outline of the proof:
\begin{enumerate}
 \item Use a compactness argument to show that every subsequence $(\mu_{n_k})$ of $(\mu_n)$ has a weakly convergent subsequence.
 \item Suppose, for the sake of contradiction, $(\mu_n)$ does not converge weakly to $\mu$. Then there is a subsequence $(\mu_{n_k})$ such that no subsequence of \emph{that} can converge weakly to $\mu$.
 \item By 1, this particular sequence $(\mu_{n_k})$ converges weakly to some measure, $\nu$.
 \item By the ``only if'' direction, already proved, $\ph_{n_k}$ converges pointwise to the characteristic function of $\nu$.
 \item Since, by hypothesis, $\ph_n(t)$ converges to $\ph(t)$ for every $t$, the characteristic function of $\nu$ must be $\ph$.
 \item By the Uniqueness Theorem, this implies that $\nu = \mu$, contrary to the choice of $(\mu_{n_k})$ in 2.
\end{enumerate}

The necessary compactness principle is a consequence of the \emph{Helly Selection Theorem}, which we now describe. Although the proof of this theorem and its consequence take up only a page-and-a-half in Billingsley's textbook, these were among the most subtle components of our formalization. The theorem states the following:
\begin{theorem}
Let $(f_n)_{n \in \NN}$ be a uniformly bounded sequence of nondecreasing, right continuous functions. Then there are a subsequence $(f_{n_k})_{k \in \NN}$ and a nondecreasing, right-continuous function $F$ such that $\lim_k f_{n_k}(x) = F(x)$ at continuity points of $F$.
\end{theorem}
In our formalization, this is expressed as follows:
\Snippet{helly}
Saying that $(f_{n_k})$ is a subsequence of $(f_n)$ means that the map $k \mapsto n_k$ is strictly increasing. The statement is represented formally by explicitly asserting the existence of the strictly increasing function $s : \NN \to \NN$ which returns, for each $k$, the value $n_k$. The proof involves a diagonalization argument: for each rational $r$, we thin the sequence to guarantee convergence at $r$, and then take a ``diagonal limit'' to construct the required subsequence and limit. To that end, we used a general framework for such diagonalization arguments, provided by Fabian Immler.

To describe the relevant corollary, we need to introduce a definition. A sequence $(\mu_n)$ of real measures is said to be \emph{tight} if, for every $\varepsilon > 0$, there is a finite interval $(a, b]$ such that $\mu_n(a, b] > 1 - \varepsilon$ for all $n$. Roughly, a sequence of probability measures is tight if no mass ``escapes to infinity;'' the sequence $(\mu_n)$, where $\mu_n$ is a unit mass at $n$, is an example of a sequence that is \emph{not} tight. Helly's theorem can be used to show that if $(\mu_n)$ is a tight sequence of measures, then for every subsequence $(\mu_{n_k})$ there is a further subsequence $(\mu_{n_{k(j)}})$ and a probability measure $\mu$ such that $(\mu_{n_{k(j)}})$ converges weakly to $\mu$ as $j$ approaches infinity.
\Snippet{thight}

In the Isabelle 2016 distribution, the Helly Selection Theorem and its corollary are proved in \texttt{Helly\_Selection}, and the L\'evy Continuity Theorem is proved in \texttt{Levy}. Immler's general framework for diagonal arguments can be found in the theory \texttt{Diagonal\_Sequence} in \texttt{HOL-Library}.

\subsection{The Central Limit Theorem}

Proving the Central Limit Theorem is now just a matter of putting the pieces together. Let $(X_n)$ be a sequence of random variables, all of which have the same distribution $\mu$ and finite variance $\sigma^2 > 0$. Without loss of generality (subtracting a common offset) we can assume that each $X_n$ has mean $0$. Let
\[
 S'_n = \sum_{i < n} X_i / \sqrt {n \sigma^2}
\]
be the normalized sums. Our goal is to show that the distributions of $S'_n$ converge weakly to the standard normal distribution.

For each $n$, let $\ph_n$ be the characteristic function of $S'_n$. By the L\'evy continuity theorem, it suffices to show that $\ph_n$ approaches the characteristic function of the standard normal distribution pointwise. In other words, we need to show that for every $t$, $\ph_n(t)$ approaches $e^{-t^2/2}$.

Since each $X_i$ has the same distribution, all the $X_i$'s have the same characteristic function; call it $\psi$. By the key property of characteristic functions, the characteristic function of the sum $S'_n$ is the product of the characteristic functions of the components, so
\begin{align*}
 \ph_n(t) & = \prod_{j = 1}^n \int e^{itX_j} / \sqrt{n \sigma^2} \, d\mu \\
   & = \prod_{j = 1}^n \psi(t / \sqrt{n \sigma^2}) \\
   & = (\psi(t / \sqrt{n \sigma^2}))^n.
\end{align*}
Now some of the explicit calculations described in Section~\ref{subsection:calculus} can be used to show that $\psi(t)$ is well-approximated by
\[
1 + it \int X \, d\mu + \frac{t^2}{2} \int X^2 d\mu,
\]
which is equal to $1 - t^2\sigma^2 / 2$, since we are assuming $X$ has mean $0$ and variance $\sigma^2$. Plugging $t / \sqrt{n \sigma^2}$ in for $t$, we obtain an approximation to $\psi(t / \sqrt{n \sigma^2})$, and substituting that in the expression for $\ph_n(t)$, we see that $\ph_n(t)$ is approximated by $(1 - \frac{t^2}{2n})^n$. This last expression approaches $e^{-t^2/2}$ as $t$ approaches infinity, as required.

The Central Limit Theorem is found in the file \texttt{Central\_Limit\_Theorem}. The formal version of the proof we have just sketched is given in its entirety in the appendix. (As above, we derive the mean zero case first, and then derive Theorem~\ref{theorem:clt} as a corollary.)

\section{Reflections}
\label{section:reflections}

We are by no means the first to formalize substantial portions of analysis in an interactive theorem prover. Both HOL4 and HOL Light have extensive theories of multivariate real analysis, and HOL Light has a substantial theory of complex analysis as well \cite{harrison:07a}. The real analysis library in HOL Light played an important part in the Flyspeck project formalization of Thomas Hales' proof of the Kepler conjecture \cite{hales:et:al:15}, and Isabelle's real analysis library has also been used to formalize properties of dynamical systems \cite{immler:traut:16}. Substantial portions of measure theory and measure-theoretic probability have been formalized in HOL4 \cite{mhamdi:et:al:11,qasim:et:al:16}. The C-CoRN and Coquelicot projects \cite{krebbers:spitters:11,boldo:et:al:12} provide a libraries for real analysis based on dependent type theory for the Coq proof assistant. It would take us too far afield to discuss all this work and compare all the other approaches to ours, so, instead, we will focus on our own formalization efforts and try to convey some of the lessons we learned. We also refer the reader to an article by Boldo, Lelay, and Melquiond \cite{boldo:et:al:16}, which provides a thoughtful and thorough survey of approaches to formalizing real analysis.

\subsection{Dealing with partial functions}
\label{subsection:partial}

In the logical framework of type theory, where every function is assumed to be total, one often has to deal with partial functions, such as limits, derivatives, integrals, and so on. One common way of proceeding is to represent partial functions as relations. For example, in Isabelle one can write \isa{f~sums~l} to indicate that the finite partial sums $\sum_{i < n} f(i)$ converge to $l$ as $n$ approaches infinity. Another option is to make the function in question total by assigning an arbitrary value at inputs where it would otherwise be undefined, and use a predicate to pick out the ``real'' values. For example, \isa{summable f} is defined to mean that there exists an \isa{l} such that \isa{f sums l}, and \isa{suminf f} specifies that value of \isa{l}, if it exists, and $0$ otherwise. It is not hard to see that the expression \isa{f sums l} is then equivalent to the conjunction $\isa{summable f} \wedge \isa{suminf f} = \isa{l}$. The expression \isa{suminf f} allows us to refer to the value of the infinite sum, as we do when we write $\sum_{i = 0}^\infty f(i)$ in ordinary mathematics. In particular, this expression can occur in a more complicated expression; for example, we can write $\isa{suminf f} + \isa{suminf g}$. But one typically also wants to know that \isa{summable f} holds, since otherwise the value of \isa{suminf f} may be meaningless.

In Isabelle, many partial functions related to analysis, such as limits and derivatives, are represented in this way, with a relation, a predicate asserting the existence of a value, and a function that returns an arbitrary value when the predicate fails. An exception is the notion of a measure on a measure space: while the expression $\isa{s} \in \isa{sets M}$ expresses that $s$ is a measurable set for the measure \isa{M}, in which case, \isa{measure M s} is the measure of \isa{s}, there is no relation between a set and its measure.

In the measure theory library, however, integration is handled in the usual way: we have \isa{has\_bochner\_integral M f x} to express that \isa{f} has Bochner integral \isa{x} with respect to the measure space \isa{M}, \isa{integrable M f} to say that \isa{f} is integrable, and notation $\isa{integral$^L$ M f}$ for the value of the integral, when it exists (the superscripted \isa{L} is a holdover from Lebesgue measure). The library tends to favor the latter representations, however, with one theorem asserting the value of an integral, and another theorem asserting integrability. For example, we have:
\Snippet{integraladd}
\Snippet{integrableadd}
This sometimes got us into trouble. A couple of times, we used theorems in the library, only to realize that the accompanying integrability assertions were missing; we then had to revise the library to provide these additional assertions. We often made the same mistake in our own developments, and in proofs we often found that we had to carry out parallel calculations: after calculating an integral, we had to go back and prove that the expression we began with was in fact integrable. In some cases, Isabelle's automation could dispel integrability claims for us, but typically in those cases the calculations could also be carried out automatically. On the other hand, writing formulas with integrals rather than the \isa{has\_integral} predicate makes them look much more like the formulas one finds in an ordinary mathematical textbook.

We do not know the ideal solution to the problem. In a dependent type theory with propositions as types, one could require \isa{integrable f} as a ``precondition'' --- a hidden argument --- to an expression \isa{integral f}. Coq's constructive C-CoRN library uses such an approach \cite{krebbers:spitters:11}, but the Coquelicot project uses a classical axiomatization of the real numbers to totalize limits \cite{boldo:et:al:12}. In simple type theory, it seems that one has to choose between using a relational version or using a function together with a definedness predicate. In the latter case, one has to take care to keep the two pieces of information close together.

\subsection{Strategies for limit proofs}

It is not always obvious how to carry out limit proofs at the right level of formal abstraction. With measure theory, it is often advantageous to adopt an order-theoretic point of view: instead of proving that a function approaches a certain point in a $\varepsilon$-environment, it is sometimes preferable to do this separately for an upper and lower bound. This is contrary to what is done in Billingsley~\cite{billingsley:95}, where many proofs, like that of the Helly selection theorem, are performed by choosing $\varepsilon$'s. That approach works when the domain is the set of real numbers or at least a metric space. But we often needed to use the extended real number structures $\eRR$ or $\ennRR$, for two reasons: (1) we reason about measures which are not necessarily finite, and (2) we reason about $\liminf$'s and $\limsup$'s, which are defined on complete lattices but not on the real numbers. Working with $\varepsilon$'s typically requires us to compute differences, which is difficult on $\eRR$ or $\ennRR$, where subtraction and addition are not as well behaved as they are on $\RR$. For example, if we use metric limits to prove a property on a neighborhood of $f~x$, we may obtain an $\varepsilon > 0$, but we do not immediately have the property $f~x < f~x + \varepsilon$, since this fails for $f~x = \infty$. Using the order topology on extended real number structures, however, we still have limits and filters, but instead of obtaining a $\varepsilon$-neighborhood of a specific point, we get upper and lower bounds on values for which the property still holds. In the previous example, using an order-theoretic limit we instead get an upper bound $y$ with $f~x < y$, and this moreover implies $f~x < \infty$.

Even using $\varepsilon$-proofs in a metric-space setting can be formally inconvenient. Many textbook proofs adopt a style whereby various $\varepsilon$'s are obtained in the course of the argument, for example, as diameters of neighborhoods in open sets or neighborhoods in the range of continuous functions. The properties of all these $\varepsilon$'s are combined by computing a minimal $\varepsilon$ value and proving it correct. Textbooks often elide such details by taking an $\varepsilon$ that is ``sufficiently small.''

Formally, it is often better to avoid the uses of $\varepsilon$ values entirely, showing instead that the required properties give rise to a set in the relevant filter. For example, suppose that the functions $f$ and $g_i$, for $i < n$, are continuous at $x$, and that $1 < f~x$ and $f~y \le g_i~y$ around $x$. Suppose further that we want to obtain a neighborhood of $x$ where $f$ is above $1$ and below all the $g_i$s. One approach is to obtain $\varepsilon$ and $\varepsilon_i$'s where $1 < f~y$ where $f~y \le g_i~y$ in the corresponding balls around $x$, and then to compute the minimum of these values, with the special case where $n = 0$. But working with these $\varepsilon$'s is completely auxiliary to our original goal. Instead, we can easily show that $\{ y \mid 1 < f~y \} \cap \bigcap_{i < n} \{ y \mid f~y \le g_i~y \}$ is in the neighborhood filter at $x$, simply using the fact that a filter is closed under finite intersections.

\subsection{Strategies for integrals}

Measure theory gives us two different integrals on measure spaces, the nonnegative Lebesgue integral and the Bochner integral, as described in Section~\ref{subsection:bochner}. The distinction is clear: the Lebesgue integral only requires a measurable function, and handles functions into the nonnegative extended reals $\ennRR$, while the Bochner integral requires an integrable function, but handles functions into arbitrary second-countable Banach spaces. There are two important advantages to using the Lebesgue integral: (1)~measurability is compositional, supporting different measurable spaces, while this is not the case for integrability, and (2)~the extended nonnegative reals include $\infty$, and so no integrability condition is needed for the integral to be closed under addition and constant multiplication. The property \isa{integrable\_iff\_bounded}, which states that a function is integrable if and only if it is bounded, provides a key way to prove integrability. Similarly, a function is integrable if it is measurable and has integrable upper and lower bounds.

A small trick that results in more convenient proof rules for the Bochner integral is to fix the value of \isa{integral $\mu$ $f$} to $0$ for a non-integrable function $f$. In interactive theorem proving, this is a common trick to totalize a function taking values in a numeric domain. Exploiting this fact avoids some auxiliary integrability rules. For the constant multiplication rules (i.e. multiplication with a real or complex value, scalar multiplication and the inner vector product) we get a rule of the form
\begin{quote}
\isa{integral $\mu$ ($\lambda x.~c * f~x$) = $c$ * integral $\mu$ $f$}
\end{quote}
without any assumptions. Also, integrability is invariant under the transformations \isa{distr} and \isa{density}:
\begin{quote}
\isa{integral (distr $\mu$ $\nu$ $f$) $g$ = integral $\mu$ $(g \circ f)$}
\end{quote}
with the assumption that $f$ and $g$ are measurable. Similarly, we allow the affine transformation under the integral:
\begin{quote}
\isa{integral lborel $f$ = $|c|$ * integral lborel $(\lambda x.~f~(t + c * x))$}
\end{quote}
with the only assumption $c \not= 0$.

Of course, we still have the problems mentioned in Subsection~\ref{subsection:partial}, i.e.~we need to prove integrability separately. (Without a proof of integrability, the results may not mean what we think they mean, and they are generally unusable.) But in many analytical proofs, integrability is proved separately anyhow, with a proof that may have little to do with the calculation of the integral value.

\subsection{Cleanup and length}

It is impossible to give a meaningful estimate of the time involved in the formalization, as the work was carried out intermittently over a long period of time, and includes time spent learning to use Isabelle by the third author, Serafin, who was an undergraduate student at Carnegie Mellon at the time. When we first obtained a proof of the CLT, we reported that our repository contained about 13,000 lines \cite{avigad:hoelzl:serafin:14}. This included all the general infrastructure and additions to Isabelle's libraries as well as the core parts of the proof. Since then, we have cleaned up and refactored most of our proof scripts, and many of them have been shortened considerably. Subsequently, H\"olzl implemented the Bochner integral, eliminating the need for a separate notion of integration for functions from the reals to the complex numbers. In addition, many of the supporting theorems and facts have been moved to other parts of the Isabelle libraries.

The proof of the CLT is now part of the Isabelle distribution. One interesting observation is that, in terms of the number of lines, the majority of effort went into developing the background and general infrastructure. For example, some of our longest files involve general facts about integration:
\begin{center}
\begin{tabular}{lr}
\verb=Bochner_Integral=: & $3,066$ lines \\
\verb=Set_Integral=: & $602$ lines \\
\verb=Interval_Integral=: & $1,123$ lines
\end{tabular}
\end{center}
In addition, the construction of Lebesgue-Stieltjes measure, described in Section~\ref{subsection:distribution:functions}, is found in \verb=Lebesgue_Measure= and requires about $270$ lines. Some of the key background for the formalization is contained in the following files:
\begin{center}
\begin{tabular}{lr}
 \verb=Distribution_Functions=: & $259$ lines \\
 \verb=Weak_Convergence=: & $422$ lines \\
 \verb=Sinc_Integral=: & $403$ lines
\end{tabular}
\end{center}
Also, general facts about the standard normal distribution take about about $380$ lines in \texttt{Distributions}. The core development of characteristic functions and their properties, and the proof of the CLT, is found in the following files:
\begin{center}
\begin{tabular}{lr}
 \verb=Characteristic_Functions=: & $554$ lines \\
 \verb=Helly_Selection=: & $298$ lines \\
 \verb=Levy=: & $542$ lines \\
 \verb=Central_Limit_Theorem=: & $144$ lines
\end{tabular}
\end{center}
Once all the background information was in place, many of our proofs followed those in Billingsley quite closely. This allows for some direct comparisons. The increases in length are most dramatic in technical proofs where there are one-step arguments in the text that are indeed straightforward to verify, but nonetheless require long and tedious arguments. This includes verifying straightforward continuity claims, filling in implicit limit arguments, finding explicit choices of $\varepsilon$ sufficiently small to make a proof go through, and so on. Thus the Helly selection theorem, only 10 lines in Billingsley's text, is 127 lines in our formalization. Billingsley derives two corollaries from that, each with a proof of 9 lines; our formal versions are 102 and 18 lines, respectively. The L\'evy inversion theorem runs only 15 lines in Billingsley, and about 170 lines in our formalization. Billingsley observes that the uniqueness theorem follows from the inversion theorem with four lines of proof, which translates to 74 lines in our formal version.

\subsection{Future directions}

The version of the Central Limit Theorem we proved is not the most general version that is presented in Billingsley's book. With some more calculational effort one could formalize the Lindeberg central limit theorem, which relaxes the requirement that the random variables that are summed be identically distributed; we only need to assume that they do not deviate too much in distribution, as made precise by the {\em Lindeberg condition} \cite[p.~359]{billingsley:95}. Even the condition that the variables being summed are independent can be weakened to a condition of weak dependence, as outlined in \cite[p.~363]{billingsley:95}. Other generalizations include the CLT for random vectors \cite[p.~385]{billingsley:95}, and various versions of the CLT for martingales \cite[pp.~475--478]{billingsley:95}. There are many additional refinements and generalizations of the Central Limit Theorem in the mathematical literature.

Supporting automation can always be improved, and it was at times frustrating that automated tools would get stuck on seemingly trivial matters like determining whether an instance of zero should be interpreted as a real or an nonnegative extended real. As we remarked in Section~\ref{subsection:calculus}, carrying out ordinary calculations with integrals was often the most painful part of the formalization. It would be especially useful to have better automated support for such calculations, either implementing features of computer algebra systems in a proof-producing framework, or reconstructing formal proofs of such results from suitable certificates.


\input{clt.bbl}
\section*{Appendix}

\Snippet{cltproof}

\end{document}

%% file: snippets-generated.tex
\DefineSnippet{cltproof}{
\isacommand{theorem}\isamarkupfalse%
\ {\isacharparenleft}\isakeyword{in}\ prob{\isacharunderscore}space{\isacharparenright}\ central{\isacharunderscore}limit{\isacharunderscore}theorem{\isacharunderscore}zero{\isacharunderscore}mean{\isacharcolon}\isanewline
\ \ \isakeyword{fixes}\ X\ {\isacharcolon}{\isacharcolon}\ {\isachardoublequoteopen}nat\ {\isasymRightarrow}\ {\isacharprime}a\ {\isasymRightarrow}\ real{\isachardoublequoteclose}\isanewline
\ \ \ \ \isakeyword{and}\ {\isasymmu}\ {\isacharcolon}{\isacharcolon}\ {\isachardoublequoteopen}real\ measure{\isachardoublequoteclose}\isanewline
\ \ \ \ \isakeyword{and}\ {\isasymsigma}\ {\isacharcolon}{\isacharcolon}\ real\isanewline
\ \ \ \ \isakeyword{and}\ S\ {\isacharcolon}{\isacharcolon}\ {\isachardoublequoteopen}nat\ {\isasymRightarrow}\ {\isacharprime}a\ {\isasymRightarrow}\ real{\isachardoublequoteclose}\isanewline
\ \ \isakeyword{assumes}\ X{\isacharunderscore}indep{\isacharcolon}\ {\isachardoublequoteopen}indep{\isacharunderscore}vars\ {\isacharparenleft}{\isasymlambda}i{\isachardot}\ borel{\isacharparenright}\ X\ UNIV{\isachardoublequoteclose}\isanewline
\ \ \ \ \isakeyword{and}\ X{\isacharunderscore}integrable{\isacharcolon}\ {\isachardoublequoteopen}{\isasymAnd}n{\isachardot}\ integrable\ M\ {\isacharparenleft}X\ n{\isacharparenright}{\isachardoublequoteclose}\isanewline
\ \ \ \ \isakeyword{and}\ X{\isacharunderscore}mean{\isacharunderscore}{\isadigit{0}}{\isacharcolon}\ {\isachardoublequoteopen}{\isasymAnd}n{\isachardot}\ expectation\ {\isacharparenleft}X\ n{\isacharparenright}\ {\isacharequal}\ {\isadigit{0}}{\isachardoublequoteclose}\isanewline
\ \ \ \ \isakeyword{and}\ {\isasymsigma}{\isacharunderscore}pos{\isacharcolon}\ {\isachardoublequoteopen}{\isasymsigma}\ {\isachargreater}\ {\isadigit{0}}{\isachardoublequoteclose}\isanewline
\ \ \ \ \isakeyword{and}\ X{\isacharunderscore}square{\isacharunderscore}integrable{\isacharcolon}\ {\isachardoublequoteopen}{\isasymAnd}n{\isachardot}\ integrable\ M\ {\isacharparenleft}{\isasymlambda}x{\isachardot}\ {\isacharparenleft}X\ n\ x{\isacharparenright}\isactrlsup {\isadigit{2}}{\isacharparenright}{\isachardoublequoteclose}\isanewline
\ \ \ \ \isakeyword{and}\ X{\isacharunderscore}variance{\isacharcolon}\ {\isachardoublequoteopen}{\isasymAnd}n{\isachardot}\ variance\ {\isacharparenleft}X\ n{\isacharparenright}\ {\isacharequal}\ {\isasymsigma}\isactrlsup {\isadigit{2}}{\isachardoublequoteclose}\isanewline
\ \ \ \ \isakeyword{and}\ X{\isacharunderscore}distrib{\isacharcolon}\ {\isachardoublequoteopen}{\isasymAnd}n{\isachardot}\ distr\ M\ borel\ {\isacharparenleft}X\ n{\isacharparenright}\ {\isacharequal}\ {\isasymmu}{\isachardoublequoteclose}\isanewline
\ \ \isakeyword{defines}\ {\isachardoublequoteopen}S\ n\ {\isasymequiv}\ {\isasymlambda}x{\isachardot}\ {\isasymSum}i{\isacharless}n{\isachardot}\ X\ i\ x{\isachardoublequoteclose}\isanewline
\ \ \isakeyword{shows}\ {\isachardoublequoteopen}weak{\isacharunderscore}conv{\isacharunderscore}m\isanewline
\ \ \ \ {\isacharparenleft}{\isasymlambda}n{\isachardot}\ distr\ M\ borel\ {\isacharparenleft}{\isasymlambda}x{\isachardot}\ S\ n\ x\ {\isacharslash}\ sqrt\ {\isacharparenleft}n\ {\isacharasterisk}\ {\isasymsigma}\isactrlsup {\isadigit{2}}{\isacharparenright}{\isacharparenright}{\isacharparenright}\isanewline
\ \ \ \ std{\isacharunderscore}normal{\isacharunderscore}distribution{\isachardoublequoteclose}\isanewline
\isadelimproof
\endisadelimproof
\isatagproof
\isacommand{proof}\isamarkupfalse%
\ {\isacharminus}\isanewline
\ \ \isacommand{let}\isamarkupfalse%
\ {\isacharquery}S{\isacharprime}\ {\isacharequal}\ {\isachardoublequoteopen}{\isasymlambda}n\ x{\isachardot}\ S\ n\ x\ {\isacharslash}\ sqrt\ {\isacharparenleft}real\ n\ {\isacharasterisk}\ {\isasymsigma}\isactrlsup {\isadigit{2}}{\isacharparenright}{\isachardoublequoteclose}\isanewline
\ \ \ \ \isakeyword{and}\ {\isacharquery}m\ {\isacharequal}\ {\isachardoublequoteopen}{\isasymlambda}x{\isachardot}\ min\ {\isacharparenleft}{\isadigit{6}}\ {\isacharasterisk}\ x\isactrlsup {\isadigit{2}}{\isacharparenright}{\isachardoublequoteclose}\isanewline
\ \ \isacommand{define}\isamarkupfalse%
\ {\isasymphi}\ \isakeyword{where}\ {\isachardoublequoteopen}{\isasymphi}\ n\ {\isacharequal}\ char\ {\isacharparenleft}distr\ M\ borel\ {\isacharparenleft}{\isacharquery}S{\isacharprime}\ n{\isacharparenright}{\isacharparenright}{\isachardoublequoteclose}\ \isakeyword{for}\ n\isanewline
\ \ \isacommand{define}\isamarkupfalse%
\ {\isasympsi}\ \isakeyword{where}\ {\isachardoublequoteopen}{\isasympsi}\ n\ t\ {\isacharequal}\ char\ {\isasymmu}\ {\isacharparenleft}t\ {\isacharslash}\ sqrt\ {\isacharparenleft}{\isasymsigma}\isactrlsup {\isadigit{2}}\ {\isacharasterisk}\ n{\isacharparenright}{\isacharparenright}{\isachardoublequoteclose}\ \isakeyword{for}\ n\ t\isanewline
\isanewline
\ \ \isacommand{have}\isamarkupfalse%
\ X{\isacharunderscore}rv\ {\isacharbrackleft}simp{\isacharcomma}\ measurable{\isacharbrackright}{\isacharcolon}\ {\isachardoublequoteopen}{\isasymAnd}n{\isachardot}\ random{\isacharunderscore}variable\ borel\ {\isacharparenleft}X\ n{\isacharparenright}{\isachardoublequoteclose}\isanewline
\ \ \ \ \isacommand{using}\isamarkupfalse%
\ X{\isacharunderscore}indep\ \isacommand{unfolding}\isamarkupfalse%
\ indep{\isacharunderscore}vars{\isacharunderscore}def{\isadigit{2}}\ \isacommand{by}\isamarkupfalse%
\ simp\isanewline
\ \ \isacommand{interpret}\isamarkupfalse%
\ {\isasymmu}{\isacharcolon}\ real{\isacharunderscore}distribution\ {\isasymmu}\isanewline
\ \ \ \ \isacommand{by}\isamarkupfalse%
\ {\isacharparenleft}subst\ X{\isacharunderscore}distrib\ {\isacharbrackleft}symmetric{\isacharcomma}\ of\ {\isadigit{0}}{\isacharbrackright}{\isacharcomma}\ \isanewline
\ \ \ \ \ \ \ \ rule\ real{\isacharunderscore}distribution{\isacharunderscore}distr{\isacharcomma}\ simp{\isacharparenright}\isanewline
\isanewline
\ \ \isacommand{have}\isamarkupfalse%
\ {\isasymmu}{\isacharunderscore}integrable\ {\isacharbrackleft}simp{\isacharbrackright}{\isacharcolon}\ {\isachardoublequoteopen}integrable\ {\isasymmu}\ {\isacharparenleft}{\isasymlambda}x{\isachardot}\ x{\isacharparenright}{\isachardoublequoteclose}\isanewline
\ \ \ \ \isakeyword{and}\ {\isasymmu}{\isacharunderscore}mean{\isacharunderscore}integrable\ {\isacharbrackleft}simp{\isacharbrackright}{\isacharcolon}\ {\isachardoublequoteopen}{\isasymmu}{\isachardot}expectation\ {\isacharparenleft}{\isasymlambda}x{\isachardot}\ x{\isacharparenright}\ {\isacharequal}\ {\isadigit{0}}{\isachardoublequoteclose}\isanewline
\ \ \ \ \isakeyword{and}\ {\isasymmu}{\isacharunderscore}square{\isacharunderscore}integrable\ {\isacharbrackleft}simp{\isacharbrackright}{\isacharcolon}\ {\isachardoublequoteopen}integrable\ {\isasymmu}\ {\isacharparenleft}{\isasymlambda}x{\isachardot}\ x{\isacharcircum}{\isadigit{2}}{\isacharparenright}{\isachardoublequoteclose}\isanewline
\ \ \ \ \isakeyword{and}\ {\isasymmu}{\isacharunderscore}variance\ {\isacharbrackleft}simp{\isacharbrackright}{\isacharcolon}\ {\isachardoublequoteopen}{\isasymmu}{\isachardot}expectation\ {\isacharparenleft}{\isasymlambda}x{\isachardot}\ x{\isacharcircum}{\isadigit{2}}{\isacharparenright}\ {\isacharequal}\ {\isasymsigma}\isactrlsup {\isadigit{2}}{\isachardoublequoteclose}\isanewline
\ \ \ \ \isacommand{using}\isamarkupfalse%
\ assms\ \isacommand{by}\isamarkupfalse%
\ {\isacharparenleft}simp{\isacharunderscore}all\ add{\isacharcolon}\ X{\isacharunderscore}distrib\ {\isacharbrackleft}symmetric{\isacharcomma}\ of\ {\isadigit{0}}{\isacharbrackright}\isanewline
\ \ \ \ \ \ \ \ \ \ \ \ \ \ \ \ \ \ \ \ \ \ \ \ \ \ \ \ \ \ \ \ \ \ integrable{\isacharunderscore}distr{\isacharunderscore}eq\ integral{\isacharunderscore}distr{\isacharparenright}\isanewline
\isanewline
\ \ \isacommand{let}\isamarkupfalse%
\ {\isacharquery}I\ {\isacharequal}\ {\isachardoublequoteopen}{\isasymlambda}n\ t{\isachardot}\ LINT\ x{\isacharbar}{\isasymmu}{\isachardot}\ {\isacharquery}m\ x\ {\isacharparenleft}{\isasymbar}t\ {\isacharslash}\ sqrt\ {\isacharparenleft}{\isasymsigma}\isactrlsup {\isadigit{2}}\ {\isacharasterisk}\ n{\isacharparenright}{\isasymbar}\ {\isacharasterisk}\ {\isasymbar}x{\isasymbar}\ {\isacharcircum}\ {\isadigit{3}}{\isacharparenright}{\isachardoublequoteclose}\isanewline
\ \ \isacommand{have}\isamarkupfalse%
\ main{\isacharcolon}\ {\isachardoublequoteopen}{\isasymforall}\isactrlsub F\ n\ in\ sequentially{\isachardot}\isanewline
\ \ \ \ \ \ cmod\ {\isacharparenleft}{\isasymphi}\ n\ t\ {\isacharminus}\ {\isacharparenleft}{\isadigit{1}}\ {\isacharplus}\ {\isacharparenleft}{\isacharminus}{\isacharparenleft}t{\isacharcircum}{\isadigit{2}}{\isacharparenright}\ {\isacharslash}\ {\isadigit{2}}{\isacharparenright}\ {\isacharslash}\ n{\isacharparenright}{\isacharcircum}n{\isacharparenright}\ {\isasymle}\ \isanewline
\ \ \ \ \ \ \ \ t\isactrlsup {\isadigit{2}}\ {\isacharslash}\ {\isacharparenleft}{\isadigit{6}}\ {\isacharasterisk}\ {\isasymsigma}\isactrlsup {\isadigit{2}}{\isacharparenright}\ {\isacharasterisk}\ {\isacharquery}I\ n\ t{\isachardoublequoteclose}\isanewline
\ \ \ \ \ \ \isakeyword{for}\ t\isanewline
\ \ \isacommand{proof}\isamarkupfalse%
\ {\isacharparenleft}rule\ eventually{\isacharunderscore}sequentiallyI{\isacharparenright}\isanewline
\ \ \ \ \isacommand{fix}\isamarkupfalse%
\ n\ {\isacharcolon}{\isacharcolon}\ nat\isanewline
\ \ \ \ \isacommand{assume}\isamarkupfalse%
\ {\isachardoublequoteopen}n\ {\isasymge}\ nat\ {\isacharparenleft}ceiling\ {\isacharparenleft}t{\isacharcircum}{\isadigit{2}}\ {\isacharslash}\ {\isadigit{4}}{\isacharparenright}{\isacharparenright}{\isachardoublequoteclose}\isanewline
\ \ \ \ \isacommand{hence}\isamarkupfalse%
\ n{\isacharcolon}\ {\isachardoublequoteopen}n\ {\isasymge}\ t{\isacharcircum}{\isadigit{2}}\ {\isacharslash}\ {\isadigit{4}}{\isachardoublequoteclose}\ \isacommand{by}\isamarkupfalse%
\ {\isacharparenleft}subst\ nat{\isacharunderscore}ceiling{\isacharunderscore}le{\isacharunderscore}eq\ {\isacharbrackleft}symmetric{\isacharbrackright}{\isacharparenright}\isanewline
\ \ \ \ \isacommand{let}\isamarkupfalse%
\ {\isacharquery}t\ {\isacharequal}\ {\isachardoublequoteopen}t\ {\isacharslash}\ sqrt\ {\isacharparenleft}{\isasymsigma}\isactrlsup {\isadigit{2}}\ {\isacharasterisk}\ n{\isacharparenright}{\isachardoublequoteclose}\isanewline
\isanewline
\ \ \ \ \isacommand{define}\isamarkupfalse%
\ {\isasympsi}{\isacharprime}\ \isakeyword{where}\ {\isachardoublequoteopen}{\isasympsi}{\isacharprime}\ n\ i\ {\isacharequal}\ char\ {\isacharparenleft}distr\ M\ borel\isanewline
\ \ \ \ \ \ {\isacharparenleft}{\isasymlambda}x{\isachardot}\ X\ i\ x\ {\isacharslash}\ sqrt\ {\isacharparenleft}{\isasymsigma}\isactrlsup {\isadigit{2}}\ {\isacharasterisk}\ n{\isacharparenright}{\isacharparenright}{\isacharparenright}{\isachardoublequoteclose}\ \isakeyword{for}\ n\ i\isanewline
\ \ \ \ \isacommand{have}\isamarkupfalse%
\ {\isacharasterisk}{\isacharcolon}\ {\isachardoublequoteopen}{\isasymAnd}n\ i\ t{\isachardot}\ {\isasympsi}{\isacharprime}\ n\ i\ t\ {\isacharequal}\ {\isasympsi}\ n\ t{\isachardoublequoteclose}\isanewline
\ \ \ \ \ \ \isacommand{unfolding}\isamarkupfalse%
\ {\isasympsi}{\isacharunderscore}def\ {\isasympsi}{\isacharprime}{\isacharunderscore}def\ char{\isacharunderscore}def\isanewline
\ \ \ \ \ \ \isacommand{by}\isamarkupfalse%
\ {\isacharparenleft}subst\ X{\isacharunderscore}distrib\ {\isacharbrackleft}symmetric{\isacharbrackright}{\isacharparenright}\ {\isacharparenleft}auto\ simp{\isacharcolon}\ integral{\isacharunderscore}distr{\isacharparenright}\isanewline
\isanewline
\ \ \ \ \isacommand{have}\isamarkupfalse%
\ {\isachardoublequoteopen}{\isasymphi}\ n\ t\ {\isacharequal}\ char\ {\isacharparenleft}distr\ M\ borel\isanewline
\ \ \ \ \ \ \ {\isacharparenleft}{\isasymlambda}x{\isachardot}\ {\isasymSum}i{\isacharless}n{\isachardot}\ X\ i\ x\ {\isacharslash}\ sqrt\ {\isacharparenleft}{\isasymsigma}\isactrlsup {\isadigit{2}}\ {\isacharasterisk}\ real\ n{\isacharparenright}{\isacharparenright}{\isacharparenright}\ t{\isachardoublequoteclose}\isanewline
\ \ \ \ \ \ \isacommand{by}\isamarkupfalse%
\ {\isacharparenleft}auto\ simp{\isacharcolon}\ {\isasymphi}{\isacharunderscore}def\ S{\isacharunderscore}def\ sum{\isacharunderscore}divide{\isacharunderscore}distrib\ ac{\isacharunderscore}simps{\isacharparenright}\isanewline
\ \ \ \ \isacommand{also}\isamarkupfalse%
\ \isacommand{have}\isamarkupfalse%
\ {\isachardoublequoteopen}{\isasymdots}\ {\isacharequal}\ {\isacharparenleft}{\isasymProd}\ i\ {\isacharless}\ n{\isachardot}\ {\isasympsi}{\isacharprime}\ n\ i\ t{\isacharparenright}{\isachardoublequoteclose}\isanewline
\ \ \ \ \ \ \isacommand{unfolding}\isamarkupfalse%
\ {\isasympsi}{\isacharprime}{\isacharunderscore}def\isanewline
\ \ \ \ \ \ \isacommand{apply}\isamarkupfalse%
\ {\isacharparenleft}rule\ char{\isacharunderscore}distr{\isacharunderscore}sum{\isacharparenright}\isanewline
\ \ \ \ \ \ \isacommand{apply}\isamarkupfalse%
\ {\isacharparenleft}rule\ indep{\isacharunderscore}vars{\isacharunderscore}compose{\isadigit{2}}{\isacharbrackleft}\isakeyword{where}\ X{\isacharequal}X{\isacharbrackright}{\isacharparenright}\isanewline
\ \ \ \ \ \ \isacommand{apply}\isamarkupfalse%
\ {\isacharparenleft}rule\ indep{\isacharunderscore}vars{\isacharunderscore}subset{\isacharparenright}\isanewline
\ \ \ \ \ \ \isacommand{apply}\isamarkupfalse%
\ {\isacharparenleft}rule\ X{\isacharunderscore}indep{\isacharparenright}\isanewline
\ \ \ \ \ \ \isacommand{apply}\isamarkupfalse%
\ auto\isanewline
\ \ \ \ \ \ \isacommand{done}\isamarkupfalse%
\isanewline
\ \ \ \ \isacommand{also}\isamarkupfalse%
\ \isacommand{have}\isamarkupfalse%
\ {\isachardoublequoteopen}{\isasymdots}\ {\isacharequal}\ {\isacharparenleft}{\isasympsi}\ n\ t{\isacharparenright}{\isacharcircum}n{\isachardoublequoteclose}\isanewline
\ \ \ \ \ \ \isacommand{by}\isamarkupfalse%
\ {\isacharparenleft}auto\ simp\ add{\isacharcolon}\ {\isacharasterisk}\ prod{\isacharunderscore}constant{\isacharparenright}\isanewline
\ \ \ \ \isacommand{finally}\isamarkupfalse%
\ \isacommand{have}\isamarkupfalse%
\ {\isasymphi}{\isacharunderscore}eq{\isacharcolon}\ {\isachardoublequoteopen}{\isasymphi}\ n\ t\ {\isacharequal}\ {\isacharparenleft}{\isasympsi}\ n\ t{\isacharparenright}{\isacharcircum}n{\isachardoublequoteclose}\ \isacommand{{\isachardot}}\isamarkupfalse%
\isanewline
\isanewline
\ \ \ \ \isacommand{have}\isamarkupfalse%
\ {\isachardoublequoteopen}norm\ {\isacharparenleft}{\isasympsi}\ n\ t\ {\isacharminus}\ {\isacharparenleft}{\isadigit{1}}\ {\isacharminus}\ {\isacharquery}t{\isacharcircum}{\isadigit{2}}\ {\isacharasterisk}\ {\isasymsigma}\isactrlsup {\isadigit{2}}\ {\isacharslash}\ {\isadigit{2}}{\isacharparenright}{\isacharparenright}\ {\isasymle}\isanewline
\ \ \ \ \ \ \ \ {\isacharquery}t\isactrlsup {\isadigit{2}}\ {\isacharslash}\ {\isadigit{6}}\ {\isacharasterisk}\ {\isacharquery}I\ n\ t{\isachardoublequoteclose}\isanewline
\ \ \ \ \ \ \isacommand{unfolding}\isamarkupfalse%
\ {\isasympsi}{\isacharunderscore}def\ \isacommand{by}\isamarkupfalse%
\ {\isacharparenleft}rule\ {\isasymmu}{\isachardot}char{\isacharunderscore}approx{\isadigit{3}}{\isacharcomma}\ auto{\isacharparenright}\isanewline
\ \ \ \ \isacommand{also}\isamarkupfalse%
\ \isacommand{have}\isamarkupfalse%
\ {\isachardoublequoteopen}{\isacharquery}t{\isacharcircum}{\isadigit{2}}\ {\isacharasterisk}\ {\isasymsigma}\isactrlsup {\isadigit{2}}\ {\isacharequal}\ t{\isacharcircum}{\isadigit{2}}\ {\isacharslash}\ n{\isachardoublequoteclose}\isanewline
\ \ \ \ \ \ \isacommand{using}\isamarkupfalse%
\ {\isasymsigma}{\isacharunderscore}pos\ \isacommand{by}\isamarkupfalse%
\ {\isacharparenleft}simp\ add{\isacharcolon}\ power{\isacharunderscore}divide{\isacharparenright}\isanewline
\ \ \ \ \isacommand{also}\isamarkupfalse%
\ \isacommand{have}\isamarkupfalse%
\ {\isachardoublequoteopen}t{\isacharcircum}{\isadigit{2}}\ {\isacharslash}\ n\ {\isacharslash}\ {\isadigit{2}}\ {\isacharequal}\ {\isacharparenleft}t{\isacharcircum}{\isadigit{2}}\ {\isacharslash}\ {\isadigit{2}}{\isacharparenright}\ {\isacharslash}\ n{\isachardoublequoteclose}\isanewline
\ \ \ \ \ \ \isacommand{by}\isamarkupfalse%
\ simp\isanewline
\ \ \ \ \isacommand{finally}\isamarkupfalse%
\ \isacommand{have}\isamarkupfalse%
\ {\isacharasterisk}{\isacharasterisk}{\isacharcolon}\ {\isachardoublequoteopen}norm\ {\isacharparenleft}{\isasympsi}\ n\ t\ {\isacharminus}\ {\isacharparenleft}{\isadigit{1}}\ {\isacharplus}\ {\isacharparenleft}{\isacharminus}{\isacharparenleft}t{\isacharcircum}{\isadigit{2}}{\isacharparenright}\ {\isacharslash}\ {\isadigit{2}}{\isacharparenright}\ {\isacharslash}\ n{\isacharparenright}{\isacharparenright}\ {\isasymle}\isanewline
\ \ \ \ \ \ \ \ {\isacharquery}t\isactrlsup {\isadigit{2}}\ {\isacharslash}\ {\isadigit{6}}\ {\isacharasterisk}\ {\isacharquery}I\ n\ t{\isachardoublequoteclose}\isanewline
\ \ \ \ \ \ \isacommand{by}\isamarkupfalse%
\ simp\isanewline
\isanewline
\ \ \ \ \isacommand{have}\isamarkupfalse%
\ {\isachardoublequoteopen}norm\ {\isacharparenleft}{\isasymphi}\ n\ t\ {\isacharminus}\ {\isacharparenleft}of{\isacharunderscore}real\ {\isacharparenleft}{\isadigit{1}}\ {\isacharplus}\ {\isacharparenleft}{\isacharminus}{\isacharparenleft}t{\isacharcircum}{\isadigit{2}}{\isacharparenright}\ {\isacharslash}\ {\isadigit{2}}{\isacharparenright}\ {\isacharslash}\ n{\isacharparenright}{\isacharparenright}{\isacharcircum}n{\isacharparenright}\ {\isasymle}\isanewline
\ \ \ \ \ \ \ \ n\ {\isacharasterisk}\ norm\ {\isacharparenleft}{\isasympsi}\ n\ t\ {\isacharminus}\ {\isacharparenleft}of{\isacharunderscore}real\ {\isacharparenleft}{\isadigit{1}}\ {\isacharplus}\ {\isacharparenleft}{\isacharminus}{\isacharparenleft}t{\isacharcircum}{\isadigit{2}}{\isacharparenright}\ {\isacharslash}\ {\isadigit{2}}{\isacharparenright}\ {\isacharslash}\ n{\isacharparenright}{\isacharparenright}{\isacharparenright}{\isachardoublequoteclose}\isanewline
\ \ \ \ \ \ \isacommand{using}\isamarkupfalse%
\ n\ \isacommand{unfolding}\isamarkupfalse%
\ {\isasymphi}{\isacharunderscore}eq\ {\isasympsi}{\isacharunderscore}def\isanewline
\ \ \ \ \ \ \isacommand{by}\isamarkupfalse%
\ {\isacharparenleft}auto\ intro{\isacharbang}{\isacharcolon}\ norm{\isacharunderscore}power{\isacharunderscore}diff\ {\isasymmu}{\isachardot}cmod{\isacharunderscore}char{\isacharunderscore}le{\isacharunderscore}{\isadigit{1}}\ abs{\isacharunderscore}leI\isanewline
\ \ \ \ \ \ \ \ \ \ \ \ \ \ \ simp\ del{\isacharcolon}\ of{\isacharunderscore}real{\isacharunderscore}diff\isanewline
\ \ \ \ \ \ \ \ \ \ \ \ \ \ \ simp{\isacharcolon}\ of{\isacharunderscore}real{\isacharunderscore}diff{\isacharbrackleft}symmetric{\isacharbrackright}\ divide{\isacharunderscore}le{\isacharunderscore}eq{\isacharparenright}\isanewline
\ \ \ \ \isacommand{also}\isamarkupfalse%
\ \isacommand{have}\isamarkupfalse%
\ {\isachardoublequoteopen}{\isasymdots}\ {\isasymle}\ n\ {\isacharasterisk}\ {\isacharparenleft}{\isacharquery}t\isactrlsup {\isadigit{2}}\ {\isacharslash}\ {\isadigit{6}}\ {\isacharasterisk}\ {\isacharquery}I\ n\ t{\isacharparenright}{\isachardoublequoteclose}\isanewline
\ \ \ \ \ \ \isacommand{by}\isamarkupfalse%
\ {\isacharparenleft}rule\ mult{\isacharunderscore}left{\isacharunderscore}mono\ {\isacharbrackleft}OF\ {\isacharasterisk}{\isacharasterisk}{\isacharbrackright}{\isacharcomma}\ simp{\isacharparenright}\isanewline
\ \ \ \ \isacommand{also}\isamarkupfalse%
\ \isacommand{have}\isamarkupfalse%
\ {\isachardoublequoteopen}{\isasymdots}\ {\isacharequal}\ {\isacharparenleft}t\isactrlsup {\isadigit{2}}\ {\isacharslash}\ {\isacharparenleft}{\isadigit{6}}\ {\isacharasterisk}\ {\isasymsigma}\isactrlsup {\isadigit{2}}{\isacharparenright}\ {\isacharasterisk}\ {\isacharquery}I\ n\ t{\isacharparenright}{\isachardoublequoteclose}\isanewline
\ \ \ \ \ \ \isacommand{using}\isamarkupfalse%
\ {\isasymsigma}{\isacharunderscore}pos\ \isacommand{by}\isamarkupfalse%
\ {\isacharparenleft}simp\ add{\isacharcolon}\ field{\isacharunderscore}simps\ min{\isacharunderscore}absorb{\isadigit{2}}{\isacharparenright}\isanewline
\ \ \ \ \isacommand{finally}\isamarkupfalse%
\ \isacommand{show}\isamarkupfalse%
\ {\isachardoublequoteopen}norm\ {\isacharparenleft}{\isasymphi}\ n\ t\ {\isacharminus}\ {\isacharparenleft}{\isadigit{1}}\ {\isacharplus}\ {\isacharparenleft}{\isacharminus}{\isacharparenleft}t{\isacharcircum}{\isadigit{2}}{\isacharparenright}\ {\isacharslash}\ {\isadigit{2}}{\isacharparenright}\ {\isacharslash}\ n{\isacharparenright}{\isacharcircum}n{\isacharparenright}\ {\isasymle}\isanewline
\ \ \ \ \ \ \ \ {\isacharparenleft}t\isactrlsup {\isadigit{2}}\ {\isacharslash}\ {\isacharparenleft}{\isadigit{6}}\ {\isacharasterisk}\ {\isasymsigma}\isactrlsup {\isadigit{2}}{\isacharparenright}\ {\isacharasterisk}\ {\isacharquery}I\ n\ t{\isacharparenright}{\isachardoublequoteclose}\isanewline
\ \ \ \ \ \ \isacommand{by}\isamarkupfalse%
\ simp\isanewline
\ \ \isacommand{qed}\isamarkupfalse%
\isanewline
\isanewline
\ \ \isacommand{show}\isamarkupfalse%
\ {\isacharquery}thesis\isanewline
\ \ \isacommand{proof}\isamarkupfalse%
\ {\isacharparenleft}rule\ levy{\isacharunderscore}continuity{\isacharparenright}\isanewline
\ \ \ \ \isacommand{fix}\isamarkupfalse%
\ t\isanewline
\ \ \ \ \isacommand{have}\isamarkupfalse%
\ {\isachardoublequoteopen}{\isacharparenleft}{\isasymlambda}n{\isachardot}\ {\isacharquery}m\ x\ {\isacharparenleft}{\isasymbar}t{\isasymbar}\ {\isacharasterisk}\ {\isasymbar}x{\isasymbar}\ {\isacharcircum}\ {\isadigit{3}}\ {\isacharslash}\ {\isasymbar}sqrt\ {\isacharparenleft}{\isasymsigma}\isactrlsup {\isadigit{2}}\ {\isacharasterisk}\ real\ n{\isacharparenright}{\isasymbar}{\isacharparenright}{\isacharparenright}\isanewline
\ \ \ \ \ \ \ \ {\isasymlonglonglongrightarrow}\ {\isadigit{0}}{\isachardoublequoteclose}\ \isakeyword{for}\ x\isanewline
\ \ \ \ \ \ \isacommand{using}\isamarkupfalse%
\ {\isasymsigma}{\isacharunderscore}pos\isanewline
\ \ \ \ \ \ \isacommand{by}\isamarkupfalse%
\ {\isacharparenleft}auto\ simp{\isacharcolon}\ real{\isacharunderscore}sqrt{\isacharunderscore}mult\ min{\isacharunderscore}absorb{\isadigit{2}}\isanewline
\ \ \ \ \ \ \ \ \ \ \ \ \ \ \ intro{\isacharbang}{\isacharcolon}\ tendsto{\isacharunderscore}min{\isacharbrackleft}THEN\ tendsto{\isacharunderscore}eq{\isacharunderscore}rhs{\isacharbrackright}\isanewline
\ \ \ \ \ \ \ \ \ \ \ \ \ \ \ \ \ \ \ \ \ \ \ sqrt{\isacharunderscore}at{\isacharunderscore}top{\isacharbrackleft}THEN\ filterlim{\isacharunderscore}compose{\isacharbrackright}\isanewline
\ \ \ \ \ \ \ \ \ \ \ \ \ \ \ \ \ \ \ \ \ \ \ filterlim{\isacharunderscore}tendsto{\isacharunderscore}pos{\isacharunderscore}mult{\isacharunderscore}at{\isacharunderscore}top\isanewline
\ \ \ \ \ \ \ \ \ \ \ \ \ \ \ \ \ \ \ \ \ \ \ filterlim{\isacharunderscore}at{\isacharunderscore}top{\isacharunderscore}imp{\isacharunderscore}at{\isacharunderscore}infinity\isanewline
\ \ \ \ \ \ \ \ \ \ \ \ \ \ \ \ \ \ \ \ \ \ \ tendsto{\isacharunderscore}divide{\isacharunderscore}{\isadigit{0}}\isanewline
\ \ \ \ \ \ \ \ \ \ \ \ \ \ \ \ \ \ \ \ \ \ \ filterlim{\isacharunderscore}real{\isacharunderscore}sequentially{\isacharparenright}\isanewline
\ \ \ \ \isacommand{then}\isamarkupfalse%
\ \isacommand{have}\isamarkupfalse%
\ {\isachardoublequoteopen}{\isacharparenleft}{\isasymlambda}n{\isachardot}\ {\isacharquery}I\ n\ t{\isacharparenright}\ {\isasymlonglonglongrightarrow}\ {\isacharparenleft}LINT\ x{\isacharbar}{\isasymmu}{\isachardot}\ {\isadigit{0}}{\isacharparenright}{\isachardoublequoteclose}\isanewline
\ \ \ \ \ \ \isacommand{by}\isamarkupfalse%
\ {\isacharparenleft}intro\ integral{\isacharunderscore}dominated{\isacharunderscore}convergence\ {\isacharbrackleft}\isakeyword{where}\ \isanewline
\ \ \ \ \ \ \ \ \ \ \ \ w\ {\isacharequal}\ {\isachardoublequoteopen}{\isasymlambda}x{\isachardot}\ {\isadigit{6}}\ {\isacharasterisk}\ x{\isacharcircum}{\isadigit{2}}{\isachardoublequoteclose}{\isacharbrackright}{\isacharparenright}\ auto\isanewline
\ \ \ \ \isacommand{then}\isamarkupfalse%
\ \isacommand{have}\isamarkupfalse%
\ {\isacharasterisk}{\isacharcolon}\ {\isachardoublequoteopen}{\isacharparenleft}{\isasymlambda}n{\isachardot}\ t\isactrlsup {\isadigit{2}}\ {\isacharslash}\ {\isacharparenleft}{\isadigit{6}}\ {\isacharasterisk}\ {\isasymsigma}\isactrlsup {\isadigit{2}}{\isacharparenright}\ {\isacharasterisk}\ {\isacharquery}I\ n\ t{\isacharparenright}\ {\isasymlonglonglongrightarrow}\ {\isadigit{0}}{\isachardoublequoteclose}\isanewline
\ \ \ \ \ \ \isacommand{by}\isamarkupfalse%
\ {\isacharparenleft}simp\ only{\isacharcolon}\ integral{\isacharunderscore}zero\ tendsto{\isacharunderscore}mult{\isacharunderscore}right{\isacharunderscore}zero{\isacharparenright}\isanewline
\isanewline
\ \ \ \ \isacommand{have}\isamarkupfalse%
\ {\isachardoublequoteopen}{\isacharparenleft}{\isasymlambda}n{\isachardot}\ complex{\isacharunderscore}of{\isacharunderscore}real\ {\isacharparenleft}{\isacharparenleft}{\isadigit{1}}\ {\isacharplus}\ {\isacharparenleft}{\isacharminus}{\isacharparenleft}t{\isacharcircum}{\isadigit{2}}{\isacharparenright}\ {\isacharslash}\ {\isadigit{2}}{\isacharparenright}\ {\isacharslash}\ n{\isacharparenright}{\isacharcircum}n{\isacharparenright}{\isacharparenright}\ {\isasymlonglonglongrightarrow}\isanewline
\ \ \ \ \ \ \ \ complex{\isacharunderscore}of{\isacharunderscore}real\ {\isacharparenleft}exp\ {\isacharparenleft}{\isacharminus}{\isacharparenleft}t{\isacharcircum}{\isadigit{2}}{\isacharparenright}\ {\isacharslash}\ {\isadigit{2}}{\isacharparenright}{\isacharparenright}{\isachardoublequoteclose}\isanewline
\ \ \ \ \ \ \isacommand{by}\isamarkupfalse%
\ {\isacharparenleft}rule\ isCont{\isacharunderscore}tendsto{\isacharunderscore}compose\ {\isacharbrackleft}OF\ {\isacharunderscore}\ tendsto{\isacharunderscore}exp{\isacharunderscore}limit{\isacharunderscore}sequentially{\isacharbrackright}{\isacharparenright}\isanewline
\ \ \ \ \ \ \ \ \ auto\isanewline
\ \ \ \ \isacommand{then}\isamarkupfalse%
\ \isacommand{have}\isamarkupfalse%
\ {\isachardoublequoteopen}{\isacharparenleft}{\isasymlambda}n{\isachardot}\ {\isasymphi}\ n\ t{\isacharparenright}\ {\isasymlonglonglongrightarrow}\ of{\isacharunderscore}real\ {\isacharparenleft}exp\ {\isacharparenleft}{\isacharminus}{\isacharparenleft}t{\isacharcircum}{\isadigit{2}}{\isacharparenright}\ {\isacharslash}\ {\isadigit{2}}{\isacharparenright}{\isacharparenright}{\isachardoublequoteclose}\isanewline
\ \ \ \ \ \ \isacommand{by}\isamarkupfalse%
\ {\isacharparenleft}rule\ Lim{\isacharunderscore}transform{\isacharparenright}\isanewline
\ \ \ \ \ \ \ \ \ {\isacharparenleft}rule\ Lim{\isacharunderscore}null{\isacharunderscore}comparison\ {\isacharbrackleft}OF\ main\ {\isacharasterisk}{\isacharbrackright}{\isacharparenright}\isanewline
\ \ \ \ \isacommand{then}\isamarkupfalse%
\ \isacommand{show}\isamarkupfalse%
\ {\isachardoublequoteopen}{\isacharparenleft}{\isasymlambda}n{\isachardot}\ char\ {\isacharparenleft}distr\ M\ borel\ {\isacharparenleft}{\isacharquery}S{\isacharprime}\ n{\isacharparenright}{\isacharparenright}\ t{\isacharparenright}\ {\isasymlonglonglongrightarrow}\isanewline
\ \ \ \ \ \ \ \ char\ std{\isacharunderscore}normal{\isacharunderscore}distribution\ t{\isachardoublequoteclose}\isanewline
\ \ \ \ \ \ \isacommand{by}\isamarkupfalse%
\ {\isacharparenleft}simp\ add{\isacharcolon}\ {\isasymphi}{\isacharunderscore}def\ char{\isacharunderscore}std{\isacharunderscore}normal{\isacharunderscore}distribution{\isacharparenright}\isanewline
\ \ \isacommand{qed}\isamarkupfalse%
\ {\isacharparenleft}auto\ intro{\isacharbang}{\isacharcolon}\ real{\isacharunderscore}dist{\isacharunderscore}normal{\isacharunderscore}dist\ simp{\isacharcolon}\ S{\isacharunderscore}def{\isacharparenright}\isanewline
\isacommand{qed}\isamarkupfalse%
\endisatagproof
{\isafoldproof}%
\isadelimproof
\isanewline
\endisadelimproof
\isanewline
\isacommand{theorem}\isamarkupfalse%
\ {\isacharparenleft}\isakeyword{in}\ prob{\isacharunderscore}space{\isacharparenright}\ central{\isacharunderscore}limit{\isacharunderscore}theorem{\isacharcolon}\isanewline
\ \ \isakeyword{fixes}\ X\ {\isacharcolon}{\isacharcolon}\ {\isachardoublequoteopen}nat\ {\isasymRightarrow}\ {\isacharprime}a\ {\isasymRightarrow}\ real{\isachardoublequoteclose}\isanewline
\ \ \ \ \isakeyword{and}\ {\isasymmu}\ {\isacharcolon}{\isacharcolon}\ {\isachardoublequoteopen}real\ measure{\isachardoublequoteclose}\isanewline
\ \ \ \ \isakeyword{and}\ c\ {\isasymsigma}\ {\isacharcolon}{\isacharcolon}\ real\isanewline
\ \ \ \ \isakeyword{and}\ S\ {\isacharcolon}{\isacharcolon}\ {\isachardoublequoteopen}nat\ {\isasymRightarrow}\ {\isacharprime}a\ {\isasymRightarrow}\ real{\isachardoublequoteclose}\isanewline
\ \ \isakeyword{assumes}\ X{\isacharunderscore}indep{\isacharcolon}\ {\isachardoublequoteopen}indep{\isacharunderscore}vars\ {\isacharparenleft}{\isasymlambda}i{\isachardot}\ borel{\isacharparenright}\ X\ UNIV{\isachardoublequoteclose}\isanewline
\ \ \ \ \isakeyword{and}\ X{\isacharunderscore}integrable{\isacharcolon}\ {\isachardoublequoteopen}{\isasymAnd}n{\isachardot}\ integrable\ M\ {\isacharparenleft}X\ n{\isacharparenright}{\isachardoublequoteclose}\isanewline
\ \ \ \ \isakeyword{and}\ X{\isacharunderscore}mean{\isacharcolon}\ {\isachardoublequoteopen}{\isasymAnd}n{\isachardot}\ expectation\ {\isacharparenleft}X\ n{\isacharparenright}\ {\isacharequal}\ c{\isachardoublequoteclose}\isanewline
\ \ \ \ \isakeyword{and}\ {\isasymsigma}{\isacharunderscore}pos{\isacharcolon}\ {\isachardoublequoteopen}{\isasymsigma}\ {\isachargreater}\ {\isadigit{0}}{\isachardoublequoteclose}\isanewline
\ \ \ \ \isakeyword{and}\ X{\isacharunderscore}square{\isacharunderscore}integrable{\isacharcolon}\ {\isachardoublequoteopen}{\isasymAnd}n{\isachardot}\ integrable\ M\ {\isacharparenleft}{\isasymlambda}x{\isachardot}\ {\isacharparenleft}X\ n\ x{\isacharparenright}\isactrlsup {\isadigit{2}}{\isacharparenright}{\isachardoublequoteclose}\isanewline
\ \ \ \ \isakeyword{and}\ X{\isacharunderscore}variance{\isacharcolon}\ {\isachardoublequoteopen}{\isasymAnd}n{\isachardot}\ variance\ {\isacharparenleft}X\ n{\isacharparenright}\ {\isacharequal}\ {\isasymsigma}\isactrlsup {\isadigit{2}}{\isachardoublequoteclose}\isanewline
\ \ \ \ \isakeyword{and}\ X{\isacharunderscore}distrib{\isacharcolon}\ {\isachardoublequoteopen}{\isasymAnd}n{\isachardot}\ distr\ M\ borel\ {\isacharparenleft}X\ n{\isacharparenright}\ {\isacharequal}\ {\isasymmu}{\isachardoublequoteclose}\isanewline
\ \ \isakeyword{defines}\ {\isachardoublequoteopen}S\ n\ x\ {\isasymequiv}\ {\isasymSum}i{\isacharless}n{\isachardot}\ X\ i\ x{\isachardoublequoteclose}\isanewline
\ \ \isakeyword{shows}\ {\isachardoublequoteopen}weak{\isacharunderscore}conv{\isacharunderscore}m\isanewline
\ \ \ \ {\isacharparenleft}{\isasymlambda}n{\isachardot}\ distr\ M\ borel\ {\isacharparenleft}{\isasymlambda}x{\isachardot}\ {\isacharparenleft}S\ n\ x\ {\isacharminus}\ n\ {\isacharasterisk}\ c{\isacharparenright}\ {\isacharslash}\ sqrt\ {\isacharparenleft}n{\isacharasterisk}{\isasymsigma}\isactrlsup {\isadigit{2}}{\isacharparenright}{\isacharparenright}{\isacharparenright}\isanewline
\ \ \ \ std{\isacharunderscore}normal{\isacharunderscore}distribution{\isachardoublequoteclose}\isanewline
\isadelimproof
\endisadelimproof
\isatagproof
\isacommand{proof}\isamarkupfalse%
\ {\isacharminus}\isanewline
\ \ \isacommand{have}\isamarkupfalse%
\ {\isachardoublequoteopen}weak{\isacharunderscore}conv{\isacharunderscore}m\isanewline
\ \ \ \ {\isacharparenleft}{\isasymlambda}n{\isachardot}\ distr\ M\ borel\ {\isacharparenleft}{\isasymlambda}x{\isachardot}\ {\isacharparenleft}{\isasymSum}i{\isacharless}n{\isachardot}\ X\ i\ x\ {\isacharminus}\ c{\isacharparenright}\ {\isacharslash}\ sqrt\ {\isacharparenleft}n\ {\isacharasterisk}\ {\isasymsigma}\isactrlsup {\isadigit{2}}{\isacharparenright}{\isacharparenright}{\isacharparenright}\isanewline
\ \ \ \ std{\isacharunderscore}normal{\isacharunderscore}distribution{\isachardoublequoteclose}\isanewline
\ \ \isacommand{proof}\isamarkupfalse%
\ {\isacharparenleft}intro\ central{\isacharunderscore}limit{\isacharunderscore}theorem{\isacharunderscore}zero{\isacharunderscore}mean{\isacharparenright}\isanewline
\ \ \ \ \isacommand{show}\isamarkupfalse%
\ {\isachardoublequoteopen}indep{\isacharunderscore}vars\ {\isacharparenleft}{\isasymlambda}i{\isachardot}\ borel{\isacharparenright}\ {\isacharparenleft}{\isasymlambda}i\ x{\isachardot}\ X\ i\ x\ {\isacharminus}\ c{\isacharparenright}\ UNIV{\isachardoublequoteclose}\isanewline
\ \ \ \ \ \ \isacommand{using}\isamarkupfalse%
\ X{\isacharunderscore}indep\ \isacommand{by}\isamarkupfalse%
\ {\isacharparenleft}rule\ indep{\isacharunderscore}vars{\isacharunderscore}compose{\isadigit{2}}{\isacharparenright}\ auto\isanewline
\ \ \ \ \isacommand{show}\isamarkupfalse%
\ {\isachardoublequoteopen}integrable\ M\ {\isacharparenleft}{\isasymlambda}x{\isachardot}\ X\ n\ x\ {\isacharminus}\ c{\isacharparenright}{\isachardoublequoteclose}\isanewline
\ \ \ \ \ \ {\isachardoublequoteopen}expectation\ {\isacharparenleft}{\isasymlambda}x{\isachardot}\ X\ n\ x\ {\isacharminus}\ c{\isacharparenright}\ {\isacharequal}\ {\isadigit{0}}{\isachardoublequoteclose}\ \isakeyword{for}\ n\isanewline
\ \ \ \ \ \ \isacommand{using}\isamarkupfalse%
\ X{\isacharunderscore}integrable\ X{\isacharunderscore}mean\ \isacommand{by}\isamarkupfalse%
\ {\isacharparenleft}auto\ simp{\isacharcolon}\ prob{\isacharunderscore}space{\isacharparenright}\isanewline
\ \ \ \ \isacommand{show}\isamarkupfalse%
\ {\isachardoublequoteopen}{\isasymsigma}\ {\isachargreater}\ {\isadigit{0}}{\isachardoublequoteclose}\ {\isachardoublequoteopen}integrable\ M\ {\isacharparenleft}{\isasymlambda}x{\isachardot}\ {\isacharparenleft}X\ n\ x\ {\isacharminus}\ c{\isacharparenright}\isactrlsup {\isadigit{2}}{\isacharparenright}{\isachardoublequoteclose}\isanewline
\ \ \ \ \ \ {\isachardoublequoteopen}variance\ {\isacharparenleft}{\isasymlambda}x{\isachardot}\ X\ n\ x\ {\isacharminus}\ c{\isacharparenright}\ {\isacharequal}\ {\isasymsigma}\isactrlsup {\isadigit{2}}{\isachardoublequoteclose}\ \isakeyword{for}\ n\isanewline
\ \ \ \ \ \ \isacommand{using}\isamarkupfalse%
\ {\isacartoucheopen}{\isadigit{0}}\ {\isacharless}\ {\isasymsigma}{\isacartoucheclose}\ X{\isacharunderscore}integrable\ X{\isacharunderscore}mean\ X{\isacharunderscore}square{\isacharunderscore}integrable\ X{\isacharunderscore}variance\isanewline
\ \ \ \ \ \ \isacommand{by}\isamarkupfalse%
\ {\isacharparenleft}auto\ simp{\isacharcolon}\ prob{\isacharunderscore}space\ power{\isadigit{2}}{\isacharunderscore}diff{\isacharparenright}\isanewline
\ \ \ \ \isacommand{show}\isamarkupfalse%
\ {\isachardoublequoteopen}distr\ M\ borel\ {\isacharparenleft}{\isasymlambda}x{\isachardot}\ X\ n\ x\ {\isacharminus}\ c{\isacharparenright}\ {\isacharequal}\ \isanewline
\ \ \ \ \ \ \ \ distr\ {\isasymmu}\ borel\ {\isacharparenleft}{\isasymlambda}x{\isachardot}\ x\ {\isacharminus}\ c{\isacharparenright}{\isachardoublequoteclose}\ \isakeyword{for}\ n\isanewline
\ \ \ \ \ \ \isacommand{unfolding}\isamarkupfalse%
\ X{\isacharunderscore}distrib{\isacharbrackleft}of\ n{\isacharcomma}\ symmetric{\isacharbrackright}\ \isacommand{using}\isamarkupfalse%
\ X{\isacharunderscore}integrable\isanewline
\ \ \ \ \ \ \isacommand{by}\isamarkupfalse%
\ {\isacharparenleft}subst\ distr{\isacharunderscore}distr{\isacharparenright}\ {\isacharparenleft}auto\ simp{\isacharcolon}\ comp{\isacharunderscore}def{\isacharparenright}\isanewline
\ \ \isacommand{qed}\isamarkupfalse%
\isanewline
\ \ \isacommand{moreover}\isamarkupfalse%
\ \isacommand{have}\isamarkupfalse%
\ {\isachardoublequoteopen}{\isacharparenleft}{\isasymSum}i{\isacharless}n{\isachardot}\ X\ i\ x\ {\isacharminus}\ c{\isacharparenright}\ {\isacharequal}\ S\ n\ x\ {\isacharminus}\ n\ {\isacharasterisk}\ c{\isachardoublequoteclose}\ \isakeyword{for}\ n\ x\isanewline
\ \ \ \ \isacommand{by}\isamarkupfalse%
\ {\isacharparenleft}simp\ add{\isacharcolon}\ sum{\isacharunderscore}subtractf\ S{\isacharunderscore}def{\isacharparenright}\isanewline
\ \ \isacommand{ultimately}\isamarkupfalse%
\ \isacommand{show}\isamarkupfalse%
\ {\isacharquery}thesis\isanewline
\ \ \ \ \isacommand{by}\isamarkupfalse%
\ simp\isanewline
\isacommand{qed}\isamarkupfalse%
\endisatagproof
{\isafoldproof}%
\isadelimproof
\endisadelimproof
}
\DefineSnippet{tendstoadd}{
\isacommand{lemma}\isamarkupfalse%
\ tendsto{\isacharunderscore}add{\isacharcolon}\isanewline
\ \ \isakeyword{fixes}\ f\ g\ {\isacharcolon}{\isacharcolon}\ {\isachardoublequoteopen}{\isacharunderscore}\ {\isasymRightarrow}\ {\isacharprime}a{\isacharcolon}{\isacharcolon}topological{\isacharunderscore}monoid{\isacharunderscore}add{\isachardoublequoteclose}\isanewline
\ \ \isakeyword{assumes}\ {\isachardoublequoteopen}{\isacharparenleft}f\ {\isasymlonglongrightarrow}\ a{\isacharparenright}\ F{\isachardoublequoteclose}\ \isakeyword{and}\ {\isachardoublequoteopen}{\isacharparenleft}g\ {\isasymlonglongrightarrow}\ b{\isacharparenright}\ F{\isachardoublequoteclose}\isanewline
\ \ \isakeyword{shows}\ {\isachardoublequoteopen}{\isacharparenleft}{\isacharparenleft}{\isasymlambda}x{\isachardot}\ f\ x\ {\isacharplus}\ g\ x{\isacharparenright}\ {\isasymlonglongrightarrow}\ a\ {\isacharplus}\ b{\isacharparenright}\ F{\isachardoublequoteclose}%
}
\DefineSnippet{skorohod}{
\isacommand{theorem}\isamarkupfalse%
\ Skorohod{\isacharcolon}\isanewline
\ \ \isakeyword{fixes}\ {\isasymmu}\ {\isacharcolon}{\isacharcolon}\ {\isachardoublequoteopen}nat\ {\isasymRightarrow}\ real\ measure{\isachardoublequoteclose}\ \isakeyword{and}\ M\ {\isacharcolon}{\isacharcolon}\ {\isachardoublequoteopen}real\ measure{\isachardoublequoteclose}\isanewline
\ \ \isakeyword{assumes}\ {\isachardoublequoteopen}{\isasymAnd}n{\isachardot}\ real{\isacharunderscore}distribution\ {\isacharparenleft}{\isasymmu}\ n{\isacharparenright}{\isachardoublequoteclose}\ {\isachardoublequoteopen}real{\isacharunderscore}distribution\ M{\isachardoublequoteclose}\isanewline
\ \ \ \ \isakeyword{and}\ {\isachardoublequoteopen}weak{\isacharunderscore}conv{\isacharunderscore}m\ {\isasymmu}\ M{\isachardoublequoteclose}\isanewline
\ \ \isakeyword{shows}\isanewline
\ \ \ \ {\isachardoublequoteopen}{\isasymexists}{\isacharparenleft}{\isasymOmega}\ {\isacharcolon}{\isacharcolon}\ real\ measure{\isacharparenright}\isanewline
\ \ \ \ \ \ {\isacharparenleft}Y{\isacharunderscore}seq\ {\isacharcolon}{\isacharcolon}\ nat\ {\isasymRightarrow}\ real\ {\isasymRightarrow}\ real{\isacharparenright}\ {\isacharparenleft}Y\ {\isacharcolon}{\isacharcolon}\ real\ {\isasymRightarrow}\ real{\isacharparenright}{\isachardot}\isanewline
\ \ \ \ prob{\isacharunderscore}space\ {\isasymOmega}\ {\isasymand}\isanewline
\ \ \ \ {\isacharparenleft}{\isasymforall}n{\isachardot}\ Y{\isacharunderscore}seq\ n\ {\isasymin}\ {\isasymOmega}\ {\isasymrightarrow}\isactrlsub M\ borel{\isacharparenright}\ {\isasymand}\isanewline
\ \ \ \ {\isacharparenleft}{\isasymforall}n{\isachardot}\ distr\ {\isasymOmega}\ borel\ {\isacharparenleft}Y{\isacharunderscore}seq\ n{\isacharparenright}\ {\isacharequal}\ {\isasymmu}\ n{\isacharparenright}\ {\isasymand}\isanewline
\ \ \ \ Y\ {\isasymin}\ {\isasymOmega}\ {\isasymrightarrow}\isactrlsub M\ lborel\ {\isasymand}\isanewline
\ \ \ \ distr\ {\isasymOmega}\ borel\ Y\ {\isacharequal}\ M\ {\isasymand}\isanewline
\ \ \ \ {\isacharparenleft}{\isasymforall}x\ {\isasymin}\ space\ {\isasymOmega}{\isachardot}\ {\isacharparenleft}{\isasymlambda}n{\isachardot}\ Y{\isacharunderscore}seq\ n\ x{\isacharparenright}\ {\isasymlonglonglongrightarrow}\ Y\ x{\isacharparenright}{\isachardoublequoteclose}%
}
\DefineSnippet{countableatoms}{
\isacommand{lemma}\isamarkupfalse%
\ countable{\isacharunderscore}atoms{\isacharcolon}\isanewline
\ \ {\isachardoublequoteopen}finite{\isacharunderscore}borel{\isacharunderscore}measure\ M\ {\isasymLongrightarrow}\ countable\ {\isacharbraceleft}x{\isachardot}\ measure\ M\ {\isacharbraceleft}x{\isacharbraceright}\ {\isachargreater}\ {\isadigit{0}}{\isacharbraceright}{\isachardoublequoteclose}%
}
\DefineSnippet{clt}{
\isacommand{theorem}\isamarkupfalse%
\ {\isacharparenleft}\isakeyword{in}\ prob{\isacharunderscore}space{\isacharparenright}\ central{\isacharunderscore}limit{\isacharunderscore}theorem{\isacharcolon}\isanewline
\ \ \isakeyword{fixes}\ X\ {\isacharcolon}{\isacharcolon}\ {\isachardoublequoteopen}nat\ {\isasymRightarrow}\ {\isacharprime}a\ {\isasymRightarrow}\ real{\isachardoublequoteclose}\isanewline
\ \ \ \ \isakeyword{and}\ {\isasymmu}\ {\isacharcolon}{\isacharcolon}\ {\isachardoublequoteopen}real\ measure{\isachardoublequoteclose}\isanewline
\ \ \ \ \isakeyword{and}\ {\isasymsigma}\ c\ {\isacharcolon}{\isacharcolon}\ real\isanewline
\ \ \ \ \isakeyword{and}\ S\ {\isacharcolon}{\isacharcolon}\ {\isachardoublequoteopen}nat\ {\isasymRightarrow}\ {\isacharprime}a\ {\isasymRightarrow}\ real{\isachardoublequoteclose}\isanewline
\ \ \isakeyword{assumes}\ X{\isacharunderscore}indep{\isacharcolon}\ {\isachardoublequoteopen}indep{\isacharunderscore}vars\ {\isacharparenleft}{\isasymlambda}i{\isachardot}\ borel{\isacharparenright}\ X\ UNIV{\isachardoublequoteclose}\isanewline
\ \ \ \ \isakeyword{and}\ X{\isacharunderscore}integrable{\isacharcolon}\ {\isachardoublequoteopen}{\isasymAnd}n{\isachardot}\ integrable\ M\ {\isacharparenleft}X\ n{\isacharparenright}{\isachardoublequoteclose}\isanewline
\ \ \ \ \isakeyword{and}\ X{\isacharunderscore}mean{\isacharcolon}\ {\isachardoublequoteopen}{\isasymAnd}n{\isachardot}\ expectation\ {\isacharparenleft}X\ n{\isacharparenright}\ {\isacharequal}\ c{\isachardoublequoteclose}\isanewline
\ \ \ \ \isakeyword{and}\ {\isasymsigma}{\isacharunderscore}pos{\isacharcolon}\ {\isachardoublequoteopen}{\isasymsigma}\ {\isachargreater}\ {\isadigit{0}}{\isachardoublequoteclose}\isanewline
\ \ \ \ \isakeyword{and}\ X{\isacharunderscore}square{\isacharunderscore}integrable{\isacharcolon}\ {\isachardoublequoteopen}{\isasymAnd}n{\isachardot}\ integrable\ M\ {\isacharparenleft}{\isasymlambda}x{\isachardot}\ {\isacharparenleft}X\ n\ x{\isacharparenright}\isactrlsup {\isadigit{2}}{\isacharparenright}{\isachardoublequoteclose}\isanewline
\ \ \ \ \isakeyword{and}\ X{\isacharunderscore}variance{\isacharcolon}\ {\isachardoublequoteopen}{\isasymAnd}n{\isachardot}\ variance\ {\isacharparenleft}X\ n{\isacharparenright}\ {\isacharequal}\ {\isasymsigma}\isactrlsup {\isadigit{2}}{\isachardoublequoteclose}\isanewline
\ \ \ \ \isakeyword{and}\ X{\isacharunderscore}distrib{\isacharcolon}\ {\isachardoublequoteopen}{\isasymAnd}n{\isachardot}\ distr\ M\ borel\ {\isacharparenleft}X\ n{\isacharparenright}\ {\isacharequal}\ {\isasymmu}{\isachardoublequoteclose}\isanewline
\ \ \isakeyword{defines}\ {\isachardoublequoteopen}S\ n\ x\ {\isasymequiv}\ {\isasymSum}i{\isacharless}n{\isachardot}\ X\ i\ x{\isachardoublequoteclose}\isanewline
\ \ \isakeyword{shows}\ {\isachardoublequoteopen}weak{\isacharunderscore}conv{\isacharunderscore}m\ \isanewline
\ \ \ \ {\isacharparenleft}{\isasymlambda}n{\isachardot}\ distr\ M\ borel\ {\isacharparenleft}{\isasymlambda}x{\isachardot}\ {\isacharparenleft}S\ n\ x\ {\isacharminus}\ n\ {\isacharasterisk}\ c{\isacharparenright}\ {\isacharslash}\ sqrt\ {\isacharparenleft}n{\isacharasterisk}{\isasymsigma}\isactrlsup {\isadigit{2}}{\isacharparenright}{\isacharparenright}{\isacharparenright}\isanewline
\ \ \ \ std{\isacharunderscore}normal{\isacharunderscore}distribution{\isachardoublequoteclose}%
}
\DefineSnippet{cdf}{
\isacommand{definition}\isamarkupfalse%
\isanewline
\ \ cdf\ {\isacharcolon}{\isacharcolon}\ {\isachardoublequoteopen}real\ measure\ {\isasymRightarrow}\ real\ {\isasymRightarrow}\ real{\isachardoublequoteclose}\isanewline
\isakeyword{where}\isanewline
\ \ {\isachardoublequoteopen}cdf\ M\ {\isasymequiv}\ {\isasymlambda}x{\isachardot}\ measure\ M\ {\isacharbraceleft}{\isachardot}{\isachardot}x{\isacharbraceright}{\isachardoublequoteclose}%
}
\DefineSnippet{cdfprop1}{
\isacommand{lemma}\isamarkupfalse%
\ {\isacharparenleft}\isakeyword{in}\ finite{\isacharunderscore}borel{\isacharunderscore}measure{\isacharparenright}\ cdf{\isacharunderscore}nondecreasing{\isacharcolon}\isanewline
\ \ {\isachardoublequoteopen}x\ {\isasymle}\ y\ {\isasymLongrightarrow}\ cdf\ M\ x\ {\isasymle}\ cdf\ M\ y{\isachardoublequoteclose}%
}
\DefineSnippet{cdfprop2}{
\isacommand{lemma}\isamarkupfalse%
\ {\isacharparenleft}\isakeyword{in}\ finite{\isacharunderscore}borel{\isacharunderscore}measure{\isacharparenright}\ cdf{\isacharunderscore}is{\isacharunderscore}right{\isacharunderscore}cont{\isacharcolon}\isanewline
\ \ {\isachardoublequoteopen}continuous\ {\isacharparenleft}at{\isacharunderscore}right\ a{\isacharparenright}\ {\isacharparenleft}cdf\ M{\isacharparenright}{\isachardoublequoteclose}%
}
\DefineSnippet{cdfprop3}{
\isacommand{lemma}\isamarkupfalse%
\ {\isacharparenleft}\isakeyword{in}\ finite{\isacharunderscore}borel{\isacharunderscore}measure{\isacharparenright}\ cdf{\isacharunderscore}lim{\isacharunderscore}at{\isacharunderscore}bot{\isacharcolon}\isanewline
\ \ {\isachardoublequoteopen}{\isacharparenleft}cdf\ M\ {\isasymlonglongrightarrow}\ {\isadigit{0}}{\isacharparenright}\ at{\isacharunderscore}bot{\isachardoublequoteclose}%
}
\DefineSnippet{cdfprop4}{
\isacommand{lemma}\isamarkupfalse%
\ {\isacharparenleft}\isakeyword{in}\ real{\isacharunderscore}distribution{\isacharparenright}\ cdf{\isacharunderscore}lim{\isacharunderscore}at{\isacharunderscore}top{\isacharunderscore}prob{\isacharcolon}\isanewline
\ \ {\isachardoublequoteopen}{\isacharparenleft}cdf\ M\ {\isasymlonglongrightarrow}\ {\isadigit{1}}{\isacharparenright}\ at{\isacharunderscore}top{\isachardoublequoteclose}%
}
\DefineSnippet{intervalmeasure}{
\isacommand{lemma}\isamarkupfalse%
\ real{\isacharunderscore}distribution{\isacharunderscore}interval{\isacharunderscore}measure{\isacharcolon}\isanewline
\ \ \isakeyword{fixes}\ F\ {\isacharcolon}{\isacharcolon}\ {\isachardoublequoteopen}real\ {\isasymRightarrow}\ real{\isachardoublequoteclose}\isanewline
\ \ \isakeyword{assumes}\ {\isachardoublequoteopen}mono\ F{\isachardoublequoteclose}\ {\isachardoublequoteopen}{\isasymAnd}a{\isachardot}\ continuous\ {\isacharparenleft}at{\isacharunderscore}right\ a{\isacharparenright}\ F{\isachardoublequoteclose}\isanewline
\ \ \ \ \isakeyword{and}\ {\isachardoublequoteopen}{\isacharparenleft}F\ {\isasymlonglongrightarrow}\ {\isadigit{0}}{\isacharparenright}\ at{\isacharunderscore}bot{\isachardoublequoteclose}\ {\isachardoublequoteopen}{\isacharparenleft}F\ {\isasymlonglongrightarrow}\ {\isadigit{1}}{\isacharparenright}\ at{\isacharunderscore}top{\isachardoublequoteclose}\isanewline
\ \ \isakeyword{shows}\ {\isachardoublequoteopen}real{\isacharunderscore}distribution\ {\isacharparenleft}interval{\isacharunderscore}measure\ F{\isacharparenright}{\isachardoublequoteclose}%
}
\DefineSnippet{cdfunique}{
\isacommand{lemma}\isamarkupfalse%
\ cdf{\isacharunderscore}unique{\isacharcolon}\isanewline
\ \ \isakeyword{fixes}\ M{\isadigit{1}}\ M{\isadigit{2}}\isanewline
\ \ \isakeyword{assumes}\ {\isachardoublequoteopen}real{\isacharunderscore}distribution\ M{\isadigit{1}}{\isachardoublequoteclose}\ \isakeyword{and}\ {\isachardoublequoteopen}real{\isacharunderscore}distribution\ M{\isadigit{2}}{\isachardoublequoteclose}\isanewline
\ \ \ \ \isakeyword{and}\ {\isachardoublequoteopen}cdf\ M{\isadigit{1}}\ {\isacharequal}\ cdf\ M{\isadigit{2}}{\isachardoublequoteclose}\isanewline
\ \ \isakeyword{shows}\ {\isachardoublequoteopen}M{\isadigit{1}}\ {\isacharequal}\ M{\isadigit{2}}{\isachardoublequoteclose}%
}
\DefineSnippet{lebesgueintegralcountableadd}{
\isacommand{lemma}\isamarkupfalse%
\ integral{\isacharunderscore}countable{\isacharunderscore}add{\isacharcolon}\isanewline
\ \ \isakeyword{fixes}\ f\ {\isacharcolon}{\isacharcolon}\ {\isachardoublequoteopen}{\isacharunderscore}\ {\isasymRightarrow}\ {\isacharprime}a\ {\isacharcolon}{\isacharcolon}\ {\isacharbraceleft}banach{\isacharcomma}\ second{\isacharunderscore}countable{\isacharunderscore}topology{\isacharbraceright}{\isachardoublequoteclose}\isanewline
\ \ \isakeyword{assumes}\ {\isachardoublequoteopen}{\isasymAnd}i{\isacharcolon}{\isacharcolon}nat{\isachardot}\ A\ i\ {\isasymin}\ sets\ M{\isachardoublequoteclose}\isanewline
\ \ \ \ \isakeyword{and}\ {\isachardoublequoteopen}{\isasymAnd}i\ j{\isachardot}\ i\ {\isasymnoteq}\ j\ {\isasymLongrightarrow}\ A\ i\ {\isasyminter}\ A\ j\ {\isacharequal}\ {\isacharbraceleft}{\isacharbraceright}{\isachardoublequoteclose}\isanewline
\ \ \ \ \isakeyword{and}\ {\isachardoublequoteopen}set{\isacharunderscore}integrable\ M\ {\isacharparenleft}{\isasymUnion}i{\isachardot}\ A\ i{\isacharparenright}\ f{\isachardoublequoteclose}\isanewline
\ \ \isakeyword{shows}\ {\isachardoublequoteopen}LINT\ x{\isacharcolon}{\isacharparenleft}{\isasymUnion}i{\isachardot}\ A\ i{\isacharparenright}{\isacharbar}M{\isachardot}\ f\ x\ {\isacharequal}\ {\isacharparenleft}{\isasymSum}i{\isachardot}\ {\isacharparenleft}LINT\ x{\isacharcolon}{\isacharparenleft}A\ i{\isacharparenright}{\isacharbar}M{\isachardot}\ f\ x{\isacharparenright}{\isacharparenright}{\isachardoublequoteclose}%
}
\DefineSnippet{dominatedconvergence}{
\isacommand{lemma}\isamarkupfalse%
\ dominated{\isacharunderscore}convergence{\isacharcolon}\isanewline
\ \ \isakeyword{fixes}\ f\ {\isacharcolon}{\isacharcolon}\ {\isachardoublequoteopen}{\isacharprime}a\ {\isasymRightarrow}\ {\isacharprime}b{\isacharcolon}{\isacharcolon}{\isacharbraceleft}banach{\isacharcomma}\ second{\isacharunderscore}countable{\isacharunderscore}topology{\isacharbraceright}{\isachardoublequoteclose}\isanewline
\ \ \ \ \isakeyword{and}\ w\ {\isacharcolon}{\isacharcolon}\ {\isachardoublequoteopen}{\isacharprime}a\ {\isasymRightarrow}\ real{\isachardoublequoteclose}\isanewline
\ \ \isakeyword{assumes}\ {\isachardoublequoteopen}f\ {\isasymin}\ M\ {\isasymrightarrow}\isactrlsub M\ borel{\isachardoublequoteclose}\ {\isachardoublequoteopen}{\isasymAnd}i{\isachardot}\ s\ i\ {\isasymin}\ M\ {\isasymrightarrow}\isactrlsub M\ borel{\isachardoublequoteclose}\isanewline
\ \ \ \ \ {\isachardoublequoteopen}integrable\ M\ w{\isachardoublequoteclose}\isanewline
\ \ \ \ \isakeyword{and}\ {\isachardoublequoteopen}AE\ x\ in\ M{\isachardot}\ {\isacharparenleft}{\isasymlambda}i{\isachardot}\ s\ i\ x{\isacharparenright}\ {\isasymlonglonglongrightarrow}\ f\ x{\isachardoublequoteclose}\isanewline
\ \ \ \ \isakeyword{and}\ {\isachardoublequoteopen}{\isasymAnd}i{\isachardot}\ AE\ x\ in\ M{\isachardot}\ norm\ {\isacharparenleft}s\ i\ x{\isacharparenright}\ {\isasymle}\ w\ x{\isachardoublequoteclose}\isanewline
\ \ \isakeyword{shows}\ {\isachardoublequoteopen}integrable\ M\ f{\isachardoublequoteclose}\ \isakeyword{and}\ {\isachardoublequoteopen}{\isasymAnd}i{\isachardot}\ integrable\ M\ {\isacharparenleft}s\ i{\isacharparenright}{\isachardoublequoteclose}\isanewline
\ \ \ \ \isakeyword{and}\ {\isachardoublequoteopen}{\isacharparenleft}{\isasymlambda}i{\isachardot}\ LINT\ x{\isacharbar}M{\isachardot}\ s\ i\ x{\isacharparenright}\ {\isasymlonglonglongrightarrow}\ {\isacharparenleft}LINT\ x{\isacharbar}M{\isachardot}\ f\ x{\isacharparenright}{\isachardoublequoteclose}%
}
\DefineSnippet{integrablebounded}{
\isacommand{lemma}\isamarkupfalse%
\ integrable{\isacharunderscore}iff{\isacharunderscore}bounded{\isacharcolon}\isanewline
\ \ \isakeyword{fixes}\ f\ {\isacharcolon}{\isacharcolon}\ {\isachardoublequoteopen}{\isacharprime}a\ {\isasymRightarrow}\ {\isacharprime}b{\isacharcolon}{\isacharcolon}{\isacharbraceleft}banach{\isacharcomma}\ second{\isacharunderscore}countable{\isacharunderscore}topology{\isacharbraceright}{\isachardoublequoteclose}\isanewline
\ \ \isakeyword{shows}\ {\isachardoublequoteopen}integrable\ M\ f\ {\isasymlongleftrightarrow}\isanewline
\ \ \ \ f\ {\isasymin}\ M\ {\isasymrightarrow}\isactrlsub M\ borel\ {\isasymand}\ {\isacharparenleft}{\isasymintegral}\isactrlsup {\isacharplus}x{\isachardot}\ norm\ {\isacharparenleft}f\ x{\isacharparenright}\ {\isasympartial}M{\isacharparenright}\ {\isacharless}\ {\isasyminfinity}{\isachardoublequoteclose}%
}
\DefineSnippet{integralnormbound}{
\isacommand{lemma}\isamarkupfalse%
\ integral{\isacharunderscore}norm{\isacharunderscore}bound{\isacharcolon}\isanewline
\ \ \isakeyword{fixes}\ f\ {\isacharcolon}{\isacharcolon}\ {\isachardoublequoteopen}{\isacharunderscore}\ {\isasymRightarrow}\ {\isacharprime}a\ {\isacharcolon}{\isacharcolon}\ {\isacharbraceleft}banach{\isacharcomma}\ second{\isacharunderscore}countable{\isacharunderscore}topology{\isacharbraceright}{\isachardoublequoteclose}\isanewline
\ \ \isakeyword{shows}\ {\isachardoublequoteopen}integrable\ M\ f\ {\isasymLongrightarrow}\isanewline
\ \ \ \ norm\ {\isacharparenleft}LINT\ x{\isacharbar}M{\isachardot}\ f\ x{\isacharparenright}\ {\isasymle}\ {\isacharparenleft}LINT\ x{\isacharbar}M{\isachardot}\ norm\ {\isacharparenleft}f\ x{\isacharparenright}{\isacharparenright}{\isachardoublequoteclose}%
}
\DefineSnippet{ftcfinite}{
\isacommand{lemma}\isamarkupfalse%
\ interval{\isacharunderscore}integral{\isacharunderscore}FTC{\isacharunderscore}finite{\isacharcolon}\isanewline
\ \ \isakeyword{fixes}\ f\ F\ {\isacharcolon}{\isacharcolon}\ {\isachardoublequoteopen}real\ {\isasymRightarrow}\ {\isacharprime}a{\isacharcolon}{\isacharcolon}euclidean{\isacharunderscore}space{\isachardoublequoteclose}\ \isakeyword{and}\ a\ b\ {\isacharcolon}{\isacharcolon}\ real\isanewline
\ \ \isakeyword{assumes}\ f{\isacharcolon}\ {\isachardoublequoteopen}continuous{\isacharunderscore}on\ {\isacharbraceleft}min\ a\ b{\isachardot}{\isachardot}max\ a\ b{\isacharbraceright}\ f{\isachardoublequoteclose}\isanewline
\ \ \ \isakeyword{and}\ F{\isacharcolon}\ {\isachardoublequoteopen}{\isasymAnd}x{\isachardot}\ min\ a\ b\ {\isasymle}\ x\ {\isasymLongrightarrow}\ x\ {\isasymle}\ max\ a\ b\ {\isasymLongrightarrow}\isanewline
\ \ \ \ {\isacharparenleft}F\ has{\isacharunderscore}vector{\isacharunderscore}derivative\ {\isacharparenleft}f\ x{\isacharparenright}{\isacharparenright}\isanewline
\ \ \ \ \ \ {\isacharparenleft}at\ x\ within\ {\isacharbraceleft}min\ a\ b{\isachardot}{\isachardot}max\ a\ b{\isacharbraceright}{\isacharparenright}{\isachardoublequoteclose}\isanewline
\ \ \isakeyword{shows}\ {\isachardoublequoteopen}{\isacharparenleft}LBINT\ x{\isacharequal}a{\isachardot}{\isachardot}b{\isachardot}\ f\ x{\isacharparenright}\ {\isacharequal}\ F\ b\ {\isacharminus}\ F\ a{\isachardoublequoteclose}%
}
\DefineSnippet{ftcintegrable}{
\isacommand{lemma}\isamarkupfalse%
\ interval{\isacharunderscore}integral{\isacharunderscore}FTC{\isacharunderscore}integrable{\isacharcolon}\isanewline
\ \ \isakeyword{fixes}\ f\ F\ {\isacharcolon}{\isacharcolon}\ {\isachardoublequoteopen}real\ {\isasymRightarrow}\ {\isacharprime}a{\isacharcolon}{\isacharcolon}euclidean{\isacharunderscore}space{\isachardoublequoteclose}\ \isakeyword{and}\ a\ b\ {\isacharcolon}{\isacharcolon}\ ereal\isanewline
\ \ \isakeyword{assumes}\ {\isachardoublequoteopen}a\ {\isacharless}\ b{\isachardoublequoteclose}\isanewline
\ \ \ \ \isakeyword{and}\ {\isachardoublequoteopen}{\isasymAnd}x{\isachardot}\ a\ {\isacharless}\ ereal\ x\ {\isasymLongrightarrow}\ ereal\ x\ {\isacharless}\ b\ {\isasymLongrightarrow}\isanewline
\ \ \ \ \ \ {\isacharparenleft}F\ has{\isacharunderscore}vector{\isacharunderscore}derivative\ f\ x{\isacharparenright}\ {\isacharparenleft}at\ x{\isacharparenright}{\isachardoublequoteclose}\isanewline
\ \ \ \ \isakeyword{and}\ {\isachardoublequoteopen}{\isasymAnd}x{\isachardot}\ a\ {\isacharless}\ ereal\ x\ {\isasymLongrightarrow}\ ereal\ x\ {\isacharless}\ b\ {\isasymLongrightarrow}\isanewline
\ \ \ \ \ \ continuous\ {\isacharparenleft}at\ x{\isacharparenright}\ f{\isachardoublequoteclose}\isanewline
\ \ \ \ \isakeyword{and}\ {\isachardoublequoteopen}set{\isacharunderscore}integrable\ lborel\ {\isacharparenleft}einterval\ a\ b{\isacharparenright}\ f{\isachardoublequoteclose}\isanewline
\ \ \ \ \isakeyword{and}\ {\isachardoublequoteopen}{\isacharparenleft}{\isacharparenleft}F\ {\isasymcirc}\ real{\isacharunderscore}of{\isacharunderscore}ereal{\isacharparenright}\ {\isasymlonglongrightarrow}\ A{\isacharparenright}\ {\isacharparenleft}at{\isacharunderscore}right\ a{\isacharparenright}{\isachardoublequoteclose}\isanewline
\ \ \ \ \isakeyword{and}\ {\isachardoublequoteopen}{\isacharparenleft}{\isacharparenleft}F\ {\isasymcirc}\ real{\isacharunderscore}of{\isacharunderscore}ereal{\isacharparenright}\ {\isasymlonglongrightarrow}\ B{\isacharparenright}\ {\isacharparenleft}at{\isacharunderscore}left\ b{\isacharparenright}{\isachardoublequoteclose}\isanewline
\ \ \isakeyword{shows}\ {\isachardoublequoteopen}{\isacharparenleft}LBINT\ x{\isacharequal}a{\isachardot}{\isachardot}b{\isachardot}\ f\ x{\isacharparenright}\ {\isacharequal}\ B\ {\isacharminus}\ A{\isachardoublequoteclose}%
}
\DefineSnippet{ftc2}{
\isacommand{lemma}\isamarkupfalse%
\ interval{\isacharunderscore}integral{\isacharunderscore}FTC{\isadigit{2}}{\isacharcolon}\isanewline
\ \ \isakeyword{fixes}\ a\ b\ c\ x\ {\isacharcolon}{\isacharcolon}\ real\ \isakeyword{and}\ f\ {\isacharcolon}{\isacharcolon}\ {\isachardoublequoteopen}real\ {\isasymRightarrow}\ {\isacharprime}a{\isacharcolon}{\isacharcolon}euclidean{\isacharunderscore}space{\isachardoublequoteclose}\isanewline
\ \ \isakeyword{assumes}\ {\isachardoublequoteopen}a\ {\isasymle}\ c{\isachardoublequoteclose}\ {\isachardoublequoteopen}c\ {\isasymle}\ b{\isachardoublequoteclose}\ {\isachardoublequoteopen}continuous{\isacharunderscore}on\ {\isacharbraceleft}a{\isachardot}{\isachardot}b{\isacharbraceright}\ f{\isachardoublequoteclose}\isanewline
\ \ \ \ {\isachardoublequoteopen}a\ {\isasymle}\ x{\isachardoublequoteclose}\ {\isachardoublequoteopen}x\ {\isasymle}\ b{\isachardoublequoteclose}\isanewline
\ \ \isakeyword{shows}\ {\isachardoublequoteopen}{\isacharparenleft}{\isacharparenleft}{\isasymlambda}u{\isachardot}\ LBINT\ y{\isacharequal}c{\isachardot}{\isachardot}u{\isachardot}\ f\ y{\isacharparenright}\ has{\isacharunderscore}vector{\isacharunderscore}derivative\ {\isacharparenleft}f\ x{\isacharparenright}{\isacharparenright}\isanewline
\ \ \ \ {\isacharparenleft}at\ x\ within\ {\isacharbraceleft}a{\isachardot}{\isachardot}b{\isacharbraceright}{\isacharparenright}{\isachardoublequoteclose}%
}
\DefineSnippet{substfinite}{
\isacommand{lemma}\isamarkupfalse%
\ interval{\isacharunderscore}integral{\isacharunderscore}substitution{\isacharunderscore}finite{\isacharcolon}\isanewline
\ \ \isakeyword{fixes}\ a\ b\ {\isacharcolon}{\isacharcolon}\ real\ \isakeyword{and}\ f\ {\isacharcolon}{\isacharcolon}\ {\isachardoublequoteopen}real\ {\isasymRightarrow}\ {\isacharprime}a{\isacharcolon}{\isacharcolon}euclidean{\isacharunderscore}space{\isachardoublequoteclose}\isanewline
\ \ \isakeyword{assumes}\ {\isachardoublequoteopen}a\ {\isasymle}\ b{\isachardoublequoteclose}\ \isakeyword{and}\ {\isachardoublequoteopen}{\isasymAnd}x{\isachardot}\ a\ {\isasymle}\ x\ {\isasymLongrightarrow}\ x\ {\isasymle}\ b\ {\isasymLongrightarrow}\isanewline
\ \ \ \ \ \ {\isacharparenleft}g\ has{\isacharunderscore}real{\isacharunderscore}derivative\ {\isacharparenleft}g{\isacharprime}\ x{\isacharparenright}{\isacharparenright}\ {\isacharparenleft}at\ x\ within\ {\isacharbraceleft}a{\isachardot}{\isachardot}b{\isacharbraceright}{\isacharparenright}{\isachardoublequoteclose}\isanewline
\ \ \ \ \isakeyword{and}\ {\isachardoublequoteopen}continuous{\isacharunderscore}on\ {\isacharparenleft}g\ {\isacharbackquote}\ {\isacharbraceleft}a{\isachardot}{\isachardot}b{\isacharbraceright}{\isacharparenright}\ f{\isachardoublequoteclose}\ {\isachardoublequoteopen}continuous{\isacharunderscore}on\ {\isacharbraceleft}a{\isachardot}{\isachardot}b{\isacharbraceright}\ g{\isacharprime}{\isachardoublequoteclose}\isanewline
\ \ \isakeyword{shows}\ {\isachardoublequoteopen}LBINT\ x{\isacharequal}a{\isachardot}{\isachardot}b{\isachardot}\ g{\isacharprime}\ x\ {\isacharasterisk}\isactrlsub R\ f\ {\isacharparenleft}g\ x{\isacharparenright}\ {\isacharequal}\ LBINT\ y{\isacharequal}g\ a{\isachardot}{\isachardot}g\ b{\isachardot}\ f\ y{\isachardoublequoteclose}%
}
\DefineSnippet{substintegrable}{
\isacommand{lemma}\isamarkupfalse%
\ interval{\isacharunderscore}integral{\isacharunderscore}substitution{\isacharunderscore}integrable{\isacharcolon}\isanewline
\ \ \isakeyword{fixes}\ f\ {\isacharcolon}{\isacharcolon}\ {\isachardoublequoteopen}real\ {\isasymRightarrow}\ {\isacharprime}a{\isacharcolon}{\isacharcolon}euclidean{\isacharunderscore}space{\isachardoublequoteclose}\ \isakeyword{and}\ a\ b\ A\ B\ {\isacharcolon}{\isacharcolon}\ ereal\isanewline
\ \ \isakeyword{assumes}\ {\isachardoublequoteopen}a\ {\isacharless}\ b{\isachardoublequoteclose}\isanewline
\ \ \ \ \isakeyword{and}\ {\isachardoublequoteopen}{\isasymAnd}x{\isachardot}\ a\ {\isacharless}\ ereal\ x\ {\isasymLongrightarrow}\ ereal\ x\ {\isacharless}\ b\ {\isasymLongrightarrow}\isanewline
\ \ \ \ \ \ DERIV\ g\ x\ {\isacharcolon}{\isachargreater}\ g{\isacharprime}\ x{\isachardoublequoteclose}\isanewline
\ \ \ \ \isakeyword{and}\ {\isachardoublequoteopen}{\isasymAnd}x{\isachardot}\ a\ {\isacharless}\ ereal\ x\ {\isasymLongrightarrow}\ ereal\ x\ {\isacharless}\ b\ {\isasymLongrightarrow}\isanewline
\ \ \ \ \ \ continuous\ {\isacharparenleft}at\ {\isacharparenleft}g\ x{\isacharparenright}{\isacharparenright}\ f{\isachardoublequoteclose}\isanewline
\ \ \ \ \isakeyword{and}\ {\isachardoublequoteopen}{\isasymAnd}x{\isachardot}\ a\ {\isacharless}\ ereal\ x\ {\isasymLongrightarrow}\ ereal\ x\ {\isacharless}\ b\ {\isasymLongrightarrow}\isanewline
\ \ \ \ \ \ continuous\ {\isacharparenleft}at\ x{\isacharparenright}\ g{\isacharprime}{\isachardoublequoteclose}\isanewline
\ \ \ \ \isakeyword{and}\ {\isachardoublequoteopen}{\isasymAnd}x{\isachardot}\ a\ {\isasymle}\ ereal\ x\ {\isasymLongrightarrow}\ ereal\ x\ {\isasymle}\ b\ {\isasymLongrightarrow}\ {\isadigit{0}}\ {\isasymle}\ g{\isacharprime}\ x{\isachardoublequoteclose}\isanewline
\ \ \ \ \isakeyword{and}\ {\isachardoublequoteopen}{\isacharparenleft}{\isacharparenleft}ereal\ {\isasymcirc}\ g\ {\isasymcirc}\ real{\isacharunderscore}of{\isacharunderscore}ereal{\isacharparenright}\ {\isasymlonglongrightarrow}\ A{\isacharparenright}\ {\isacharparenleft}at{\isacharunderscore}right\ a{\isacharparenright}{\isachardoublequoteclose}\isanewline
\ \ \ \ \isakeyword{and}\ {\isachardoublequoteopen}{\isacharparenleft}{\isacharparenleft}ereal\ {\isasymcirc}\ g\ {\isasymcirc}\ real{\isacharunderscore}of{\isacharunderscore}ereal{\isacharparenright}\ {\isasymlonglongrightarrow}\ B{\isacharparenright}\ {\isacharparenleft}at{\isacharunderscore}left\ b{\isacharparenright}{\isachardoublequoteclose}\isanewline
\ \ \ \ \isakeyword{and}\ {\isachardoublequoteopen}set{\isacharunderscore}integrable\ lborel\isanewline
\ \ \ \ \ {\isacharparenleft}einterval\ a\ b{\isacharparenright}\ {\isacharparenleft}{\isasymlambda}x{\isachardot}\ g{\isacharprime}\ x\ {\isacharasterisk}\isactrlsub R\ f\ {\isacharparenleft}g\ x{\isacharparenright}{\isacharparenright}{\isachardoublequoteclose}\isanewline
\ \ \ \ \isakeyword{and}\ {\isachardoublequoteopen}set{\isacharunderscore}integrable\ lborel\isanewline
\ \ \ \ \ {\isacharparenleft}einterval\ A\ B{\isacharparenright}\ {\isacharparenleft}{\isasymlambda}x{\isachardot}\ f\ x{\isacharparenright}{\isachardoublequoteclose}\isanewline
\ \ \isakeyword{shows}\ {\isachardoublequoteopen}{\isacharparenleft}LBINT\ x{\isacharequal}A{\isachardot}{\isachardot}B{\isachardot}\ f\ x{\isacharparenright}\ {\isacharequal}\ {\isacharparenleft}LBINT\ x{\isacharequal}a{\isachardot}{\isachardot}b{\isachardot}\ g{\isacharprime}\ x\ {\isacharasterisk}\isactrlsub R\ f\ {\isacharparenleft}g\ x{\isacharparenright}{\isacharparenright}{\isachardoublequoteclose}%
}
\DefineSnippet{charprop1}{
\isacommand{lemma}\isamarkupfalse%
\ {\isacharparenleft}\isakeyword{in}\ real{\isacharunderscore}distribution{\isacharparenright}\ continuous{\isacharunderscore}char{\isacharcolon}\isanewline
\ \ {\isachardoublequoteopen}continuous\ {\isacharparenleft}at\ t{\isacharparenright}\ {\isacharparenleft}char\ M{\isacharparenright}{\isachardoublequoteclose}%
}
\DefineSnippet{charprop2}{
\isacommand{lemma}\isamarkupfalse%
\ {\isacharparenleft}\isakeyword{in}\ real{\isacharunderscore}distribution{\isacharparenright}\ char{\isacharunderscore}zero{\isacharcolon}\isanewline
\ \ {\isachardoublequoteopen}char\ M\ {\isadigit{0}}\ {\isacharequal}\ {\isadigit{1}}{\isachardoublequoteclose}%
}
\DefineSnippet{charprop3}{
\isacommand{lemma}\isamarkupfalse%
\ {\isacharparenleft}\isakeyword{in}\ real{\isacharunderscore}distribution{\isacharparenright}\ cmod{\isacharunderscore}char{\isacharunderscore}le{\isacharunderscore}{\isadigit{1}}{\isacharcolon}\isanewline
\ \ {\isachardoublequoteopen}norm\ {\isacharparenleft}char\ M\ t{\isacharparenright}\ {\isasymle}\ {\isadigit{1}}{\isachardoublequoteclose}%
}
\DefineSnippet{chardistrsum}{
\isacommand{lemma}\isamarkupfalse%
\ {\isacharparenleft}\isakeyword{in}\ prob{\isacharunderscore}space{\isacharparenright}\ char{\isacharunderscore}distr{\isacharunderscore}sum{\isacharcolon}\isanewline
\ \ \isakeyword{assumes}\ {\isachardoublequoteopen}indep{\isacharunderscore}var\ borel\ X{\isadigit{1}}\ borel\ X{\isadigit{2}}{\isachardoublequoteclose}\isanewline
\ \ \isakeyword{shows}\ {\isachardoublequoteopen}char\ {\isacharparenleft}distr\ M\ borel\ {\isacharparenleft}{\isasymlambda}{\isasymomega}{\isachardot}\ X{\isadigit{1}}\ {\isasymomega}\ {\isacharplus}\ X{\isadigit{2}}\ {\isasymomega}{\isacharparenright}{\isacharparenright}\ t\ {\isacharequal}\isanewline
\ \ \ \ char\ {\isacharparenleft}distr\ M\ borel\ X{\isadigit{1}}{\isacharparenright}\ t\ {\isacharasterisk}\ char\ {\isacharparenleft}distr\ M\ borel\ X{\isadigit{2}}{\isacharparenright}\ t{\isachardoublequoteclose}\isanewline
\isadelimproof
\endisadelimproof
\isatagproof
\isacommand{proof}\isamarkupfalse%
\ {\isacharminus}\isanewline
\ \ \isacommand{have}\isamarkupfalse%
\ {\isacharbrackleft}measurable{\isacharbrackright}{\isacharcolon}\isanewline
\ \ \ \ \ \ {\isachardoublequoteopen}random{\isacharunderscore}variable\ borel\ X{\isadigit{1}}{\isachardoublequoteclose}\ {\isachardoublequoteopen}random{\isacharunderscore}variable\ borel\ X{\isadigit{2}}{\isachardoublequoteclose}\isanewline
\ \ \ \ \isacommand{using}\isamarkupfalse%
\ assms\ \isacommand{by}\isamarkupfalse%
\ {\isacharparenleft}auto\ dest{\isacharcolon}\ indep{\isacharunderscore}var{\isacharunderscore}rv{\isadigit{1}}\ indep{\isacharunderscore}var{\isacharunderscore}rv{\isadigit{2}}{\isacharparenright}\isanewline
\isanewline
\ \ \isacommand{have}\isamarkupfalse%
\ {\isachardoublequoteopen}char\ {\isacharparenleft}distr\ M\ borel\ {\isacharparenleft}{\isasymlambda}{\isasymomega}{\isachardot}\ X{\isadigit{1}}\ {\isasymomega}\ {\isacharplus}\ X{\isadigit{2}}\ {\isasymomega}{\isacharparenright}{\isacharparenright}\ t\ {\isacharequal}\isanewline
\ \ \ \ \ \ {\isacharparenleft}LINT\ x{\isacharbar}M{\isachardot}\ iexp\ {\isacharparenleft}t\ {\isacharasterisk}\ {\isacharparenleft}X{\isadigit{1}}\ x\ {\isacharplus}\ X{\isadigit{2}}\ x{\isacharparenright}{\isacharparenright}{\isacharparenright}{\isachardoublequoteclose}\isanewline
\ \ \ \ \isacommand{by}\isamarkupfalse%
\ {\isacharparenleft}simp\ add{\isacharcolon}\ char{\isacharunderscore}def\ integral{\isacharunderscore}distr{\isacharparenright}\isanewline
\ \ \isacommand{also}\isamarkupfalse%
\ \isacommand{have}\isamarkupfalse%
\ {\isachardoublequoteopen}{\isasymdots}\ {\isacharequal}\isanewline
\ \ \ \ \ \ {\isacharparenleft}LINT\ x{\isacharbar}M{\isachardot}\ iexp\ {\isacharparenleft}t\ {\isacharasterisk}\ {\isacharparenleft}X{\isadigit{1}}\ x{\isacharparenright}{\isacharparenright}\ {\isacharasterisk}\ iexp\ {\isacharparenleft}t\ {\isacharasterisk}\ {\isacharparenleft}X{\isadigit{2}}\ x{\isacharparenright}{\isacharparenright}{\isacharparenright}{\isachardoublequoteclose}\isanewline
\ \ \ \ \isacommand{by}\isamarkupfalse%
\ {\isacharparenleft}simp\ add{\isacharcolon}\ field{\isacharunderscore}simps\ exp{\isacharunderscore}add{\isacharparenright}\isanewline
\ \ \isacommand{also}\isamarkupfalse%
\ \isacommand{have}\isamarkupfalse%
\ {\isachardoublequoteopen}{\isasymdots}\ {\isacharequal}\isanewline
\ \ \ \ \ \ {\isacharparenleft}LINT\ x{\isacharbar}M{\isachardot}\ iexp\ {\isacharparenleft}t\ {\isacharasterisk}\ {\isacharparenleft}X{\isadigit{1}}\ x{\isacharparenright}{\isacharparenright}{\isacharparenright}\ {\isacharasterisk}\ {\isacharparenleft}LINT\ x{\isacharbar}M{\isachardot}\ iexp\ {\isacharparenleft}t\ {\isacharasterisk}\ {\isacharparenleft}X{\isadigit{2}}\ x{\isacharparenright}{\isacharparenright}{\isacharparenright}{\isachardoublequoteclose}\isanewline
\ \ \ \ \isacommand{by}\isamarkupfalse%
\ {\isacharparenleft}auto\ intro{\isacharbang}{\isacharcolon}\ indep{\isacharunderscore}var{\isacharunderscore}compose{\isacharbrackleft}unfolded\ comp{\isacharunderscore}def{\isacharcomma}\ OF\ assms{\isacharbrackright}\isanewline
\ \ \ \ \ \ \ \ \ \ \ \ \ \ \ \ \ \ \ \ \ integrable{\isacharunderscore}iexp\ indep{\isacharunderscore}var{\isacharunderscore}lebesgue{\isacharunderscore}integral{\isacharparenright}\isanewline
\ \ \isacommand{also}\isamarkupfalse%
\ \isacommand{have}\isamarkupfalse%
\ {\isachardoublequoteopen}{\isasymdots}\ {\isacharequal}\isanewline
\ \ \ \ \ \ char\ {\isacharparenleft}distr\ M\ borel\ X{\isadigit{1}}{\isacharparenright}\ t\ {\isacharasterisk}\ char\ {\isacharparenleft}distr\ M\ borel\ X{\isadigit{2}}{\isacharparenright}\ t{\isachardoublequoteclose}\isanewline
\ \ \ \ \isacommand{by}\isamarkupfalse%
\ {\isacharparenleft}simp\ add{\isacharcolon}\ char{\isacharunderscore}def\ integral{\isacharunderscore}distr{\isacharparenright}\isanewline
\ \ \isacommand{finally}\isamarkupfalse%
\ \isacommand{show}\isamarkupfalse%
\ {\isacharquery}thesis\ \isacommand{{\isachardot}}\isamarkupfalse%
\isanewline
\isacommand{qed}\isamarkupfalse%
\endisatagproof
{\isafoldproof}%
\isadelimproof
\endisadelimproof
}
\DefineSnippet{chardistrsetsum}{
\isacommand{lemma}\isamarkupfalse%
\ {\isacharparenleft}\isakeyword{in}\ prob{\isacharunderscore}space{\isacharparenright}\ char{\isacharunderscore}distr{\isacharunderscore}setsum{\isacharcolon}\isanewline
\ \ {\isachardoublequoteopen}indep{\isacharunderscore}vars\ {\isacharparenleft}{\isasymlambda}i{\isachardot}\ borel{\isacharparenright}\ X\ A\ {\isasymLongrightarrow}\isanewline
\ \ \ \ char\ {\isacharparenleft}distr\ M\ borel\ {\isacharparenleft}{\isasymlambda}{\isasymomega}{\isachardot}\ {\isasymSum}i{\isasymin}A{\isachardot}\ X\ i\ {\isasymomega}{\isacharparenright}{\isacharparenright}\ t\ {\isacharequal}\isanewline
\ \ \ \ {\isacharparenleft}{\isasymProd}i{\isasymin}A{\isachardot}\ char\ {\isacharparenleft}distr\ M\ borel\ {\isacharparenleft}X\ i{\isacharparenright}{\isacharparenright}\ t{\isacharparenright}{\isachardoublequoteclose}\isanewline
\isadelimproof
\endisadelimproof
\isatagproof
\isacommand{proof}\isamarkupfalse%
\ {\isacharparenleft}induct\ A\ rule{\isacharcolon}\ infinite{\isacharunderscore}finite{\isacharunderscore}induct{\isacharparenright}\isanewline
\ \ \isacommand{case}\isamarkupfalse%
\ {\isacharparenleft}insert\ x\ F{\isacharparenright}\ \isacommand{then}\isamarkupfalse%
\ \isacommand{show}\isamarkupfalse%
\ {\isacharquery}case\isanewline
\ \ \ \ \isacommand{using}\isamarkupfalse%
\ indep{\isacharunderscore}vars{\isacharunderscore}subset{\isacharbrackleft}of\ {\isachardoublequoteopen}{\isasymlambda}{\isacharunderscore}{\isachardot}\ borel{\isachardoublequoteclose}\ X\ {\isachardoublequoteopen}insert\ x\ F{\isachardoublequoteclose}\ F{\isacharbrackright}\isanewline
\ \ \ \ \isacommand{by}\isamarkupfalse%
\ {\isacharparenleft}auto\ simp\ add{\isacharcolon}\ char{\isacharunderscore}distr{\isacharunderscore}sum\ indep{\isacharunderscore}vars{\isacharunderscore}sum{\isacharparenright}\isanewline
\isacommand{qed}\isamarkupfalse%
\ {\isacharparenleft}simp{\isacharunderscore}all\ add{\isacharcolon}\ char{\isacharunderscore}def\ integral{\isacharunderscore}distr\ prob{\isacharunderscore}space\ del{\isacharcolon}\ distr{\isacharunderscore}const{\isacharparenright}%
\endisatagproof
{\isafoldproof}%
\isadelimproof
\endisadelimproof
}
\DefineSnippet{charapprox}{
\isacommand{lemma}\isamarkupfalse%
\ {\isacharparenleft}\isakeyword{in}\ prob{\isacharunderscore}space{\isacharparenright}\ char{\isacharunderscore}approx{\isadigit{3}}{\isacharprime}{\isacharcolon}\isanewline
\ \ \isakeyword{fixes}\ {\isasymmu}\ {\isacharcolon}{\isacharcolon}\ {\isachardoublequoteopen}real\ measure{\isachardoublequoteclose}\ \isakeyword{and}\ X\isanewline
\ \ \isakeyword{assumes}\ {\isachardoublequoteopen}random{\isacharunderscore}variable\ borel\ X{\isachardoublequoteclose}\isanewline
\ \ \ \ \isakeyword{and}\ {\isachardoublequoteopen}integrable\ M\ X{\isachardoublequoteclose}\ {\isachardoublequoteopen}integrable\ M\ {\isacharparenleft}{\isasymlambda}x{\isachardot}\ {\isacharparenleft}X\ x{\isacharparenright}{\isacharcircum}{\isadigit{2}}{\isacharparenright}{\isachardoublequoteclose}\isanewline
\ \ \ \ \isakeyword{and}\ {\isachardoublequoteopen}expectation\ X\ {\isacharequal}\ {\isadigit{0}}{\isachardoublequoteclose}\isanewline
\ \ \ \ \isakeyword{and}\ {\isachardoublequoteopen}variance\ X\ {\isacharequal}\ {\isasymsigma}{\isadigit{2}}{\isachardoublequoteclose}\isanewline
\ \ \ \ \isakeyword{and}\ {\isachardoublequoteopen}{\isasymmu}\ {\isacharequal}\ distr\ M\ borel\ X{\isachardoublequoteclose}\isanewline
\ \ \isakeyword{shows}\ {\isachardoublequoteopen}cmod\ {\isacharparenleft}char\ {\isasymmu}\ t\ {\isacharminus}\ {\isacharparenleft}{\isadigit{1}}\ {\isacharminus}\ t{\isacharcircum}{\isadigit{2}}\ {\isacharasterisk}\ {\isasymsigma}{\isadigit{2}}\ {\isacharslash}\ {\isadigit{2}}{\isacharparenright}{\isacharparenright}\ {\isasymle}\isanewline
\ \ \ \ {\isacharparenleft}t{\isacharcircum}{\isadigit{2}}\ {\isacharslash}\ {\isadigit{6}}{\isacharparenright}\ {\isacharasterisk}\ expectation\ {\isacharparenleft}{\isasymlambda}x{\isachardot}\ min\ {\isacharparenleft}{\isadigit{6}}\ {\isacharasterisk}\ {\isacharparenleft}X\ x{\isacharparenright}{\isacharcircum}{\isadigit{2}}{\isacharparenright}\ {\isacharparenleft}{\isasymbar}t{\isasymbar}\ {\isacharasterisk}\ {\isasymbar}X\ x{\isasymbar}{\isacharcircum}{\isadigit{3}}{\isacharparenright}{\isacharparenright}{\isachardoublequoteclose}%
}
\DefineSnippet{momenteven}{
\isacommand{lemma}\isamarkupfalse%
\ std{\isacharunderscore}normal{\isacharunderscore}moment{\isacharunderscore}even{\isacharcolon}\isanewline
\ \ {\isachardoublequoteopen}has{\isacharunderscore}bochner{\isacharunderscore}integral\ lborel\isanewline
\ \ \ \ {\isacharparenleft}{\isasymlambda}x{\isachardot}\ std{\isacharunderscore}normal{\isacharunderscore}density\ x\ {\isacharasterisk}\ x\ {\isacharcircum}\ {\isacharparenleft}{\isadigit{2}}\ {\isacharasterisk}\ k{\isacharparenright}{\isacharparenright}\isanewline
\ \ \ \ {\isacharparenleft}fact\ {\isacharparenleft}{\isadigit{2}}\ {\isacharasterisk}\ k{\isacharparenright}\ {\isacharslash}\ {\isacharparenleft}{\isadigit{2}}{\isacharcircum}k\ {\isacharasterisk}\ fact\ k{\isacharparenright}{\isacharparenright}{\isachardoublequoteclose}%
}
\DefineSnippet{momentodd}{
\isacommand{lemma}\isamarkupfalse%
\ std{\isacharunderscore}normal{\isacharunderscore}moment{\isacharunderscore}odd{\isacharcolon}\isanewline
\ \ {\isachardoublequoteopen}has{\isacharunderscore}bochner{\isacharunderscore}integral\ lborel\isanewline
\ \ \ \ {\isacharparenleft}{\isasymlambda}x{\isachardot}\ std{\isacharunderscore}normal{\isacharunderscore}density\ x\ {\isacharasterisk}\ x{\isacharcircum}{\isacharparenleft}{\isadigit{2}}\ {\isacharasterisk}\ k\ {\isacharplus}\ {\isadigit{1}}{\isacharparenright}{\isacharparenright}\ {\isadigit{0}}{\isachardoublequoteclose}%
}
\DefineSnippet{levyunique}{
\isacommand{theorem}\isamarkupfalse%
\ Levy{\isacharunderscore}uniqueness{\isacharcolon}\isanewline
\ \ \isakeyword{fixes}\ M{\isadigit{1}}\ M{\isadigit{2}}\ {\isacharcolon}{\isacharcolon}\ {\isachardoublequoteopen}real\ measure{\isachardoublequoteclose}\isanewline
\ \ \isakeyword{assumes}\ {\isachardoublequoteopen}real{\isacharunderscore}distribution\ M{\isadigit{1}}{\isachardoublequoteclose}\ {\isachardoublequoteopen}real{\isacharunderscore}distribution\ M{\isadigit{2}}{\isachardoublequoteclose}\isanewline
\ \ \ \ \isakeyword{and}\ {\isachardoublequoteopen}char\ M{\isadigit{1}}\ {\isacharequal}\ char\ M{\isadigit{2}}{\isachardoublequoteclose}\isanewline
\ \ \isakeyword{shows}\ {\isachardoublequoteopen}M{\isadigit{1}}\ {\isacharequal}\ M{\isadigit{2}}{\isachardoublequoteclose}%
}
\DefineSnippet{levycont}{
\isacommand{theorem}\isamarkupfalse%
\ Levy{\isacharunderscore}continuity{\isacharcolon}\isanewline
\ \ \isakeyword{fixes}\ M\ {\isacharcolon}{\isacharcolon}\ {\isachardoublequoteopen}nat\ {\isasymRightarrow}\ real\ measure{\isachardoublequoteclose}\ \isakeyword{and}\ M{\isacharprime}\ {\isacharcolon}{\isacharcolon}\ {\isachardoublequoteopen}real\ measure{\isachardoublequoteclose}\isanewline
\ \ \isakeyword{assumes}\ {\isachardoublequoteopen}{\isasymAnd}n{\isachardot}\ real{\isacharunderscore}distribution\ {\isacharparenleft}M\ n{\isacharparenright}{\isachardoublequoteclose}\isanewline
\ \ \ \ \isakeyword{and}\ {\isachardoublequoteopen}real{\isacharunderscore}distribution\ M{\isacharprime}{\isachardoublequoteclose}\isanewline
\ \ \ \ \isakeyword{and}\ {\isachardoublequoteopen}{\isasymAnd}t{\isachardot}\ {\isacharparenleft}{\isasymlambda}n{\isachardot}\ char\ {\isacharparenleft}M\ n{\isacharparenright}\ t{\isacharparenright}\ {\isasymlonglonglongrightarrow}\ char\ M{\isacharprime}\ t{\isachardoublequoteclose}\isanewline
\ \ \isakeyword{shows}\ {\isachardoublequoteopen}weak{\isacharunderscore}conv{\isacharunderscore}m\ M\ M{\isacharprime}{\isachardoublequoteclose}%
}
\DefineSnippet{levycont1}{
\isacommand{theorem}\isamarkupfalse%
\ Levy{\isacharunderscore}continuity{\isadigit{1}}{\isacharcolon}\isanewline
\ \ {\isachardoublequoteopen}{\isacharparenleft}{\isasymAnd}n{\isachardot}\ real{\isacharunderscore}distribution\ {\isacharparenleft}M\ n{\isacharparenright}{\isacharparenright}\ {\isasymLongrightarrow}\ real{\isacharunderscore}distribution\ M{\isacharprime}\ {\isasymLongrightarrow}\isanewline
\ \ \ \ weak{\isacharunderscore}conv{\isacharunderscore}m\ M\ M{\isacharprime}\ {\isasymLongrightarrow}\isanewline
\ \ \ \ {\isacharparenleft}{\isasymlambda}n{\isachardot}\ char\ {\isacharparenleft}M\ n{\isacharparenright}\ t{\isacharparenright}\ {\isasymlonglonglongrightarrow}\ char\ M{\isacharprime}\ t{\isachardoublequoteclose}\isanewline
\isadelimproof
\ \ %
\endisadelimproof
\isatagproof
\isacommand{unfolding}\isamarkupfalse%
\ char{\isacharunderscore}def\isanewline
\ \ \isacommand{by}\isamarkupfalse%
\ {\isacharparenleft}rule\ weak{\isacharunderscore}conv{\isacharunderscore}imp{\isacharunderscore}integral{\isacharunderscore}bdd{\isacharunderscore}continuous{\isacharunderscore}conv{\isacharparenright}\ auto%
\endisatagproof
{\isafoldproof}%
\isadelimproof
\endisadelimproof
}
\DefineSnippet{helly}{
\isacommand{theorem}\isamarkupfalse%
\ Helly{\isacharunderscore}selection{\isacharcolon}\isanewline
\ \ \isakeyword{fixes}\ f\ {\isacharcolon}{\isacharcolon}\ {\isachardoublequoteopen}nat\ {\isasymRightarrow}\ real\ {\isasymRightarrow}\ real{\isachardoublequoteclose}\isanewline
\ \ \isakeyword{assumes}\ {\isachardoublequoteopen}{\isasymAnd}n\ x{\isachardot}\ continuous\ {\isacharparenleft}at{\isacharunderscore}right\ x{\isacharparenright}\ {\isacharparenleft}f\ n{\isacharparenright}{\isachardoublequoteclose}\isanewline
\ \ \ \ \isakeyword{and}\ {\isachardoublequoteopen}{\isasymAnd}n{\isachardot}\ mono\ {\isacharparenleft}f\ n{\isacharparenright}{\isachardoublequoteclose}\isanewline
\ \ \ \ \isakeyword{and}\ {\isachardoublequoteopen}{\isasymAnd}n\ x{\isachardot}\ {\isasymbar}f\ n\ x{\isasymbar}\ {\isasymle}\ M{\isachardoublequoteclose}\isanewline
\ \ \isakeyword{shows}\ {\isachardoublequoteopen}{\isasymexists}s{\isachardot}\ subseq\ s\ {\isasymand}\isanewline
\ \ \ \ {\isacharparenleft}{\isasymexists}F{\isachardot}\ {\isacharparenleft}{\isasymforall}x{\isachardot}\ continuous\ {\isacharparenleft}at{\isacharunderscore}right\ x{\isacharparenright}\ F{\isacharparenright}\ {\isasymand}\isanewline
\ \ \ \ \ \ mono\ F\ {\isasymand}\ {\isacharparenleft}{\isasymforall}x{\isachardot}\ {\isasymbar}F\ x{\isasymbar}\ {\isasymle}\ M{\isacharparenright}\ {\isasymand}\isanewline
\ \ \ \ \ \ {\isacharparenleft}{\isasymforall}x{\isachardot}\ isCont\ F\ x\ {\isasymlongrightarrow}\ {\isacharparenleft}{\isasymlambda}n{\isachardot}\ f\ {\isacharparenleft}s\ n{\isacharparenright}\ x{\isacharparenright}\ {\isasymlonglonglongrightarrow}\ F\ x{\isacharparenright}{\isacharparenright}{\isachardoublequoteclose}%
}
\DefineSnippet{thight}{
\isacommand{theorem}\isamarkupfalse%
\ tight{\isacharunderscore}imp{\isacharunderscore}convergent{\isacharunderscore}subsubsequence{\isacharcolon}\isanewline
\ \ \isakeyword{assumes}\ {\isasymmu}{\isacharcolon}\ {\isachardoublequoteopen}tight\ {\isasymmu}{\isachardoublequoteclose}\ {\isachardoublequoteopen}subseq\ s{\isachardoublequoteclose}\isanewline
\ \ \isakeyword{shows}\ {\isachardoublequoteopen}{\isasymexists}r\ M{\isachardot}\ subseq\ r\ {\isasymand}\ real{\isacharunderscore}distribution\ M\ {\isasymand}\isanewline
\ \ \ \ weak{\isacharunderscore}conv{\isacharunderscore}m\ {\isacharparenleft}{\isasymmu}\ {\isasymcirc}\ s\ {\isasymcirc}\ r{\isacharparenright}\ M{\isachardoublequoteclose}%
}
\DefineSnippet{integraladd}{
\isacommand{lemma}\isamarkupfalse%
\ integral{\isacharunderscore}add{\isacharcolon}\isanewline
\ \ {\isachardoublequoteopen}integrable\ M\ f\ {\isasymLongrightarrow}\ integrable\ M\ g\ {\isasymLongrightarrow}\isanewline
\ \ \ \ {\isacharparenleft}LINT\ x{\isacharbar}M{\isachardot}\ f\ x\ {\isacharplus}\ g\ x{\isacharparenright}\ {\isacharequal}\ {\isacharparenleft}LINT\ x{\isacharbar}M{\isachardot}\ f\ x{\isacharparenright}\ {\isacharplus}\ {\isacharparenleft}LINT\ x{\isacharbar}M{\isachardot}\ g\ x{\isacharparenright}{\isachardoublequoteclose}%
}
\DefineSnippet{integrableadd}{
\isacommand{lemma}\isamarkupfalse%
\ integrable{\isacharunderscore}add{\isacharcolon}\isanewline
\ \ {\isachardoublequoteopen}integrable\ M\ f\ {\isasymLongrightarrow}\ integrable\ M\ g\ {\isasymLongrightarrow}\isanewline
\ \ \ \ integrable\ M\ {\isacharparenleft}{\isasymlambda}x{\isachardot}\ f\ x\ {\isacharplus}\ g\ x{\isacharparenright}{\isachardoublequoteclose}%
}
\DefineSnippet{weakconv}{
\isacommand{definition}\isamarkupfalse%
\isanewline
\ \ weak{\isacharunderscore}conv\ {\isacharcolon}{\isacharcolon}\ {\isachardoublequoteopen}{\isacharparenleft}nat\ {\isasymRightarrow}\ {\isacharparenleft}real\ {\isasymRightarrow}\ real{\isacharparenright}{\isacharparenright}\ {\isasymRightarrow}\ {\isacharparenleft}real\ {\isasymRightarrow}\ real{\isacharparenright}\ {\isasymRightarrow}\ bool{\isachardoublequoteclose}\isanewline
\isakeyword{where}\isanewline
\ \ {\isachardoublequoteopen}weak{\isacharunderscore}conv\ F{\isacharunderscore}seq\ F\ {\isasymequiv}\isanewline
\ \ \ \ \ {\isasymforall}x{\isachardot}\ continuous\ {\isacharparenleft}at\ x{\isacharparenright}\ F\ {\isasymlongrightarrow}\ {\isacharparenleft}{\isasymlambda}n{\isachardot}\ F{\isacharunderscore}seq\ n\ x{\isacharparenright}\ {\isasymlonglonglongrightarrow}\ F\ x{\isachardoublequoteclose}\isanewline
\isanewline
\isacommand{definition}\isamarkupfalse%
\isanewline
\ \ weak{\isacharunderscore}conv{\isacharunderscore}m\ {\isacharcolon}{\isacharcolon}\ {\isachardoublequoteopen}{\isacharparenleft}nat\ {\isasymRightarrow}\ real\ measure{\isacharparenright}\ {\isasymRightarrow}\ real\ measure\ {\isasymRightarrow}\ bool{\isachardoublequoteclose}\isanewline
\isakeyword{where}\isanewline
\ \ {\isachardoublequoteopen}weak{\isacharunderscore}conv{\isacharunderscore}m\ M{\isacharunderscore}seq\ M\ {\isasymequiv}\ weak{\isacharunderscore}conv\ {\isacharparenleft}{\isasymlambda}n{\isachardot}\ cdf\ {\isacharparenleft}M{\isacharunderscore}seq\ n{\isacharparenright}{\isacharparenright}\ {\isacharparenleft}cdf\ M{\isacharparenright}{\isachardoublequoteclose}%
}
\DefineSnippet{distr}{
\isacommand{definition}\isamarkupfalse%
\isanewline
\ \ distr\ {\isacharcolon}{\isacharcolon}\ {\isachardoublequoteopen}{\isacharprime}a\ measure\ {\isasymRightarrow}\ {\isacharprime}b\ measure\ {\isasymRightarrow}\ {\isacharparenleft}{\isacharprime}a\ {\isasymRightarrow}\ {\isacharprime}b{\isacharparenright}\ {\isasymRightarrow}\ {\isacharprime}b\ measure{\isachardoublequoteclose}\isanewline
\isakeyword{where}\isanewline
\ \ {\isachardoublequoteopen}distr\ M\ N\ f\ {\isacharequal}\isanewline
\ \ \ \ measure{\isacharunderscore}of\ {\isacharparenleft}space\ N{\isacharparenright}\ {\isacharparenleft}sets\ N{\isacharparenright}\isanewline
\ \ \ \ \ \ {\isacharparenleft}{\isasymlambda}A{\isachardot}\ emeasure\ M\ {\isacharparenleft}f\ {\isacharminus}{\isacharbackquote}\ A\ {\isasyminter}\ space\ M{\isacharparenright}{\isacharparenright}{\isachardoublequoteclose}%
}
\DefineSnippet{density}{
\isacommand{definition}\isamarkupfalse%
\isanewline
\ \ density\ {\isacharcolon}{\isacharcolon}\ {\isachardoublequoteopen}{\isacharprime}a\ measure\ {\isasymRightarrow}\ {\isacharparenleft}{\isacharprime}a\ {\isasymRightarrow}\ ennreal{\isacharparenright}\ {\isasymRightarrow}\ {\isacharprime}a\ measure{\isachardoublequoteclose}\isanewline
\isakeyword{where}\isanewline
\ \ {\isachardoublequoteopen}density\ M\ f\ {\isacharequal}\isanewline
\ \ \ \ measure{\isacharunderscore}of\ {\isacharparenleft}space\ M{\isacharparenright}\ {\isacharparenleft}sets\ M{\isacharparenright}\ \isanewline
\ \ \ \ \ \ {\isacharparenleft}{\isasymlambda}A{\isachardot}\ {\isasymintegral}\isactrlsup {\isacharplus}\ x{\isachardot}\ f\ x\ {\isacharasterisk}\ indicator\ A\ x\ {\isasympartial}M{\isacharparenright}{\isachardoublequoteclose}%
}
\DefineSnippet{probspace}{
\isacommand{locale}\isamarkupfalse%
\ prob{\isacharunderscore}space\ {\isacharequal}\isanewline
\ \ \isakeyword{fixes}\ M\ {\isacharcolon}{\isacharcolon}\ {\isachardoublequoteopen}{\isacharprime}a\ measure{\isachardoublequoteclose}\ \isakeyword{assumes}\ {\isachardoublequoteopen}emeasure\ M\ {\isacharparenleft}space\ M{\isacharparenright}\ {\isacharequal}\ {\isadigit{1}}{\isachardoublequoteclose}\isanewline
\isakeyword{begin}\isanewline
\ \ \isacommand{abbreviation}\isamarkupfalse%
\ {\isachardoublequoteopen}events\ {\isasymequiv}\ sets\ M{\isachardoublequoteclose}\isanewline
\ \ \isacommand{abbreviation}\isamarkupfalse%
\ {\isachardoublequoteopen}prob\ {\isasymequiv}\ measure\ M{\isachardoublequoteclose}\isanewline
\ \ \isacommand{abbreviation}\isamarkupfalse%
\ {\isachardoublequoteopen}random{\isacharunderscore}variable\ M{\isacharprime}\ X\ {\isasymequiv}\ X\ {\isasymin}\ M\ {\isasymrightarrow}\isactrlsub M\ M{\isacharprime}{\isachardoublequoteclose}\isanewline
\ \ \isacommand{abbreviation}\isamarkupfalse%
\ {\isachardoublequoteopen}expectation\ X\ {\isasymequiv}\ {\isacharparenleft}LINT\ x{\isacharbar}M{\isachardot}\ X\ x{\isacharparenright}{\isachardoublequoteclose}\isanewline
\ \ \isacommand{abbreviation}\isamarkupfalse%
\ {\isachardoublequoteopen}variance\ X\ {\isasymequiv}\ {\isacharparenleft}LINT\ x{\isacharbar}M{\isachardot}\ {\isacharparenleft}X\ x\ {\isacharminus}\ expectation\ X{\isacharparenright}\isactrlsup {\isadigit{2}}{\isacharparenright}{\isachardoublequoteclose}\isanewline
\isacommand{end}\isamarkupfalse%
}
\DefineSnippet{indepsets}{
\isacommand{definition}\isamarkupfalse%
\ {\isacharparenleft}\isakeyword{in}\ prob{\isacharunderscore}space{\isacharparenright}\isanewline
\ \ indep{\isacharunderscore}sets\ {\isacharcolon}{\isacharcolon}\ {\isachardoublequoteopen}{\isacharparenleft}{\isacharprime}i\ {\isasymRightarrow}\ {\isacharprime}a\ set\ set{\isacharparenright}\ {\isasymRightarrow}\ {\isacharprime}i\ set\ {\isasymRightarrow}\ bool{\isachardoublequoteclose}\isanewline
\isakeyword{where}\isanewline
\ \ {\isachardoublequoteopen}indep{\isacharunderscore}sets\ F\ I\ {\isasymlongleftrightarrow}\isanewline
\ \ \ \ {\isacharparenleft}{\isasymforall}i{\isasymin}I{\isachardot}\ F\ i\ {\isasymsubseteq}\ events{\isacharparenright}\ {\isasymand}\isanewline
\ \ \ \ {\isacharparenleft}{\isasymforall}J{\isasymsubseteq}I{\isachardot}\ J\ {\isasymnoteq}\ {\isacharbraceleft}{\isacharbraceright}\ {\isasymlongrightarrow}\ finite\ J\ {\isasymlongrightarrow}\ {\isacharparenleft}{\isasymforall}A{\isasymin}{\isacharparenleft}{\isasymPi}\ i{\isasymin}J{\isachardot}\ F\ i{\isacharparenright}{\isachardot}\isanewline
\ \ \ \ \ \ \ \ \ prob\ {\isacharparenleft}{\isasymInter}j{\isasymin}J{\isachardot}\ A\ j{\isacharparenright}\ {\isacharequal}\ {\isacharparenleft}{\isasymProd}j{\isasymin}J{\isachardot}\ prob\ {\isacharparenleft}A\ j{\isacharparenright}{\isacharparenright}{\isacharparenright}{\isacharparenright}{\isachardoublequoteclose}%
}
\DefineSnippet{indepevents}{
\isacommand{definition}\isamarkupfalse%
\ {\isacharparenleft}\isakeyword{in}\ prob{\isacharunderscore}space{\isacharparenright}\isanewline
\ \ {\isachardoublequoteopen}indep{\isacharunderscore}events\ A\ I\ {\isasymlongleftrightarrow}\ indep{\isacharunderscore}sets\ {\isacharparenleft}{\isasymlambda}i{\isachardot}\ {\isacharbraceleft}A\ i{\isacharbraceright}{\isacharparenright}\ I{\isachardoublequoteclose}%
}
\DefineSnippet{indepvars}{
\isacommand{definition}\isamarkupfalse%
\ {\isacharparenleft}\isakeyword{in}\ prob{\isacharunderscore}space{\isacharparenright}\isanewline
\ \ indep{\isacharunderscore}vars\ {\isacharcolon}{\isacharcolon}\isanewline
\ \ \ \ {\isachardoublequoteopen}{\isacharparenleft}{\isacharprime}i\ {\isasymRightarrow}\ {\isacharprime}b\ measure{\isacharparenright}\ {\isasymRightarrow}\ {\isacharparenleft}{\isacharprime}i\ {\isasymRightarrow}\ {\isacharprime}a\ {\isasymRightarrow}\ {\isacharprime}b{\isacharparenright}\ {\isasymRightarrow}\ {\isacharprime}i\ set\ {\isasymRightarrow}\ bool{\isachardoublequoteclose}\isanewline
\isakeyword{where}\isanewline
\ \ {\isachardoublequoteopen}indep{\isacharunderscore}vars\ M{\isacharprime}\ X\ I\ {\isasymlongleftrightarrow}\isanewline
\ \ {\isacharparenleft}{\isasymforall}i{\isasymin}I{\isachardot}\ random{\isacharunderscore}variable\ {\isacharparenleft}M{\isacharprime}\ i{\isacharparenright}\ {\isacharparenleft}X\ i{\isacharparenright}{\isacharparenright}\ {\isasymand}\isanewline
\ \ indep{\isacharunderscore}sets\ {\isacharparenleft}{\isasymlambda}i{\isachardot}\ {\isacharbraceleft}\ X\ i\ {\isacharminus}{\isacharbackquote}\ A\ {\isasyminter}\ space\ M\ {\isacharbar}\ A{\isachardot}\ A\ {\isasymin}\ sets\ {\isacharparenleft}M{\isacharprime}\ i{\isacharparenright}{\isacharbraceright}{\isacharparenright}\ I{\isachardoublequoteclose}%
}
\DefineSnippet{char}{
\isacommand{definition}\isamarkupfalse%
\isanewline
\ \ char\ {\isacharcolon}{\isacharcolon}\ {\isachardoublequoteopen}real\ measure\ {\isasymRightarrow}\ real\ {\isasymRightarrow}\ complex{\isachardoublequoteclose}\isanewline
\isakeyword{where}\isanewline
\ \ {\isachardoublequoteopen}char\ M\ t\ {\isacharequal}\ LINT\ x{\isacharbar}M{\isachardot}\ iexp\ {\isacharparenleft}t\ {\isacharasterisk}\ x{\isacharparenright}{\isachardoublequoteclose}%
}
\DefineSnippet{normaldistr}{
\isacommand{definition}\isamarkupfalse%
\isanewline
\ \ normal{\isacharunderscore}density\ {\isacharcolon}{\isacharcolon}\ {\isachardoublequoteopen}real\ {\isasymRightarrow}\ real\ {\isasymRightarrow}\ real\ {\isasymRightarrow}\ real{\isachardoublequoteclose}\isanewline
\isakeyword{where}\isanewline
\ \ {\isachardoublequoteopen}normal{\isacharunderscore}density\ {\isasymmu}\ {\isasymsigma}\ x\ {\isacharequal}\isanewline
\ \ \ \ \ {\isadigit{1}}\ {\isacharslash}\ sqrt\ {\isacharparenleft}{\isadigit{2}}\ {\isacharasterisk}\ pi\ {\isacharasterisk}\ {\isasymsigma}\isactrlsup {\isadigit{2}}{\isacharparenright}\ {\isacharasterisk}\ exp\ {\isacharparenleft}{\isacharminus}{\isacharparenleft}x\ {\isacharminus}\ {\isasymmu}{\isacharparenright}\isactrlsup {\isadigit{2}}{\isacharslash}\ {\isacharparenleft}{\isadigit{2}}\ {\isacharasterisk}\ {\isasymsigma}\isactrlsup {\isadigit{2}}{\isacharparenright}{\isacharparenright}{\isachardoublequoteclose}\isanewline
\isanewline
\isacommand{abbreviation}\isamarkupfalse%
\isanewline
\ \ std{\isacharunderscore}normal{\isacharunderscore}density\ {\isacharcolon}{\isacharcolon}\ {\isachardoublequoteopen}real\ {\isasymRightarrow}\ real{\isachardoublequoteclose}\isanewline
\isakeyword{where}\isanewline
\ \ {\isachardoublequoteopen}std{\isacharunderscore}normal{\isacharunderscore}density\ {\isasymequiv}\ normal{\isacharunderscore}density\ {\isadigit{0}}\ {\isadigit{1}}{\isachardoublequoteclose}\isanewline
\isanewline
\isacommand{abbreviation}\isamarkupfalse%
\isanewline
\ \ std{\isacharunderscore}normal{\isacharunderscore}distribution\ {\isacharcolon}{\isacharcolon}\ {\isachardoublequoteopen}real\ measure{\isachardoublequoteclose}\isanewline
\isakeyword{where}\isanewline
\ \ {\isachardoublequoteopen}std{\isacharunderscore}normal{\isacharunderscore}distribution\ {\isasymequiv}\ density\ lborel\ std{\isacharunderscore}normal{\isacharunderscore}density{\isachardoublequoteclose}%
}